\documentclass[a4paper,fleqn,usenatbib]{mnras}

\usepackage{mathptmx}

\usepackage[T1]{fontenc}
\usepackage{ae,aecompl}

\usepackage{graphicx}	
\usepackage{amsmath}	
\usepackage{amssymb}	
\usepackage{cite}
\usepackage{caption}
\usepackage{mathtools}
\usepackage{etoolbox}
\makeatletter
\patchcmd\@combinedblfloats{\box\@outputbox}{\unvbox\@outputbox}{}{%
  \errmessage{\noexpand\@combinedblfloats could not be patched}%
}%
\makeatother
\usepackage{multirow}

\title[Evolution of Bursts in GRO J1744--28]{The Evolution of X-ray Bursts in the ``Bursting Pulsar'' GRO J1744--28}

\author[J. M. C. Court et al.]{
J.M.C. Court$^{1}$\thanks{E-mail: J.M.Court@soton.ac.uk},
D. Altamirano$^{1}$,
A.C. Albayati$^{2}$,
A. Sanna$^{3}$,
T. Belloni$^{4}$,
\newauthor T. Overton$^{5}$,
N. Degenaar$^{6}$,
R. Wijnands$^{6}$,
K. Yamaoka$^{7}$,
A. B. Hill$^{8,1}$,
\newauthor C. Knigge$^{1}$
\\
\\
$^{1}$School of Physics and Astronomy, University of Southampton, Southampton, SO17 1BJ, UK\\
$^{2}$School of Physics and Astronomy, Queen Mary University of London,  London, E1 4NS, UK\\
$^{3}$Dipartimento di Fisica, Universit\`{a} degli Studi di Cagliari, SP Monserrato-Sestu km 0.7, 09042 Monserrato, Italy\\
$^{4}$Osservatorio Astronomico di Brera, Via E. Bianchi 46, 23807 Merate (LC), Italy\\
$^{5}$Department of Physics, Royal Holloway, University of London, Egham, TW20 0EX, UK\\
$^{6}$Anton Pannekoek Institute for Astronomy, University of Amsterdam, Science Park 904, 1098 XH, Amsterdam, The Netherlands\\
$^{7}$Department of Physics, Nagoya University, Aichi 464-8602, Japan\\
$^{8}$HAL24K, Building B.3, Johan Huizingalaan 400, 1066 JS Amsterdam, The Netherlands\\
}

\date{Accepted XXX. Received YYY; in original form ZZZ}

\pubyear{2018}

\begin{document}
\label{firstpage}
\pagerange{\pageref{firstpage}--\pageref{lastpage}}
\maketitle

\begin{abstract}
GRO J1744--28, commonly known as the `Bursting Pulsar', is a low mass X-ray binary containing a neutron star and an evolved giant star. This system, together with the Rapid Burster (MXB 1730-33), are the only two systems that display  the so-called Type II X-ray bursts. These type of bursts, which last for 10s of seconds, are thought to be caused by viscous instabilities in the disk; however the Type II bursts seen in GRO J1744--28 are qualitatively very different from those seen in the archetypal Type II bursting source the Rapid Burster. To understand these differences and to create a framework for future study, we perform a study of all X-ray observations of all 3 known outbursts of the Bursting Pulsar which contained Type II bursts, including a population study of all Type II X-ray bursts seen by \textit{RXTE}. We find that the bursts from this source are best described in four distinct phenomena or `classes' and that the characteristics of the bursts evolve in a predictable way.  We compare our results with what is known for the Rapid Burster and put out results in the context of models that try to explain this phenomena.
\end{abstract}




\section{Introduction}

\par Low Mass X-ray Binaries (hereafter LMXBs) are extremely dynamic astrophysical systems, which exhibit high-amplitude X-ray variability on timescales of milliseconds to years.  In these systems a compact object accretes matter from a stellar companion, either via a stellar wind or via Roche-lobe overflow.  The donated matter spirals in towards the compact object, forming an accretion disk of matter which heats up by friction to temperatures of $\gtrsim1$\,keV.
\par LMXBs are an excellent laboratory in which to explore the behaviour of matter under extreme physical conditions.  In addition to extreme temperatures, the inner portion of an accretion disk is a region of extreme gravity, gas pressure and photon pressure.  If the primary object in the binary is a neutron star, these systems also contain regions of extreme magnetic fields.
\par Many LMXBs containing a neutron star are known to exhibit `bursts'; discrete periods of increased X-ray emission over timescales of seconds.  These bursts are generally categorised as either Type I or Type II, depending on the profile of the burst and its spectral evolution \citep{Hoffman_RB,Lewin_Bursts}.  Type I bursts are caused by accreted matter on the surface of the neutron star reaching a critical pressure and temperature which triggers runaway thermonuclear burning (see e.g. \citealp{Lewin_Bursts,Strohmayer_TypeI}).  They appear in X-ray lightcurves as a sudden increase in intensity, followed by a power-law decay \citep{intZand_Decay}, over a timescale of a few $\sim10$s of seconds.
\par Type II bursts are believed to be caused by viscous instabilities in the accretion disk \citep{Lewin_TypeII}.  However, the exact details of the mechanism responsible for Type II bursts remain unclear.  This type of bursts is more varied in its phenomenological appearance, ranging from near-Gaussian in shape over timescales of $<1$\,s to broad flat-topped lightcurve features which last for $\sim100$\,s (e.g. \citealp{Bagnoli_PopStudy}).
\par Type I X-ray bursts are seen in data from over a hundred neutron star LMXBs, while regular Type II bursts have only been unambiguously identified in two sources: the ``Rapid Burster'' MXB 1730-335 \citep{Lewin_TypeII} and the ``Bursting Pulsar'' GRO J1744--28 \citep{Kouveliotou_BP}.   Isolated Type II bursts may have also been observed in at least one additional X-ray Binary (SMC X-1, \citealp{Angelini_SMC}), but the identification of these features remains unclear.
\par The Type II bursting behaviour in the Rapid Burster has been extensively studied (see e.g. \citealp{Lewin_TypeII,Hoffman_RB}).  \citet{Bagnoli_PopStudy} performed a full population study of all Type II bursts observed in this object by the \textit{Rossi X-ray Timing Explorer} (\textit{RXTE}, \citealp{Bradt_RXTE}).  Their results suggest that gating of the accretion by a strong magnetic field plays some role in the creation of Type II bursts.  To further probe the physics behind Type II X-ray bursts, in this paper we perform a similar population study on bursts from the Bursting Pulsar.
\par The Bursting Pulsar \citep{Paciesas_BPDiscovery} is a system containing a neutron star and a G or K class evolved companion star (e.g. \citealp{Sturner_BPNature,Gosling_BPCompanion,Masetti_BPCompanion}).  The system lies at a distance of $\sim4$--$8$\,kpc in the direction of the Galactic centre (e.g. \citealp{Kouveliotou_BP,Gosling_BPCompanion,Sanna_BP}), and it is the only known pulsar that regularly displays Type II bursts.  The Bursting Pulsar accretes at a high rate: by estimating the accretion rate of the object by measuring how fast the pulsar spins up, \citealp{Sturner_BPNature} found that the Bursting Pulsar accretes at close to the Eddington limit for a neutron star.
\par Unlike in the Rapid Burster, unambiguous Type I bursts have never been observed from the Bursting Pulsar (e.g. \citealp{Giles_BP}, however see also \citealp{Lamb_TypeIBP,Doroshenko_NBFlash}).  Type II bursts were first identified upon discovery in 1995 by the Burst and Transient Source Experiment (BATSE) aboard the \textit{Compton Gamma Ray Observatory} (\textit{CGRO}, \citealp{Gehrels_CGRO}).  Additional outbursts have occurred irregularly; specifically in 1997 and 2014 \citep{Woods_OB2,Kennea_BPOutburst}.  An additional outburst may have occurred in 2017 \citep{Sanna_BPOutburst}, but it was significantly less luminous than previous outbursts and the Bursting Pulsar did not transition to the soft state (such events are referred to as `failed outbursts' or `failed state-transition outbursts', see e.g. \citealp{Sturner_Failed}).
\par Previous work by \citet{Giles_BP} indicated that Type II bursts in the 1995--1996 outburst of the Bursting Pulsar could be separated into a number of distinct populations based on peak flux.  This is a notable difference from the Rapid Burster, in which all Type II bursts have peak fluxes approximately equal to or less than object's Eddington Luminosity \citep{Tan_RBBursts}.  In this paper we expand on the work of \citet{Giles_BP} and analyze \textit{RXTE}, \textit{NuSTAR}, \textit{Chandra}, \textit{XMM-Newton}, \textit{Swift} and \textit{INTEGRAL} data to fully quantify the population of Type II bursts in the Bursting Pulsar during all outbursts in which they were observed.  We study how the bursting in this object evolves over time throughout each outburst, and we link this behaviour to the long-term evolution of the source.  We also perform basic timing, morphology and spectral analysis on bursts, to try and understand the physical processes behind these phenomena.  

\section{Data and Data Analysis}

\par Since discovery, the Bursting Pulsar has undergone three bright outbursts, which began in 1995, 1997 and 2014.  We refer to these outbursts as Outbursts 1, 2 and 3.  We do not consider the outburst in 2017 in this paper, as no Type II bursts were observed during this time, nor do we analyse data taken while the source was in quiesence.  See \citet{Daigne_BPQ}, \citet{Wijnands_BPQ} and \citet{Degenaar_BPQuiescence} for studies of the Bursting Pulsar during quiescence.
\par We analysed data from all X-ray instruments which observed the Bursting Pulsar during these outbursts.  Specifically, we analysed lightcurves, the evolution of hardness ratios as a function of time and of count rate, and performed statistical analysis of properties associated with each individual burst.

\subsection{\textit{RXTE}}

\par We analysed data from the Proportional Counter Array (PCA, \citealp{Jahoda_PCA}) aboard \textit{RXTE} corresponding to the Outbursts 1 \& 2 of the Bursting Pulsar.  This in turn corresponded to observation IDs starting with 10401-01, 20077-01, 20078-01, 20401-01 and 30075-01, between MJDs 50117 and 51225.  This resulted in a total of 743\,ks of data over 300 observations, which we have listed in Appendix \ref{app:obs}. Lightcurve data were extracted from \texttt{fits} files using \texttt{FTOOLS}\footnote{\url{https://heasarc.gsfc.nasa.gov/ftools/ftools_menu.html}}.  Errors were calculated and quoted at the 1$\,\sigma$ level.
\par We also use data from the All-Sky Monitor (ASM, \citealp{Levine_ASM}) to monitor the long-term evolution of the source.  ASM data was taken from MIT's ASM Light Curves Overview website\footnote{\url{http://xte.mit.edu/ASM_lc.html}}.

\subsubsection{Long-Term Evolution}

\par To analyse the long-term evolution of the source during its outbursts, we extracted 2--16\,keV count rates from the \textit{Standard2} data in each observation.  Following \citet{Altamirano_CrabNorm}, we normalised the intensity estimated in each observation by the intensity of the Crab nebula, using the Crab observation that is the closest in time but within the same PCA gain epoch as the observation in question (see \citealp{Jahoda_Calibrate}).

\subsubsection{Burst Identification and Analysis}

\label{sec:burst_diff}

\par To perform population studies on the Type II bursts in the Bursting Pulsar, we first extracted lightcurves from the \texttt{Standard1} data in each observation, as this data is available for all \textit{RXTE} observations.  We used our own software\footnote{\url{https://github.com/jmcourt/pantheon}, \citep{Court_PANTHEON}} to search these lightcurves and return a list of individual bursts, using the algorithm described in Appendix A of \citet{Court_IGRClasses}.  We manually cleaned spurious detections from our sample.  We defined a `burst' as an event that lasted at least 3 seconds during which the 1\,s binned count rate exceeded 3 standard deviations above the persistent emission level and reached a maximum of at least five standard devations above the persistent emission level.  We did not subtract background, as all count rate-related parameters we analyse are persistent emission subtracted, automatically removing background contribution.
\par During the analysis, we discovered a number of different ``classes'', similar to the multiple classes of burst described by \citet{Giles_BP}.  Our classes varied significantly in terms of overall structure, and as such needed to be treated separately; we show representative lightcurves from our classes in Figure \ref{fig:classes}.  These classes were separated from one another by a number of criteria including peak count rate and recurrence time (the time between peaks of consecutive bursts).
\par The vast majority of detected bursts resembled the Type II seen in the Rapid Burster (referred to as `Normal Bursts' in Section \ref{sec:Results}) in terms of shape, duration and amplitude.  We rebinned the data corresponding to these Normal Bursts to 0.5\,s.  We sampled the persistent emission before the burst, and defined the start of the burst as the first point at which count rate exceeded 5 standard deviations above the persistent emission before the burst.  The end of the burst was defined similarly, but instead sampling the persistent emission after the burst; by doing this, we avoid making the implicit assumption that the persistent emission is equal before and after the burst.  We fitted phenomenologically-motivated lightcurve models to each of these bursts (described in detail in Section \ref{sec:struc}), and used these fits to extract a number of parameters which characterise the shape and energetics of a burst (such as burst duration, total photon counts associated with a burst and persistent emission count rate).
\par Due to the high peak count rates of Normal Bursts, data were affected by dead-time (compare e.g. \textit{GRANAT} data presented in \citealp{Sazonov_BPGranat}).  We calculate the approximate Dead-Time Factors (DTFs) for a number of the brightest Normal Bursts in our sample, using 1\,s binned data, using the following formula in the \textit{RXTE} Cookbook\footnote{\url{https://heasarc.gsfc.nasa.gov/docs/xte/recipes/pca_deadtime.html}}:
\begin{equation}
\Delta=\frac{C_{Xe}+C_{Vp}+C_{Rc}+15C_{VL}}{N_{PCU}}\times10^{-5}
\end{equation}
Where $\Delta$ is the fractional detector deadtime, $C_{Xe}$ is the Good Xenon count rate, $C_{Vp}$ is the coincident event count rate, $C_{Rc}$ is the propane layer count rate, $C_{VL}$ is the very large event count rate and $N_{PCU}$ is the number of PCUs active at the time.
\par We estimate that dead-time effects reduce the peak count rates by no more than $\sim12$\%; however, due to the sharply-peaked nature of bursts from the Bursting Pulsar, the deadtime effect depends on the binning used.  Due to this ambiguity we do not correct for dead-time in Normal Bursts.  The dead-time corrections required for the count rates seen in other classes of burst are minimal, as they are orders of magnitude fainter \citep{Giles_BP}.
\par To test for correlations between parameters in a model-independent way, we used the Spearman's Rank correlation coefficient (as available in \texttt{Scipy}, \citealp{NumPy}).  This metric only tests the hypothesis that an increase in the value of one parameter is likely to correspond to an increase in the value of another parameter, and it is not affected by the shape of the monotonic correlation to be measured.  Although dead-time effects lead to artificially low count rates being reported, a higher intensity still corresponds to a higher reported count rate.   As such, using this correlation coefficient removed the effects of dead-time on our detection of any correlations.
\par To calculate the distribution of recurrence times between consecutive bursts, we considered observations containing multiple bursts.  If fewer than 25\,s of data gap exists between a pair of bursts, we considered them to be consecutive and added their recurrence time to the distribution.  We choose this maximum gap size as this is approximately the timescale over which a Normal Burst occurs.
\par To perform basic phenomelogical analysis of the spectral behaviour of these bursts, we divided our data into two energy bands when \texttt{SB\_62us\_0\_23\_500ms} and \texttt{SB\_62us\_24\_249\_500ms} mode data were available: A (PCA channels 0--23, corresponding to $\sim2$--$7$\,keV\footnote{In \textit{RXTE} gain epoch 1, corresponding to dates before MJD 50163.  This corresponds to $\sim2$--$9$\,keV in epoch 2 (MJDs 50163--50188) and $\sim2$--$10$\,keV in epoch 3 (MJDs 50188--51259).}) and B (channels 24--249, corresponding to $\sim8$--$60$\,keV\footnote{In \textit{RXTE} gain epoch 1.  This corresponds to $\sim9$--$60$\,keV in epoch 2 and $\sim10$--$60$\,keV in epoch 3.}).  The evolution of colour (defined as the ratio of the count rates in B and A) throughout a burst could then be studied.  Due to the very high count rates during Normal Bursts, we did not correct for background.  During fainter types of burst we estimate the background in different energy bands by subtracting count rates from \textit{RXTE} observation 30075-01-26-00 of this region, when the source was inactive.  Unlike using the \textit{RXTE} background model, this method subtracts the contributions from other sources in the field.  However, as it is unclear whether any of the rest of these sources are variable, the absolute values of colours we quote should be treated with caution.  We created hardness-intensity diagrams to search for evidence of hysteretic loops in hardness-intensity space.
\par Following \citet{Bagnoli_PopStudy}, we used the total number of persistent emission-subtracted counts as a proxy for fluence for all bursts other than Normal Bursts.  As the contribution of the background does not change much during a single observation, this method also automatically subtracts background counts from our results.

\subsubsection{Detecting Pulsations}

\par The Bursting Pulsar is situated in a very dense region of the sky close to the Galactic centre, and so several additional objects also fall within the 1$^\circ$ \textit{RXTE}/PCA field of view.  Therefore it is important to confirm that the variability we observe in our data does in fact originate from the Bursting Pulsar.
\par To ascertain that all bursts considered in this study are from the Bursting Pulsar, we analyse the coherent X-ray pulse at the pulsar spin frequency to confirm that the source was active.  We first corrected the photon time of arrivals (ToA) of the \textit{RXTE} PCA dataset, and barycentre this data using the \texttt{faxbary} tool (DE-405 Solar System ephemeris).  We corrected for the binary motion by using the orbital parameters reported by \citet{Finger_Pulse}.
\par For each PCA observation we investigated the presence of the $\sim 2.14$\,Hz coherent pulsation by performing an epoch-folding search of the data using 16 phase bins and starting with the spin frequency value $\nu=2.141004$ Hz, corresponding to the spin frequency measured from the 1996 outburst of the source \citep{Finger_Pulse}, with a frequency step of $10^{-5}$\,Hz for 10001 total steps. We detected X-ray coherent pulsations in all PCA observations performed during Outbursts 1 \& 2.

\subsection{\textit{Swift}}
\par In this study, we made use of data from the X-Ray Telescope (XRT, \citealp{Burrows_XRT}) and the Burst Alert Telescope (BAT, \citealp{Krimm_BAT}) aboard the Neil Gehrels Swift Observatory (\textit{Swift}, \citealp{Gehrels_Swift}).  We extracted a long-term 0.3--10\,keV \textit{Swift}/XRT lightcurve of Outburst 3 using the lightcurve generator provided by the UK Swift Science Data Centre (UKSSDC, \citealp{Evans_Swift1}).  We also make use of \textit{Swift}/BAT lightcurves from the Swift/BAT Hard X-ray Transient website\footnote{\url{https://swift.gsfc.nasa.gov/results/transients/}} (see \citealp{Krimm_BAT}).

\subsection{\textit{INTEGRAL}}

\par We also made use of data from the Imager on Board \textit{INTEGRAL} \citep{Winkler_IBIS}.  We extracted 17.3--80\,keV IBIS/ISGRI lightcurves of the Bursting Pulsar during Outburst 3 using the \textit{INTEGRAL} Heavens portal.  This is provided by the \textit{INTEGRAL} Science Data Centre \citep{Lubinski_Heavens}.

\subsection{\textit{Chandra}}

\par The Bursting Pulsar was targeted with \textit{Chandra} \citep{Weisskopf_Chandra} three times during Outburst 3 (Table \ref{tab:Chandra}).  One of these observations (OBSID 16596) was taken simultaneously with a \textit{NuSTAR} observation (80002017004).  In all three observations data were obtained with the High Energy Transmission Grating (HETG), where the incoming light was dispersed onto the ACIS-S \citep{Garmire_ACIS} array. The ACIS-S was operated in continued clocking (CC) mode to minimize the effects of pile-up. The  Chandra/HETG observations were analysed using standard tools available within \texttt{ciao} v. 4.5 \citep{Fruscione_Ciao}. We extracted 1\,s binned light curves from the \texttt{evt2} data using \texttt{dmextract}, where we combined the first order positive and negative grating data from both the Medium Energy Grating (MEG; 0.4-5 keV) and the High Energy Grating (HEG; 0.8--8\,keV).

\begin{table}
\centering
\begin{tabular}{lllllll}
\hline
\hline
\scriptsize  OBSID &\scriptsize Exposure (ks) &\scriptsize MJD &\scriptsize Reference \\
\hline
16596  	& 10 &  56719      &   \citet{Younes_Expo} \\
16605  	& 35 &   56745    &    \citet{Degenaar_BPSpec}\\
16606  	& 35 &   56747    &    \citet{Degenaar_BPSpec}\\
\hline
\hline
\end{tabular}
\caption{Information on the three \textit{Chandra} observations of the Bursting Pulsar during Outburst 3.  All other observations of the Bursting Pulsar in the Chandra archive were obtained at times that the source was in quiescence.}
\label{tab:Chandra}
\end{table}

\subsection{\textit{XMM-Newton}}

\par A single pointed \textit{XMM-Newton} observation of the Bursting Pulsar was taken during Outburst 3 on MJD 56722 (OBSID 0729560401) for 85\,ks.  We extracted a 0.5--10\,keV lightcurve from EPIC-PN at 1\,s resolution using \texttt{xmmsas} version 15.0.0.  During this observation, EPIC-PN was operating in Fast Timing mode.  We use EPIC-PN as the statistics are better than in MOS1 or MOS2.

\subsection{\textit{Suzaku}}
\par \textit{Suzaku} \citep{Mitsuda_Suzaku} observed the Bursting Pulsar once during Outburst 3 on MJD 56740 (OBSID 908004010).
\par To create a lightcurve, we reprocessed and screened data from the X-ray Imaging Spectrometer (XIS, \citealp{Koyama_XIS})using the \texttt{aepipeline} script and the latest calibration database released on June 7, 2016.  The attitude correction for the thermal wobbling was made by \texttt{aeattcor2} and \texttt{xiscoord} \citep{Uchiyama_SuzPSF}. The source was extracted within a radius of 250 pixels corresponding to 260'' from the image center.  The background was extracted from two regions near either end of the XIS chip, and subtracted from the source.

\subsection{\textit{NuSTAR}}

\par \textit{NuSTAR} \citep{Harrison_NuSTAR} consists of two grazing-incident X-ray telescopes.  These instruments only observed the Bursting Pulsar three times during its outbursts, all times in Outburst 3.  One of these observations was taken while the Bursting Pulsar was not showing X-ray bursts, and the other two are shown in Table \ref{tab:NuS}.  We extracted lightcurves from both of these observations using \texttt{nupipeline} and \texttt{nuproducts}, following standard procedures\footnote{See \url{https://www.cosmos.esa.int/web/xmm-newton/sas-threads}.}.

\begin{table}
\centering
\begin{tabular}{lllllll}
\hline
\hline
\scriptsize  OBSID &\scriptsize Exposure (ks) &\scriptsize MJD &\scriptsize Reference \\
\hline
80002017002 	& 29 & 56703 &  \citet{Dai_Hlags}  \\
80002017004 	& 9 & 56719 & \citet{Younes_Expo}\\
\hline
\hline
\end{tabular}
\caption{Information on the two \textit{NuSTAR} observations of the Bursting Pulsar during the main part of Outburst 3.}
\label{tab:NuS}
\end{table}

\section{Results}
\label{sec:Results}

\subsection{Outburst Evolution}

\par
\begin{figure*}
  \centering
  \includegraphics[width=.9\linewidth, trim={0cm 0 0cm 0},clip]{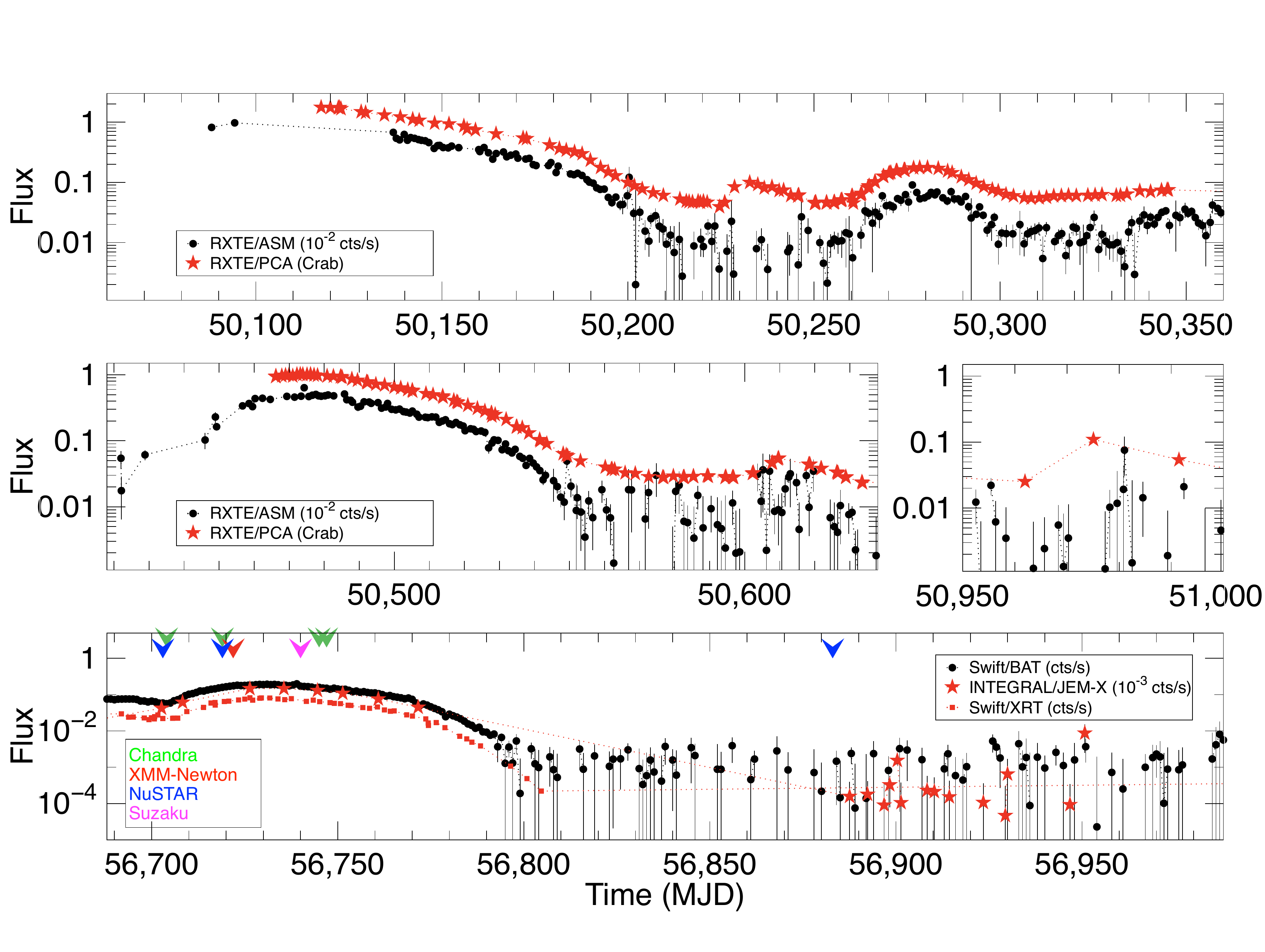}
  \caption{\small Comparisons of the three outbursts of the Bursting Pulsar reported on in this paper.  Times corresponding to pointed observations with \textit{Chandra}, \textit{NuSTAR}, \textit{Suzaku}, \textit{Swift} and \textit{XMM-Newton} are marked.}
  \label{fig:global_ob}
\end{figure*}

 We show the long-term monitoring lightcurves of Outbursts 1,2 and 3 in Figure \ref{fig:global_ob}, as well as mark the dates of pointed observations with various instruments.

\par The Bursting Pulsar was discovered already in outburst on December 12 1995 \citep{Fishman_Discovery}; BATSE data suggest that this outburst began several days earlier on December 3 \citep{Paciesas_BPDiscovery,Bildsten_Rev}.  The main outburst ended around May 10 1996 \citep{Woods_PulseBursts}.  We show the global lightcurve of this outburst in Figure \ref{fig:global_ob}, Panel 1.  As \textit{RXTE} did not observe the object before or during the peak of Outburst, we can only obtain a lower limit of $\sim1.75$\,Crab for the peak 2--16\,keV flux.
\par There are at least two major rebrightening events in the tail of Outburst 1, which can be seen clearly  in Figure \ref{fig:global_ob} centred at MJDs of $\sim50235$ and $\sim50280$.  During these rebrightening events, the 2--16\,keV flux peaked at $\sim0.10$ and $\sim0.18$\,Crab respectively.

\par Outburst 2 began on December 1 1996 and ended around April 7 1997 \citep{Woods_OB2}.  The 2--16\,keV flux peaked at 1.02\,Crab on MJD 50473; we show the global lightcurve of this outburst in Figure \ref{fig:global_ob}, Panel 2.  Type II bursts are seen in \textit{RXTE}/PCA lightcurves from Outburst 2 between MJDs 50466 and 50544.  One rebrightening event occurred during the tail of Outburst 2, centred at an MJD of $\sim50615$ with a peak 2--16\,keV flux of $\sim54$\,mCrab.  A second possible rebrightening event occurs at MJD 50975, with a peak 2--16\,keV flux of 11\,mCrab, but the cadence of \textit{RXTE}/PCA observations was too low to unambiguously confirm the existence of a reflare at this time.

\par Outburst 3 began on January 31, 2014 \citep{Negoro_OB3,Kennea_BPOutburst} and ended around April 23 (e.g. \citealp{Dai_OB3}).  The daily 0.3--10\,keV Swift/XRT rate peaked at 81\,cts\,s$^{-1}$ on MJD 56729, corresponding to 0.4\,Crab.  We show the global lightcurve of this outburst in Figure \ref{fig:global_ob}, Panel 3.
\par During the main part of Outburst 3, \textit{Swift}, \textit{XMM-Newton} and \textit{Suzaku} made one pointed observation each, \textit{Chandra} made four observations, and \textit{NuSTAR} made three observations.  The \textit{Chandra} observation on March 3 2014 was made simultaneously with one of the \textit{NuSTAR} observations (see \citealp{Younes_Expo}).  After the main part of the outburst, the source was not well-monitored, although it remained detectable by \textit{Swift}/BAT, and it is unclear whether any rebrightening events occured.  A single \textit{NuSTAR} observation was made during the outburst tail on August 14 2014.

\par As can be seen in Figure \ref{fig:global_ob}, the main section of all three outburst follow a common profile, over a timescale of $\sim150$ days.  A notable difference between outbursts 1 \& 2 is the number of rebrightening events; while we find two rebrightening events associated with Outburst 1, we only find one associated with Outburst 2 unless we assume the event at MJD 50975 is associated with the outburst.  Additionally, Outburst 2 was at least a factor $\sim1.7$ fainter at its peak than Outburst 1 (see also \citealp{Woods_OB2}), while Outburst 3 was a factor of $\gtrsim4$ fainter at peak than Outburst 1.

\subsubsection{Pulsations}

\par We found pulsations in PCA data throughout the entirety of Outbursts 1 \& 2.  This confirms that the Bursting Pulsar was active as an X-ray pulsar in all of our observations, leading us to conclude that all the types of Type II bursts we see are from the Bursting Pulsar.  


\subsubsection{Bursting Behaviour}

\par Bursts are seen in \textit{RXTE}/PCA lightcurves from the start of the Outburst 1 (e.g. \citealp{Kouveliotou_BP}).  These Type II bursts occur until around MJD 50200, as the source flux falls below $\sim0.1$\,Crab in the 2--16\,keV band.  
\par During the latter part of the first rebrightening after Outburst 1, between MJDs 50238 and 50246, we found Type II-like bursts with amplitudes $\sim2$ orders of magnitude smaller than those found during the main outburst event.  These gradually increased in frequency throughout this period of time until evolving into a period of highly structured variability which persisted until MJD 50261.
\par In Outburst 2, we found Type II bursts occuring between MJDs $\sim50466$ and $50542$.  Low-amplitude Type II-like bursts are seen during the latter stages of the main outburst, between MJDs 50562 and 50577.  These again evolve into a period of highly structured variability; this persists until MJD 50618, just after the peak of the rebrightening event.
\par High-amplitude Type II bursts were also seen in Outburst 3 (e.g. \citealp{Linares_NewBurst}).  As no soft ($\lesssim10\,$keV) X-ray instrument was monitoring the Bursting Pulsar during the latter part of Outburst 3, it is unknown whether this Outburst showed the lower-amplitude bursting behaviour seen at the end of Outbursts 1 \& 2.  Low amplitude bursting behaviour is not seen in the pointed \textit{NuStar} observation which was made during this time.

\subsection{Categorizing Bursts}

\label{sec:classes}

\par We find that bursts in the Bursting Pulsar fall into a number of discrete classes, lightcurves from which we show in Figure \ref{fig:classes}.  These classes are as follows:

\begin{figure}
  \centering
  \includegraphics[width=.9\linewidth, trim={0.7cm 0.1cm 1.5cm 1.4cm},clip]{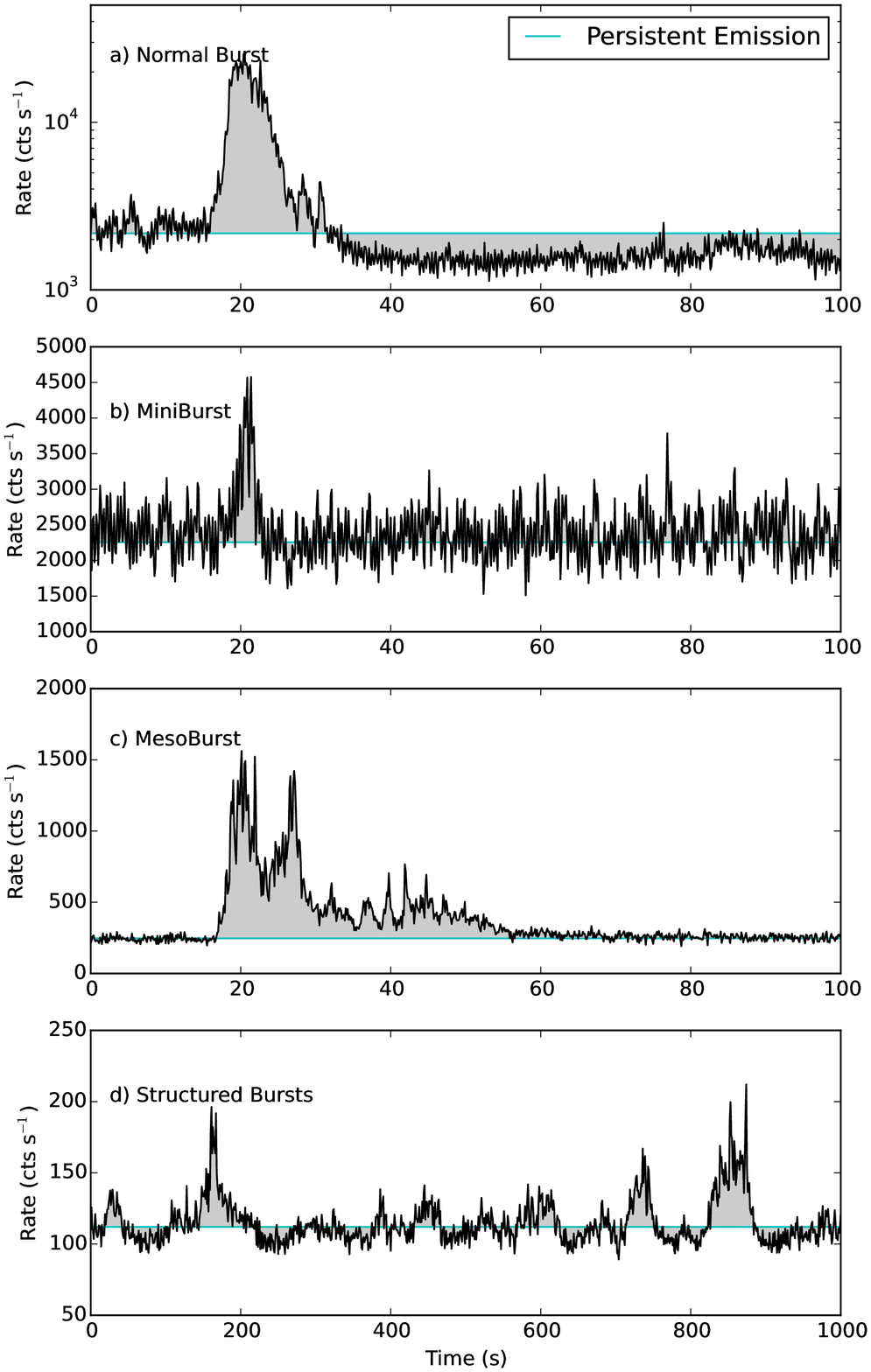}
  \caption{\small 2--49\,keV lightcurves for the four classes of bursting behaviour identified in this paper: \textbf{a)} Normal Burst, \textbf{b)} Miniburst, \textbf{c)} Mesoburst, \textbf{d)} Structured Bursts.  Note that Panel \textbf{d} is plotted with a different time scaling to the other panels so as to better show the behaviour of Structured Bursting.  On all figures the median count rate, which we use as a proxy for the persistent emission, is plotted in cyan.  Lightcurves \textbf{a}-\textbf{c} are binned to 0.125\,s, while lightcurve \textbf{d} is binned to 1\,s.}
  \label{fig:classes}
\end{figure}

\begin{itemize}
\item Normal Bursts (Figure \ref{fig:classes}, Panel a): the brightest bursts seen from this source, with peak count 1\,s binned rates of $\sim10000$\,cts\,s$^{-1}$\,PCU$^{-1}$, and recurrence timescales of order $\sim1000$\,s.  These bursts are roughly Gaussian in shape with durations of $\sim10$\,s, and are followed by a `dip' in the persistent emission count rate with a duration of order 100\,s (see also e.g. \citealp{Giles_BP}).
\item Minibursts (Figure \ref{fig:classes}, Panel b): faint bursts with 1\,s-binned peak count rates of $\sim2$ times the persistent emission count rate.  Minibursts are variable, with duration timescales between $\sim5$--50\,s.  These bursts are also sometimes followed by dips similar to those seen after Normal Bursts.
\item Mesobursts (Figure \ref{fig:classes}, Panel c): Type II-like bursts.  These bursts differ from Normal Bursts in that they do not show well-defined subsequent `dips'.  They are also fainter than Normal Bursts, with peak count 1\,s binned count rates of $\sim1000$\,cts\,s$^{-1}$\,PCU$^{-1}$.  Their burst profiles show fast rises on timescales of seconds, with slower decays and overall durations of $\sim50$\,s.  The structure of the bursts is very non-Gaussian, appearing as a small forest of peaks in lightcurves.
\item Structured Bursts (Figure \ref{fig:classes}, Panel d): the most complex class of bursting behaviour we observe from the Bursting Pulsar, consisting of patterns of flares and dips in the X-ray lightcurve.  The amplitudes of individual flares are similar to those of the faintest Mesobursts.  The recurrence timescale is of the order of the timescale of an individual flare, meaning that is it difficult to fully separate individual flares of this class.
\end{itemize}

\par
\begin{figure}
  \centering
  \includegraphics[width=.9\linewidth, trim={1.3cm 0cm 2.0cm 0cm},clip]{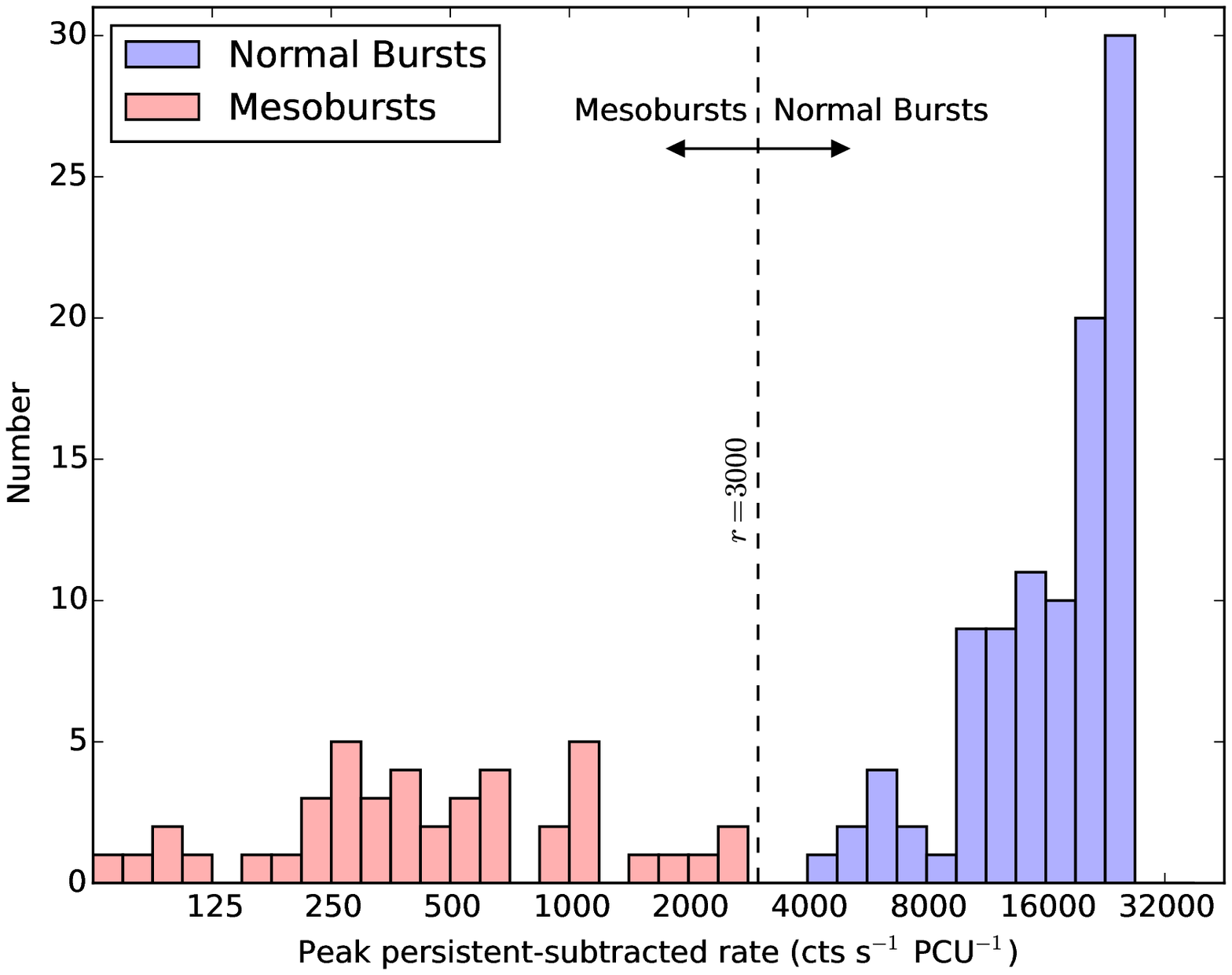}
  \includegraphics[width=.9\linewidth, trim={1.3cm 0cm 2.0cm 0cm},clip]{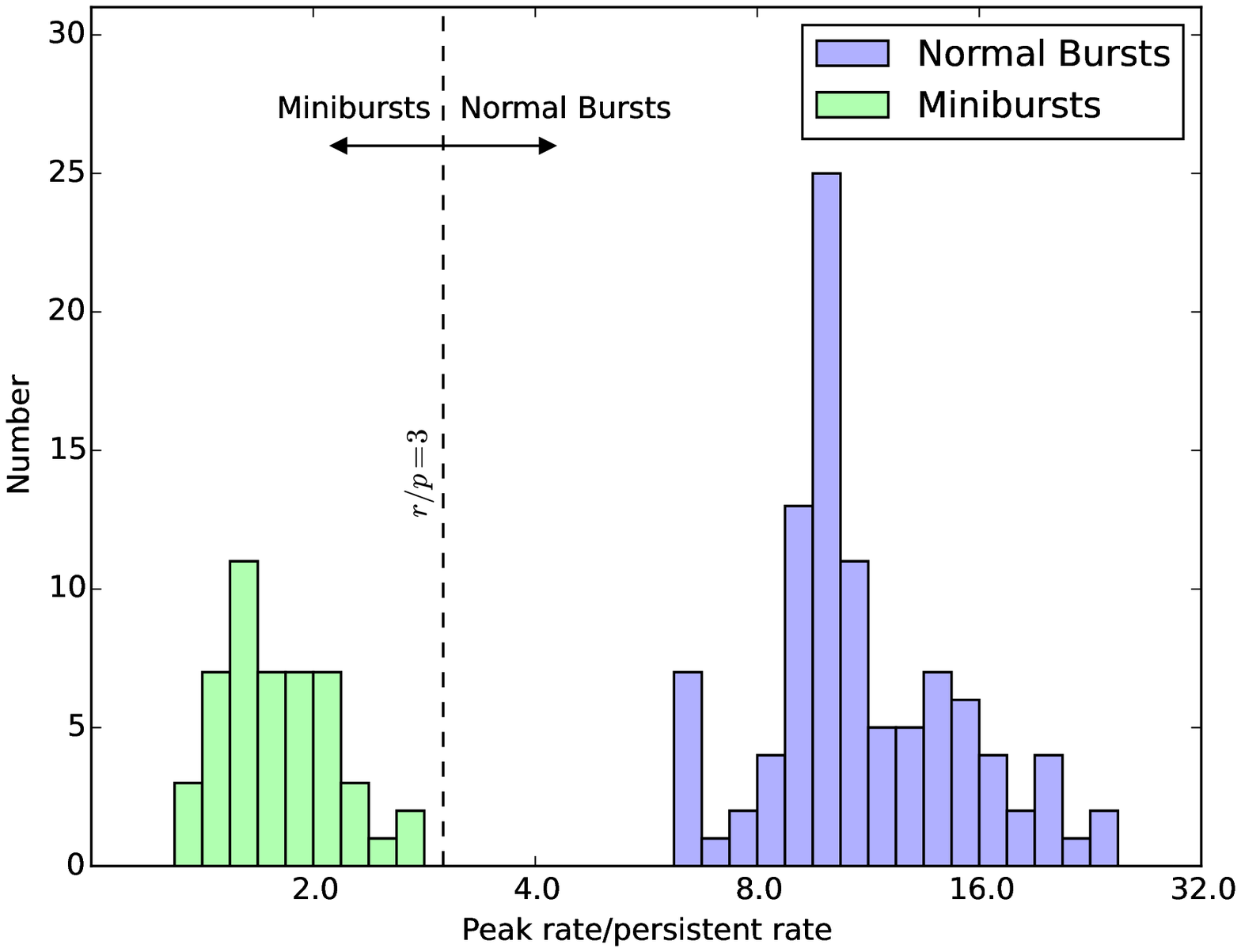}
  \caption{\small \textbf{Upper Panel:} A histogram of the peak 1\,s binned peak count rates of the joint population of all Normal and Mesobursts seen by \textit{RXTE}.  The dashed line indicates the position of the threshold above which we consider a Type II-like burst to be a Normal Burst.  The resultant split of the population into Normal and Mesobursts is indicated by blue and red shading respectively.  The skewed shape of the distribution of Normal Bursts is due to the effects of dead-time putting an effective cap on their maximum observed intensity.
 \textbf{Lower Panel:} A histogram of the peak 1\,s binned peak count rates of the joint population of all Normal and Minibursts seen by \textit{RXTE}, divided by the persistent emission count rate at that time.  The dashed line indicates the position of the threshold below which we consider a burst to be a Miniburst.  The resultant split of the population into Normal and Minibursts is indicated by blue and green shading respectively.
  Note that the $x$-axis of both plots is logarithmic, and so number density is not preserved.}
  \label{fig:jointhist}
\end{figure}

 In the upper panel of Figure \ref{fig:jointhist} we show a histogram of persistent-emission-subtracted peak count rates for all Normal and Mesobursts observed by \textit{RXTE}.  We split these two classes based on the bimodal distribution in peak count rate as well as the lack of dips in Mesobursts.
\par In the lower panel of Figure \ref{fig:jointhist}, we show the histogram of peak count rates for all Normal and Minibursts observed by \textit{RXTE} as a fraction of the persistent emission at that time.  We split these two classes based on the strongly bimodal distribution in fractional amplitude.

\par We also find 6 bursts with fast ($\sim1$\,s) rises and exponential decays that occur during the lowest flux regions of the outburst ($\lesssim50$\,mCrab).  \citet{Strohmayer_BPFieldTypeI} and \citet{Galloway_TypeI} have previously identified these bursts as being Type I X-ray bursts from another source in the \textit{RXTE} field of view.  To show that these unrelated Type I bursts would not be confused with Minibursts, we add examples of the Type I bursts to lightcurves from observations containing Minibursts.  We find that the peak count rates in Type I bursts are roughly equal to the amplitude of the noise in the persistent flux in these observations, hence they would not be detected by our algorithms.
\par
\begin{figure*}
  \centering
  \includegraphics[width=.9\linewidth, trim={9.5cm 0cm 10cm 0cm},clip]{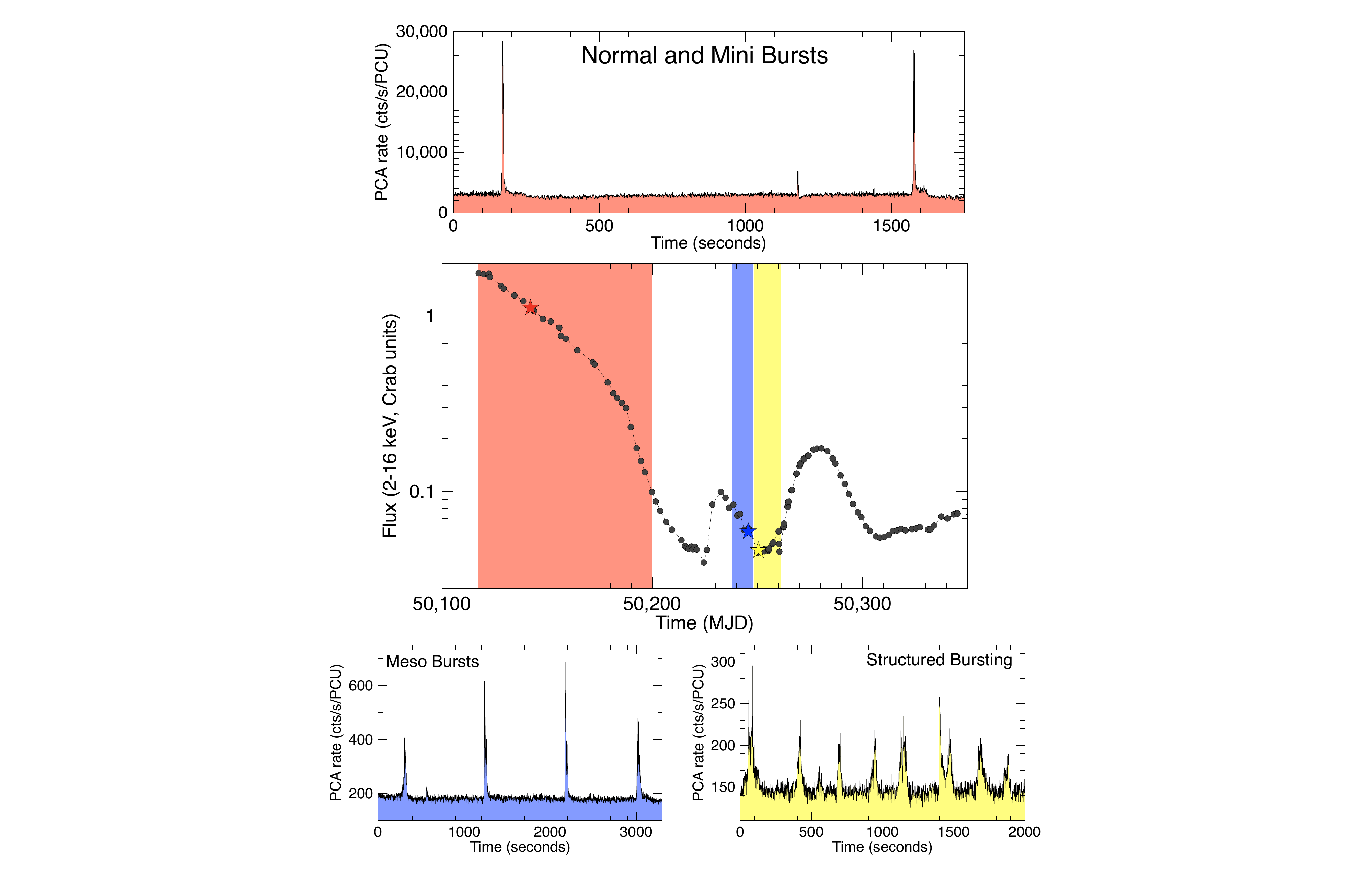}
  \caption{\small Central panel shows the global 2--16\,keV \textit{RXTE}/PCA lightcurve of the 1995--1996 outburst of the Bursting Pulsar, highlighting periods of time during which Mesobursts (blue) Structured Bursts (yellow) or Normal and Mini bursts (red) are observed.  A single Mesoburst was also observed on MJD 50253, during the period of the outburst highlighted in yellow (see Figure \ref{fig:meso_in_struc}).  Other panels show example lightcurves which contain the aforementioned types of bursting behaviour.  See section \ref{sec:classes} for a detailed treatment of burst classification.  Fluxes reported in units of Crab.}
  \label{fig:ob_evo1}
\end{figure*}

\begin{figure*}
  \centering
  \includegraphics[width=.9\linewidth, trim={9.5cm 0cm 10cm 0cm},clip]{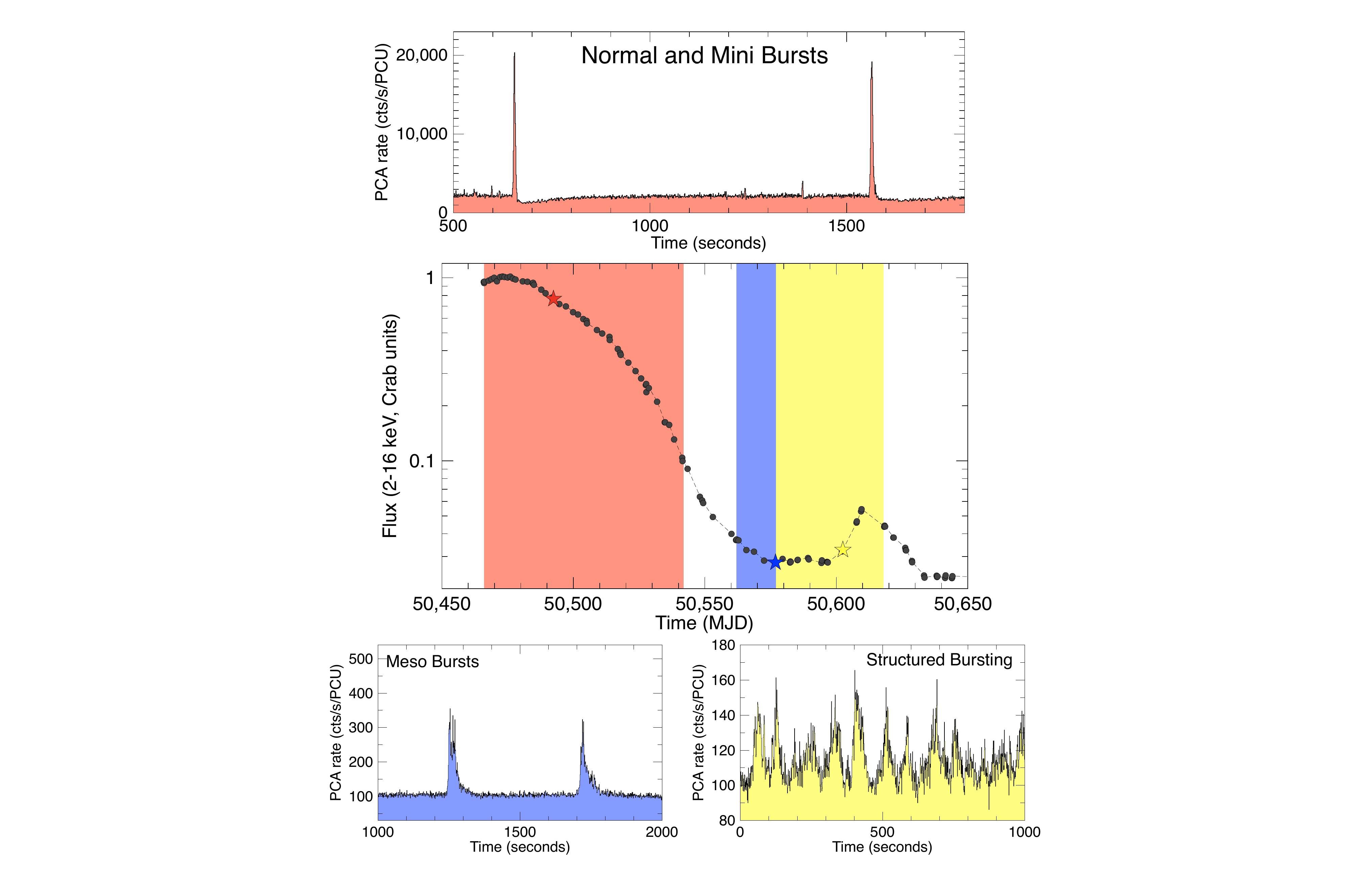}
  \caption{\small Central panel shows the global 2--16\,keV \textit{RXTE}/PCA lightcurve of the 1997--1999 outburst of the Bursting Pulsar, highlighting periods of time during which Mesobursts (blue) Structured Bursts (yellow) or Normal and Mini bursts (red) are observed.  Other panels show example lightcurves which contain the aforementioned types of bursting behaviour.}
  \label{fig:ob_evo2}
\end{figure*}

 We show when in Outbursts 1 \& 2 each type of burst was observed in Figures \ref{fig:ob_evo1} and \ref{fig:ob_evo2} respectively.  Normal Bursts and Minibursts (red) occur during the same periods of time from around the peak of an outburst until the persistent emission falls beneath $\sim0.1$\,Crab; assuming an Eddington Limit of $\sim1$\,Crab (e.g \citealp{Sazonov_BPGranat}), this corresponds to an Eddington ratio of $\sim0.1$.  After this point, bursting is not observed for a few tens of days.  Mesobursts (blue) begin at the end of a rebrightening event in Outburst 1 and during the final days of the main part of the outburst in Outburst 2.  Structured Bursts (yellow) occur during the first part of a rebrightening event in both outbursts.  Although there was a second rebrightening event after Outburst 1, neither Mesobursts nor Structured Bursts were observed at this time.  Based on this separation, as well as differences in structure, we treat each class of burst separately below.

\subsection{Normal Bursts}

\par We define Normal Bursts as the set of all bursts with a persistent-emission-subtracted peak 1\,s binned \textit{RXTE}/PCA-equivalent count rate above 3000\,cts\,s$^{-1}$\,PCU$^{-1}$.  Normal Bursts account for 99 out of the 190\footnote{This number does not include Structured Bursts as their complex structure makes them difficult to separate.} bursts identified for this study.  They are observed during all three outbursts covered in this study.  They occurred between MJDs 50117 and 50200 in Outburst 1, and between 50466 and 50542 in Outburst 2; during these intervals, \textit{RXTE} observed the source for a total of 192\,ks.  See Table \ref{tab:staretimes} to compare these with numbers for the other classes of burst identified in this study.  They occur during the same time intervals during which Minibursts are present.  In both of these outbursts, the region of Normal and Minibursts correspond to the time between the peak of the outburst and and the time that the persistent intensity falls below $\sim0.1$\,Crab.

\begin{table}
\centering
\begin{tabular}{llll}
\hline
\hline
\scriptsize  Bursting Mode &\scriptsize Bursts &\scriptsize Total Exposure (ks) &\scriptsize Duration (d) \\
\hline
Normal Bursts & 99  & 192 & 76\\
Minibursts & 48 & 192  & 76\\
Mesobursts & 43 &44 &25\\
Structured Bursts & - &80 &54 \\
\hline
\hline
\end{tabular}
\caption{Statistics on the population of bursts we use for this study, as well as the duration and integrated \textit{RXTE}/PCA exposure time of each mode of bursting.  All numbers are the sum of values for Outbursts 1 and 2.  As Normal and Minibursts happen during the same period of time in each outburst, the exposure time and mode duration for these classes of bursting are equal.}
\label{tab:staretimes}
\end{table}

\subsubsection{Recurrence Time}

\par
\begin{figure}
  \centering
  \includegraphics[width=.9\linewidth, trim={0.4cm 0 1.1cm 0},clip]{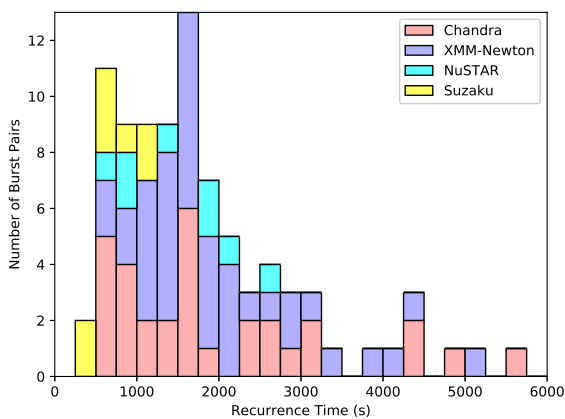}
  \caption{\small The distribution of recurrence times between consecutive Normal Bursts seen in pointed \textit{Chandra}, \textit{XMM-Newton}, \textit{NuSTAR} and \textit{Suzaku} observations of Outburst 3 of the Bursting Pulsar.  Distributions of bursts observed by different instruments are stacked on top of each other and colour coded.}
  \label{fig:sep}
\end{figure}

 Using Outburst 3 data from \textit{Chandra}, \textit{XMM-Newton}, \textit{NuSTAR} and \textit{Suzaku}, we find minimum and maximum recurrence times of $\sim345$ and $\sim5660$\,s respectively\footnote{To avoid double-counting peak pairs, we do not use \textit{NuSTAR} observation 80002017004, which was taken simultaneously with \textit{Chandra} observation 16596.}.  We show the histogram of recurrence times from Outburst 3 in Figure \ref{fig:sep}, showing which parts of the distribution were observed with which observatory.  Compared to data from \textit{Chandra} and \textit{XMM-Newton}, data from \textit{Suzaku} generally suggests shorter recurrence times.  This is likely due to \textit{Suzaku} observations consisting of a number of $\sim2$\,ks windows; as this number is of the same order of magnitude as the recurrence time between bursts, there is a strong selection effect against high recurrence times in the \textit{Suzaku} dataset.
\par From the \textit{RXTE} data we find minimum and maximum burst recurrence times of $\sim250$ and $\sim2510$\,s during Outburst 1, and minimum and maximum recurrence times of $\sim250$ and $\sim2340$\,s during Outburst 2.  As the length of an \textit{RXTE} pointing ($\lesssim3$\,ks) is also of the same order of magnitude as the recurrence time between bursts, selection effects bias us against sampling pairs of bursts with longer recurrence times, and hence this upper value is likely an underestimate.

\par To test whether consecutive bursts are independent events, we tested the hypothesis that bursts are randomly distributed in time in a Poisson distribution \citep{Poisson_Distribution}.  Assuming our hypothesis, as well as assuming that the frequency of Normal Bursts does not change during an outburst (e.g. \citealp{Aptekar_Recur}), we could concatenate different observations and the resultant distribution of burst times will still be Poissonian.  For each of Outbursts 1 \& 2, we concatenated all \textit{RXTE} data during the Normal Bursting part of the outburst into a single lightcurve.  We split our lightcurves into windows of length $w$ and counted how many bursts were in each, forming a histogram of number of bursts per window.  We fit this histogram with a Poisson probability density function, obtaining the value $\lambda$ which is the mean number of bursts in a time $w$.  $\lambda/w$ is therefore an expression of the true burst frequency per unit time, and should be independent of our choice of $w$.  We tried values of $w$ between 100 and 10000\,s for both outbursts, and found that in all cases $\lambda/w$ depends strongly on $w$.  Therefore our assumptions cannot both be valid, and we rejected the hypothesis that these bursts are from a Poisson distribution with constant $\lambda$.  This in turn suggests at least one of the following must be correct:
\begin{enumerate}
\item The average recurrence time of bursts was not constant throughout the outburst.  Or:
\item The arrival time of a given burst depends on the arrival time of the preceding burst, and therefore bursts are not independent events.
\end{enumerate}

\subsubsection{Burst Structure}

\label{sec:struc}

\par
\begin{figure}
  \centering
  \includegraphics[width=.9\linewidth, trim={0.4cm 0 1.1cm 0},clip]{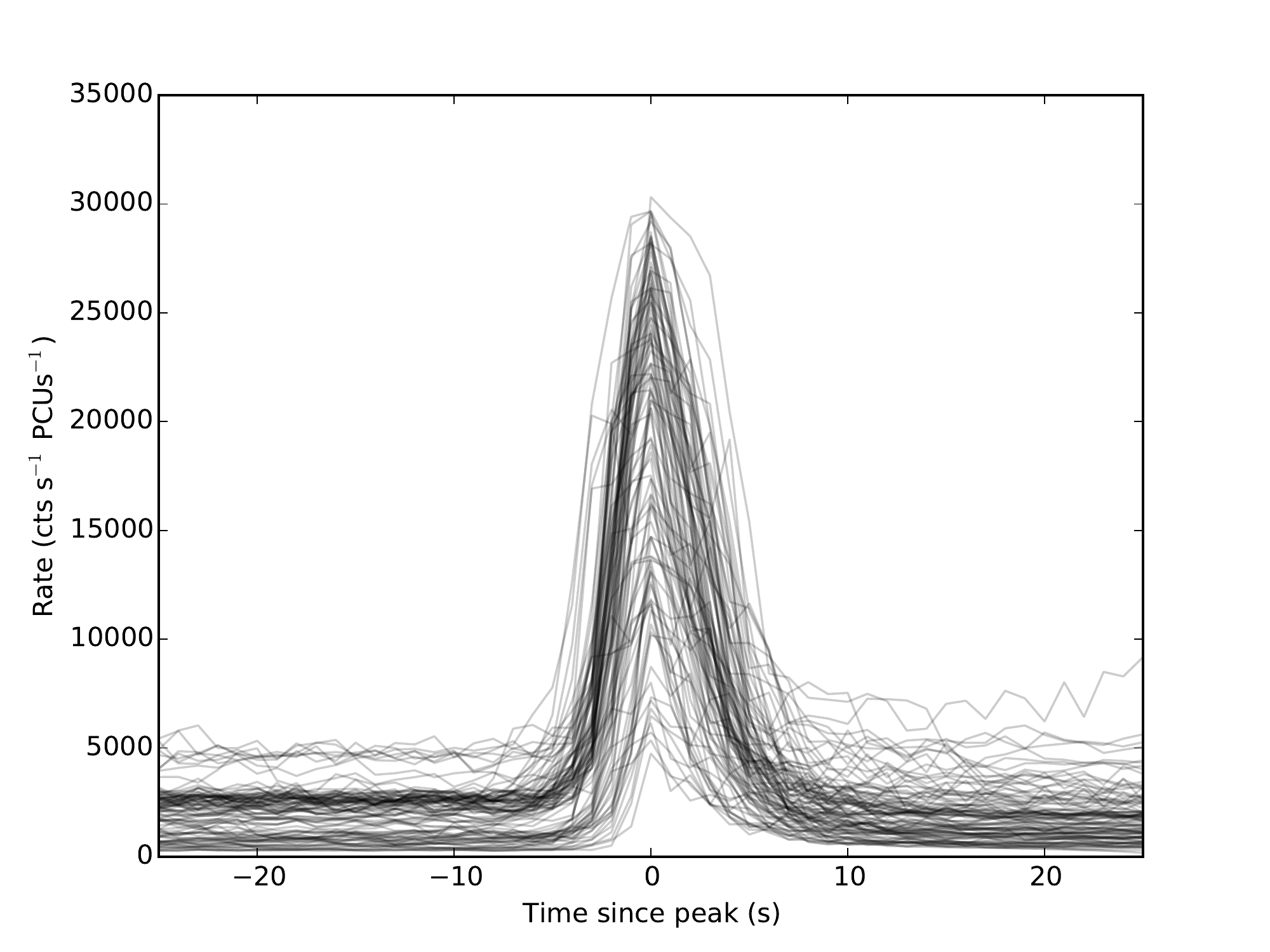}
  \includegraphics[width=.9\linewidth, trim={0.4cm 0 1.1cm 0},clip]{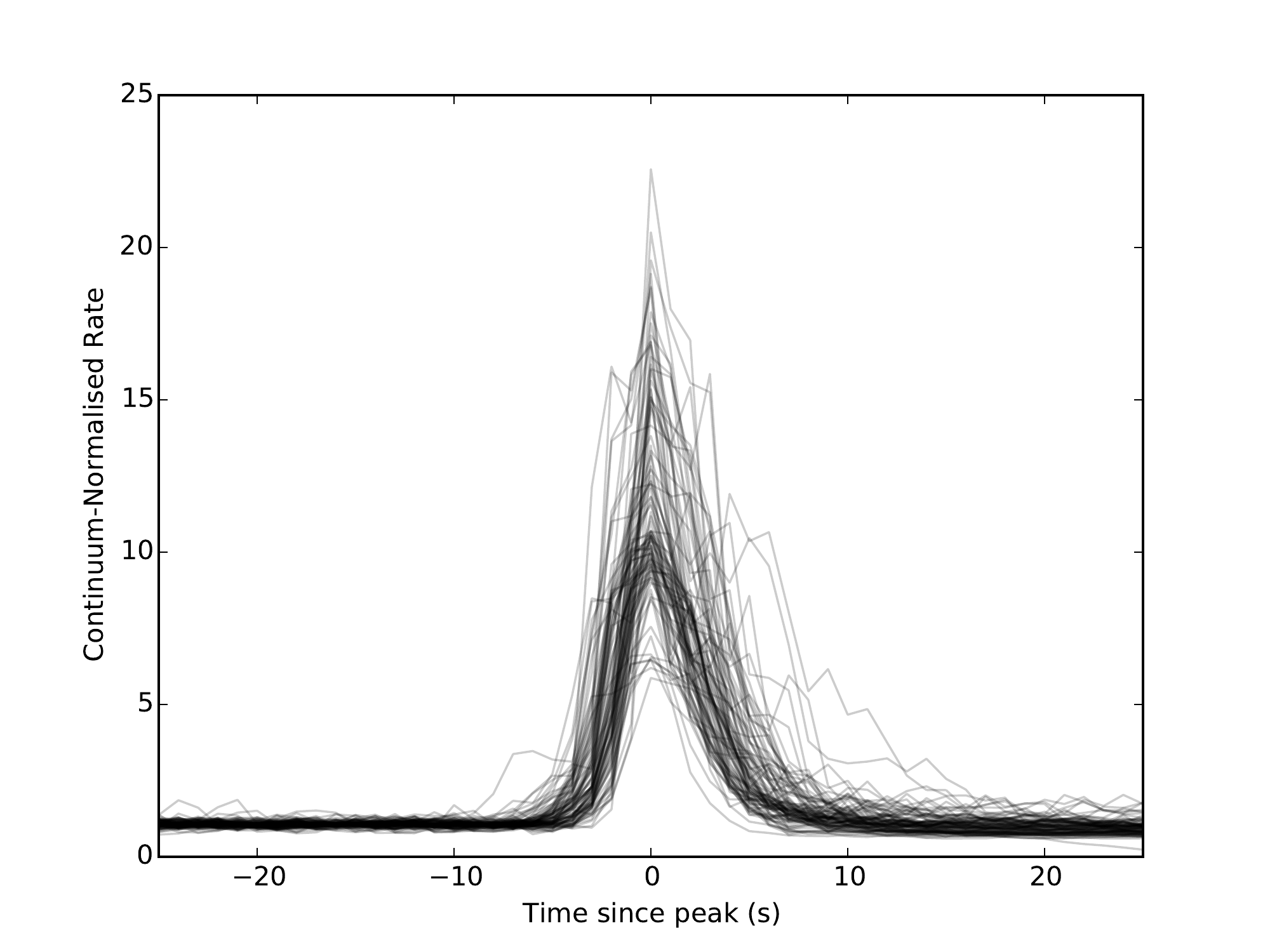}
  \caption{\small \textbf{Top:} a plot of every Normal Burst, centred by the time of its peak, overlaid on top of each other to show the existence of a common pulse profile.  \textbf{Bottom:} a plot of every Normal Burst in which count rates have been normalised by the persistent emission count rate during the observation from which each burst was observed.  As the bursts are on average closer to the average pulse profile in this metric, this suggests that the intensity of a burst is roughly dependent on the persistent emission rate.  Some persistent emission-normalised count rates may be artificially low due to dead-time effects.}
  \label{fig:norm_overlay}
\end{figure}

 In the top panel of Figure \ref{fig:norm_overlay} we show a plot of all Normal Bursts observed with \textit{RXTE} overlayed on top of one another.  We find that all Normal Bursts follow a similar burst profile with similar rise and decay timescales but varying peak intensities.  In the lower panel of Figure \ref{fig:norm_overlay} we show a plot of Normal Bursts overlaid on top of each other after being normalised by the persistent emission count rate in their respective observation.  The bursts are even closer to following a single profile in this figure, suggesting a correlation between persistent emission level in an outburst and the individual fluence of its bursts.

\par The structure of the lightcurve of a Normal Burst can be described in three well-defined parts:

\begin{enumerate}
\item The main burst: roughly approximated by a skewed Gaussian (see e.g. \citealp{Azzalini_Dist}).
\item A `plateau': a period of time after the main burst during which count rate remains relatively stable at a level above the pre-burst rate.
\item A `dip': a period during which the count rate falls below the persistent level, before exponentially decaying back up towards the pre-burst level (e.g. \citealp{Younes_Expo}).
\end{enumerate}

\par
\begin{figure}
  \centering
  \includegraphics[width=.9\linewidth, trim={0.8cm 0 1.6cm 0},clip]{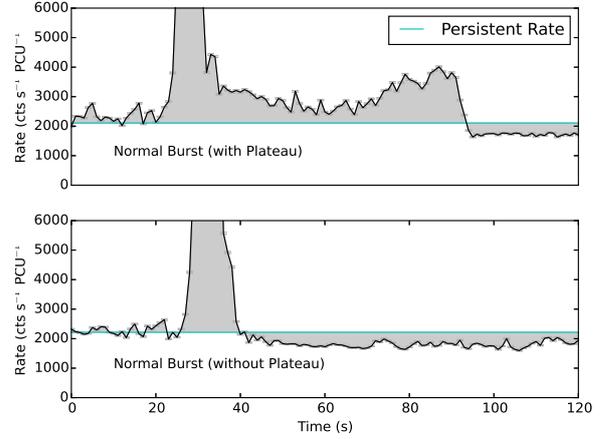}
  \caption{\small \textit{RXTE} lightcurves of Normal Bursts with (top) and without (bottom) `plateau' features, showing the burst structure in each case.  The median count rate, which we use as a proxy for the persistent emission, is plotted in cyan to highlight the presence of the count rate `dip' after each burst.}
  \label{fig:w_wo}
\end{figure}

 The dip is present after every burst in our \textit{RXTE} sample from Outbursts 1 \& 2, whereas the plateau is only seen in 39 out of 99.  We show example lightcurves of bursts with and without plateaus in Figure \ref{fig:w_wo}, which also show that the dip is present in both cases.

\par
\begin{figure}
  \centering
  \includegraphics[width=.9\linewidth, trim={1.9cm 0 2.0cm 0},clip]{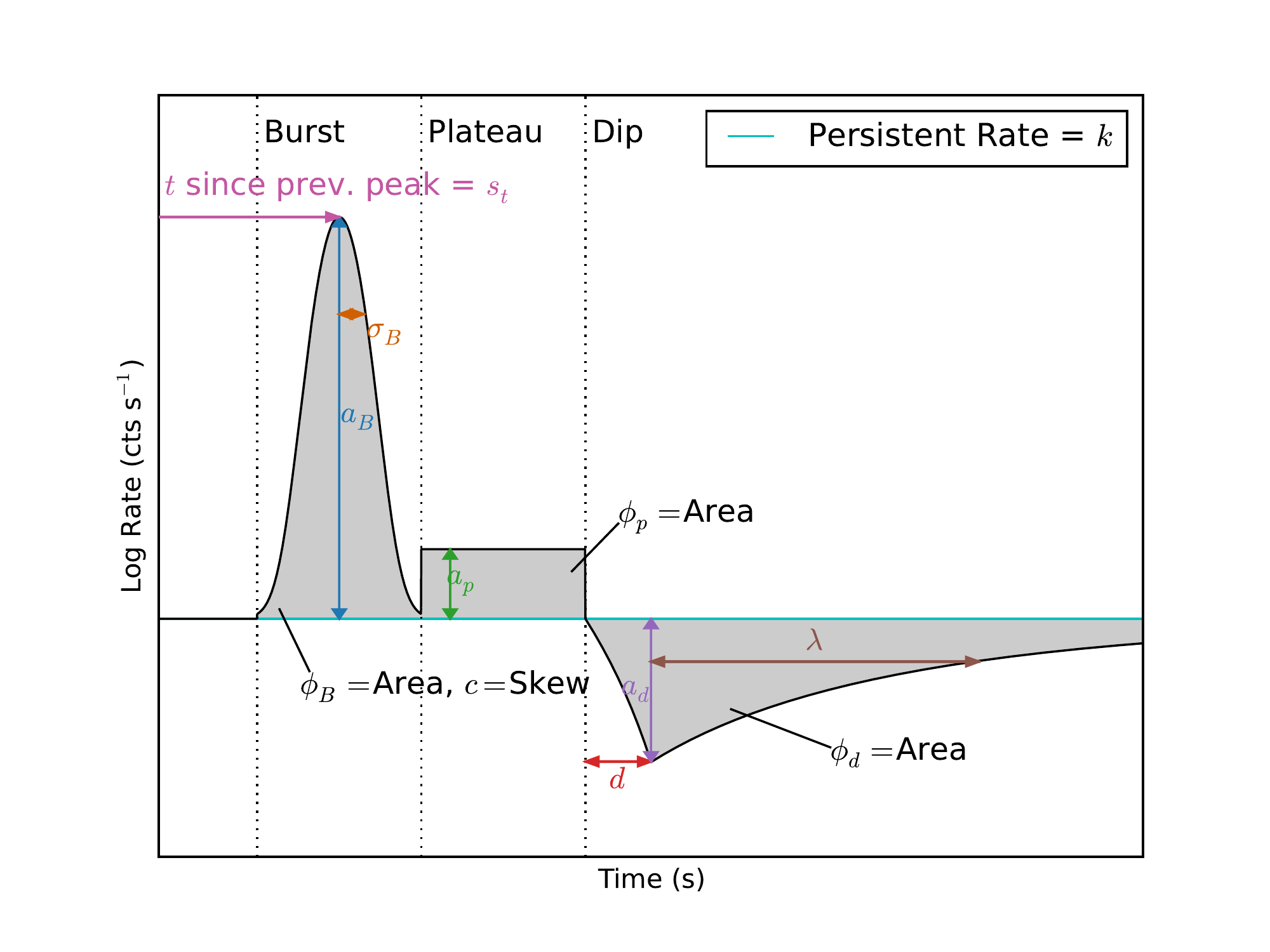}
  \caption{\small A schematic explaining the origin of the 12 Normal Burst parameters used in this study, as well as showing the functional forms of both the skewed Gaussian fit to a burst and the `dipper function' (Equation \ref{eq:dipper}) fit to a dip.  Note that we do not fit a function to the plateau, and we calculate its fluence by summing the persistent rate-subtracted counts.  Diagram is for explanation only and the burst pictured is neither based on real data nor to scale.}
  \label{fig:explain}
\end{figure}

 In order to study Normal Bursts, we fit the burst profiles with phenomenologically-motivated mathematical functions.  In Figure \ref{fig:explain} we show a schematic plot of our model, as well as annotations explaining the identities of the various parameters we use.  We fit the main burst with a skewed Gaussian, centred at $t=x_0$ with amplitude $a_b$, standard deviation $\sigma_B$ and skewness\footnote{A measure of how far the peak of the Gaussian is displaced from its centre.} $c$, added to the persistent emission rate $k$.  We fit the `dip' with the continuous piecewise function $f(t)$:

\begin{equation}
f(t)=
\begin{dcases}
k-\frac{a_d(t-t_0)}{d-t_0}, & \text{if } t\leq d\\
k-a_d\exp\left(\frac{d-t}{\lambda}\right), & \text{otherwise}
\end{dcases}
\label{eq:dipper}
\end{equation}

Where $t$ is time, $t_0$ is the start time of the dip, $a_d$ is the amplitude of the dip, $d$ is the time at the local dip minimum and $\lambda$ is the dip recovery timescale.  This function is based on the finding by \citet{Younes_Expo} that dip count rates recover exponentially, but has the added advantage that the start of the recovery phase can also be fit as an independent parameter.  Using this fit, we can estimate values for burst fluence $\phi_B$, burst scale-length $\sigma_B$, `missing' dip fluence $\phi_D$ and dip scale-length $\lambda$ and compare these with other burst parameters.  When present, we also calculate the fluence of the plateau $\phi_p$ by summing the persistent emission-subtracted counts during the region between the end of the burst (as defined in Section \ref{sec:burst_diff}) and the start of the dip.  For each pair of parameters, we do not consider datapoints when the magnitude of the error on a parameter is greater than the value of the parameter.

\par We only extract these parameters from Normal Bursts observed by \textit{RXTE} during Outbursts 1 \& 2.  This ensures that the resultant parameter distributions we extracted are not affected by differences between instruments.

\subsubsection{Parameter Distributions}

\label{sec:hists}

\par We extracted a total of ten parameters from our fit to each burst: the parameters $a_d$, $d$ and $\lambda$ of the fit to the dip, the missing fluence $\phi_D$ of the dip, the parameters $a_b$, $\sigma_B$ and $c$ of the skewed Gaussian fit to the main burst, the main burst fluence $\phi_B$, the maximum persistent emission-subtracted rate in the plateau $a_p$ and the plateau fluence $\phi_P$.
\par Using our \textit{RXTE} sample of Normal Bursts, we can construct distributions for all of the burst parameters described in Section \ref{sec:struc} for bursts in Outbursts 1 \& 2.  We give the mean and standard deviation for each parameter in each outburst in Table \ref{tab:params_perob}, and histograms for each can be found in Appendix \ref{app:hists}.

\begin{table}
\centering
\begin{tabular}{r c c c c c c}
\hline
\hline
 & \multicolumn{2}{c}{\scriptsize Outburst 1} & \multicolumn{2}{c}{\scriptsize Outburst 2} & \multicolumn{2}{c}{\scriptsize Outbursts 1\&2}  \\
 &Mean&S.D.&Mean&S.D.&Mean&S.D.\\
\hline
$\phi_B$&2.74e6&7.8e5&2.25e6&7.6e5&$2.43\mathrm{e}6$&$8.0\mathrm{e}5$\\
$a_B$&3.18e5&8.4e4&2.72e5&9.9e4&$2.90\mathrm{e}5$&$9.6\mathrm{e}4$\\
$\sigma_B$&3.39&0.35&3.42&0.59&3.41&0.52\\
$c$&2.68&1.9&2.79&2.0&2.75&2.0\\
$\phi_d$&1.74e6&1.3e6&1.17e6&3.6e5&$1.38\mathrm{e}6$&$8.7\mathrm{e}5$\\
$a_d$&550&335&536&307&541&318\\
$d$&49&46&20&22&31&36\\
$\lambda$&294&176&229&124&254&150\\
$\phi_p$&1.89e5&2.3e5&7577&5707&1.4e5&1.8e5\\
$a_p$&1289&1113&767&463&1063&928\\
\hline
\hline
\end{tabular}
\caption{A table showing the mean and standard deviation of 10 Normal Burst parameters of \textit{RXTE}-sampled bursts.  In each case, we give the values for populations from only Outburst 1, from only Outburst 2 and from the combined population from both outbursts.  Histograms for each parameter can be found in Appendix \ref{app:hists}.}
\label{tab:params_perob}
\end{table}

\par The mean value of most parameters differs by no more than $\sim50$\% between outbursts.  Notable exceptions are $d$, $\phi_p$, $\phi_d$ and $a_p$, which are $\sim2.5$, $\sim2.5$ $\sim1.5$ and $\sim1.7$ times greater in Outburst 1 than in Outburst 2 respectively.  The less significant differences between values of $\phi_B$ and $a_B$ in Outbursts 1 \& 2 are expected, as the amplitude of a burst correlates with $k$ which was generally higher in Outburst 1 than in Outburst 2.

\subsubsection{Correlations}

\label{sec:NormCorr}

\par In total, we extracted 12 parameters for each Normal Burst in our \textit{RXTE} sample: the 10 burst parameters listed in Section \ref{sec:hists}, the recurrence time $s_t$ until the next burst and the persistent emission rate $k$ at the time of the burst.  
\par As the amplitude of all 3 components in a burst scale with the persistent level, we rescaled our values of $a_b$, $a_d$, $\phi_B$, $\phi_D$ and $\phi_P$ by a factor $\frac{1}{k}$.  We show the covariance matrix with all 66 possible pairings of these normalised parameters in Figure \ref{fig:corr_n} (we present the covariance matrix of these parameters before being rescaled in Appendix \ref{app:corr}).  Using the Spearman's Rank Correlation Coefficient, we find the following $\geq5\,\sigma$ correlations which are highlighted in Figure \ref{fig:corr_n}:

\begin{itemize}
\item Persistent emission $k$ anticorrelates with normalised burst fluence $\phi_B/k$ ($>10\,\sigma$) and normalised burst amplitude $a_b/k$ ($>10\,\sigma$).
\item Normalised burst fluence $\phi_B/k$ correlates with normalised burst amplitude $a_B/k$ ($8.0\,\sigma$).
\item Normalised dip fluence $\phi_d/k$ correlates with dip recovery timescale $\lambda$ ($6.3\,\sigma$).
\item Normalised dip amplitude $a_d/k$ anticorrelates with dip falltime $d$ ($5.7\,\sigma$) and dip recovery timescale $\lambda$ ($7.1\,\sigma$).
\item Normalised plateau fluence $\phi_p/k$ correlates with normalised plateau amplitude $a_p$ ($6.4\,\sigma$).
\end{itemize}

As $\phi_B$ can be approximated to first order as a product of $a_B$ and $\sigma$, the correlation between $\phi_B$ and $a_B$ is expected as they are not independent parameters.  Similarly, the correlations between $\phi_d$ \& $\lambda$ and $\phi_p$ and $a_p$ are likely due to these pairs of parameters not being independent.

\begin{figure*}
  \centering
  \includegraphics[width=\linewidth, trim={2.1cm 2cm 3.5cm 3cm},clip]{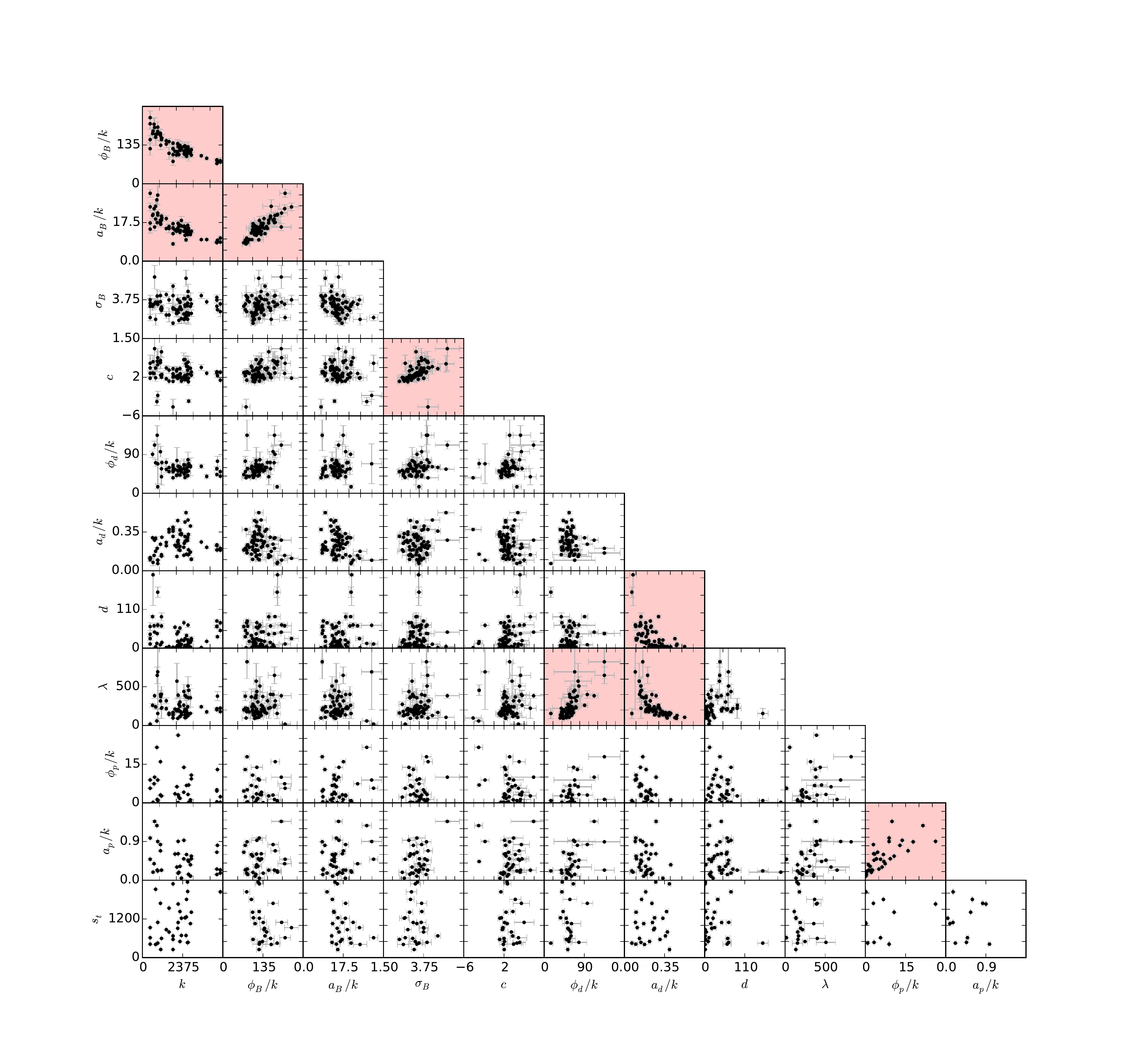}
  \caption{\small Covariance Matrix with a scatter plot of each of the 66 pairings of the 12 Normal Burst parameters listed in section \ref{sec:NormCorr}.  Amplitudes and fluences have been normalised by dividing by the persistent count rate $k$.  Pairings which show a correlation using the Spearman Rank metric with a significance $\geq5\,\sigma$ are highlighted in red.}
  \label{fig:corr_n}
\end{figure*}

\subsubsection{Colour Evolution}

\par To explore the spectral behaviour of Normal Bursts, we studied the evolution of the hardness (the ratio between count rate in the energy bands $\sim2$--$7$ and $\sim8$--$60$\,keV energy bands) as a function of count rate during the individual bursts.  These `hardness-intensity diagrams' allow us to check for spectral evolution in a model-independent way.  We do not correct them for background as the count rates in both bands are very high.
\par
\begin{figure}
  \centering
  \includegraphics[width=.9\linewidth, trim={0.4cm 1cm 1.1cm 1cm},clip]{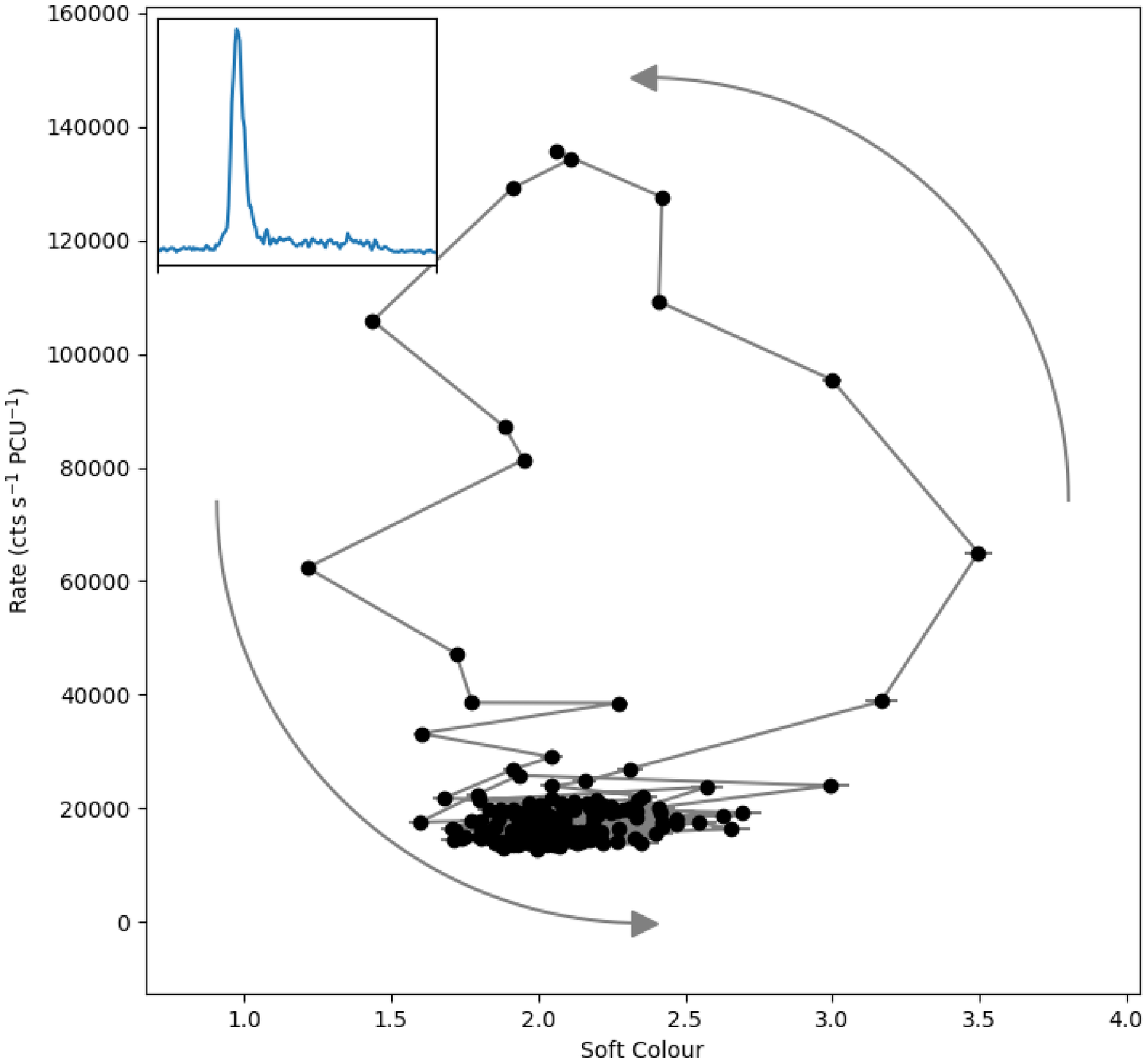}
  \caption{\small A 1\,s-binned hardness-intensity diagram of a Normal Burst from \textit{RXTE}/PCA observation 10401-01-08-00, with an inset 2--60\,keV lightcurve.  Significant colour evolution can be seen during the burst, taking the form of a loop.}
  \label{fig:loop}
\end{figure}

 We find evidence of hysteretic loops in hardness-intensity space in some, but not all, of the Normal Bursts in our sample; see Figure \ref{fig:loop} for an example of such a loop.  The existence of such a loop suggests significant spectral evolution throughout the burst.  This finding can be contrasted with results from previous studies in different energy bands (e.g. \citealp{Woods_OB2} from $\sim25$--100\,keV) which suggested no spectral evolution during Type II bursts in this source.

\subsection{Minibursts}

\par We define Minibursts as the set of all bursts with a peak 1\,s binned \textit{RXTE}/PCA-equivalent count rate of $<300\%$ of the persistent rate.  Minibursts account for 48 out of the 190 bursts identified for this study.  They are observed during all 3 Outbursts, and occur during the same times that Normal Bursts are present.  Minibursts occurred between MJDs 50117 and 50200 in Outburst 1, and between 50466 and 50542 in Outburst 2; during these intervals, \textit{RXTE} observed the source for a total of 192\,ks.  These intervals correspond to the times between the peak of each outburst and and the time that the persistent intensity falls below $\sim0.1$\,Crab.

\subsubsection{Recurrence Time}

\par There are only 10 observations with \textit{RXTE} which contain multiple Minibursts.  Using these, we find minimum and maximum Miniburst recurrence times of 116 and 1230\,s.
\par We find 17 \textit{RXTE} observations which contain both a Miniburst with a preceding Normal Burst, and find minimum and maximum Normal Burst $\rightarrow$ Miniburst recurrence times of 461 and 1801\,s.

\subsubsection{Structure}

\par
\begin{figure}
  \centering
  \includegraphics[width=.9\linewidth, trim={0cm 0 0cm 0},clip]{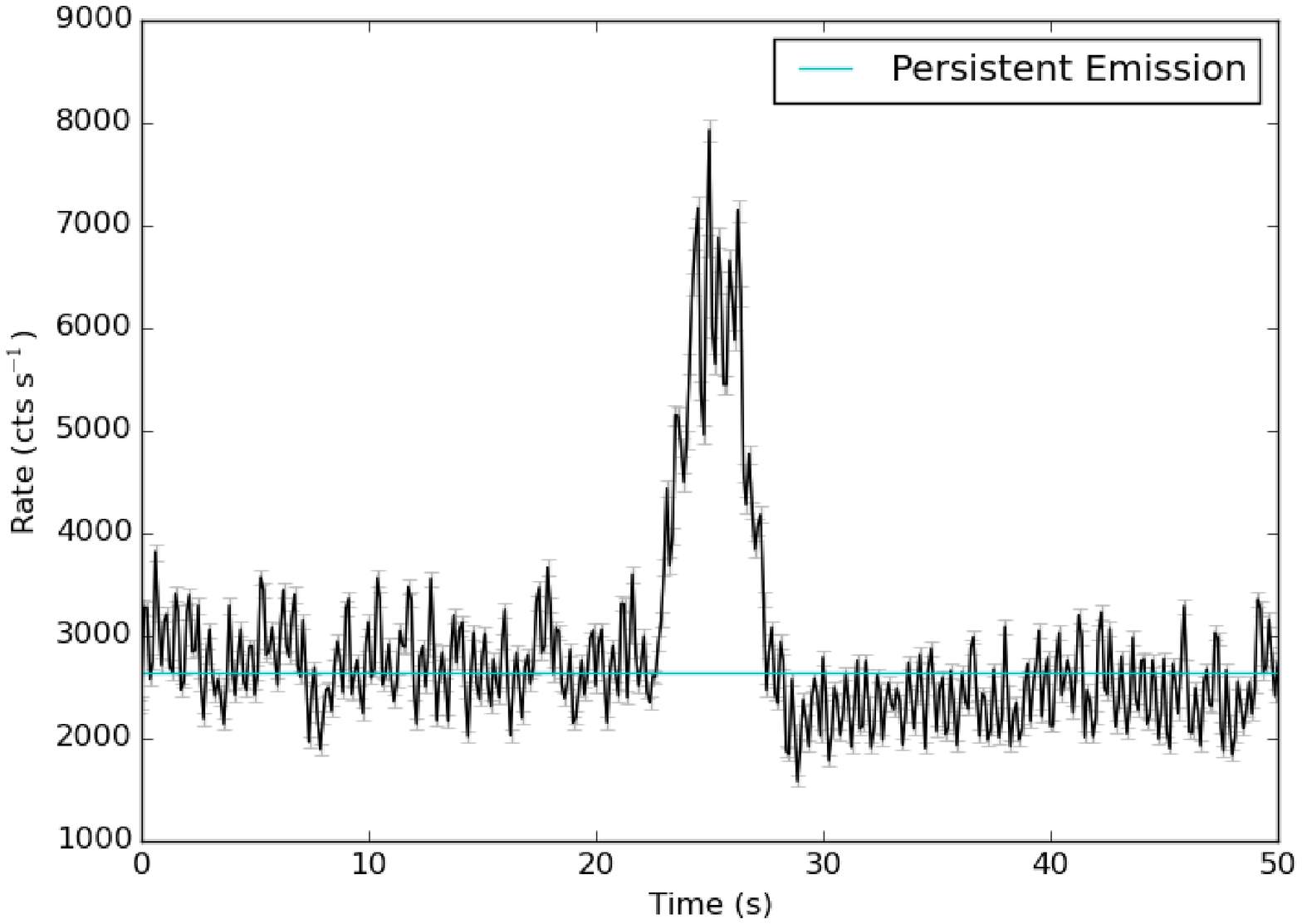}
  \caption{\small  A representative \textit{RXTE} lightcurve of a Miniburst from OBSID 20077-01-03-00 in Outburst 2.}
  \label{fig:a_mini}
\end{figure}

\begin{figure}
  \centering
  \includegraphics[width=.9\linewidth, trim={0.4cm 0 1.1cm 0},clip]{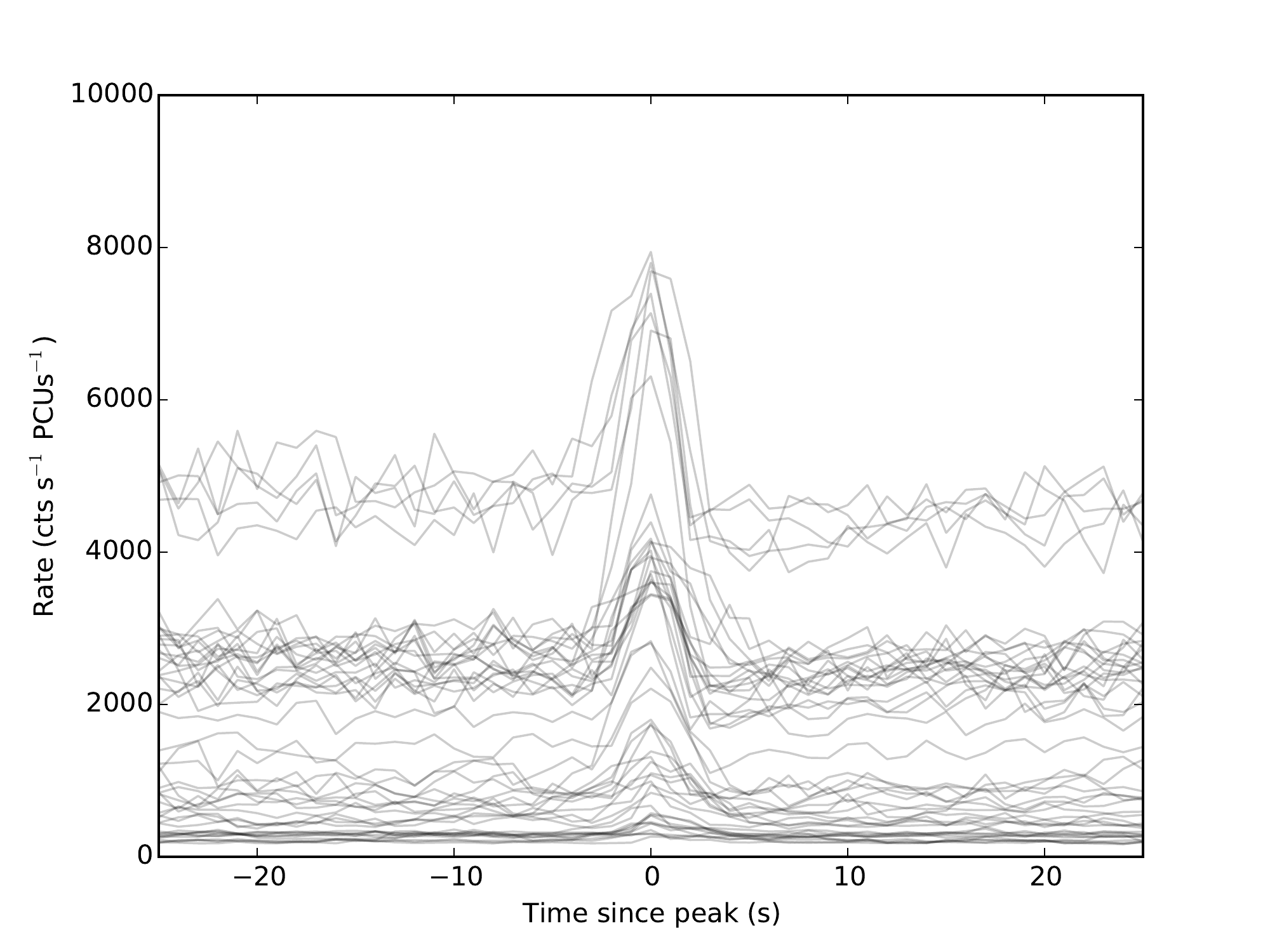}
  \includegraphics[width=.9\linewidth, trim={0.4cm 0 1.1cm 0},clip]{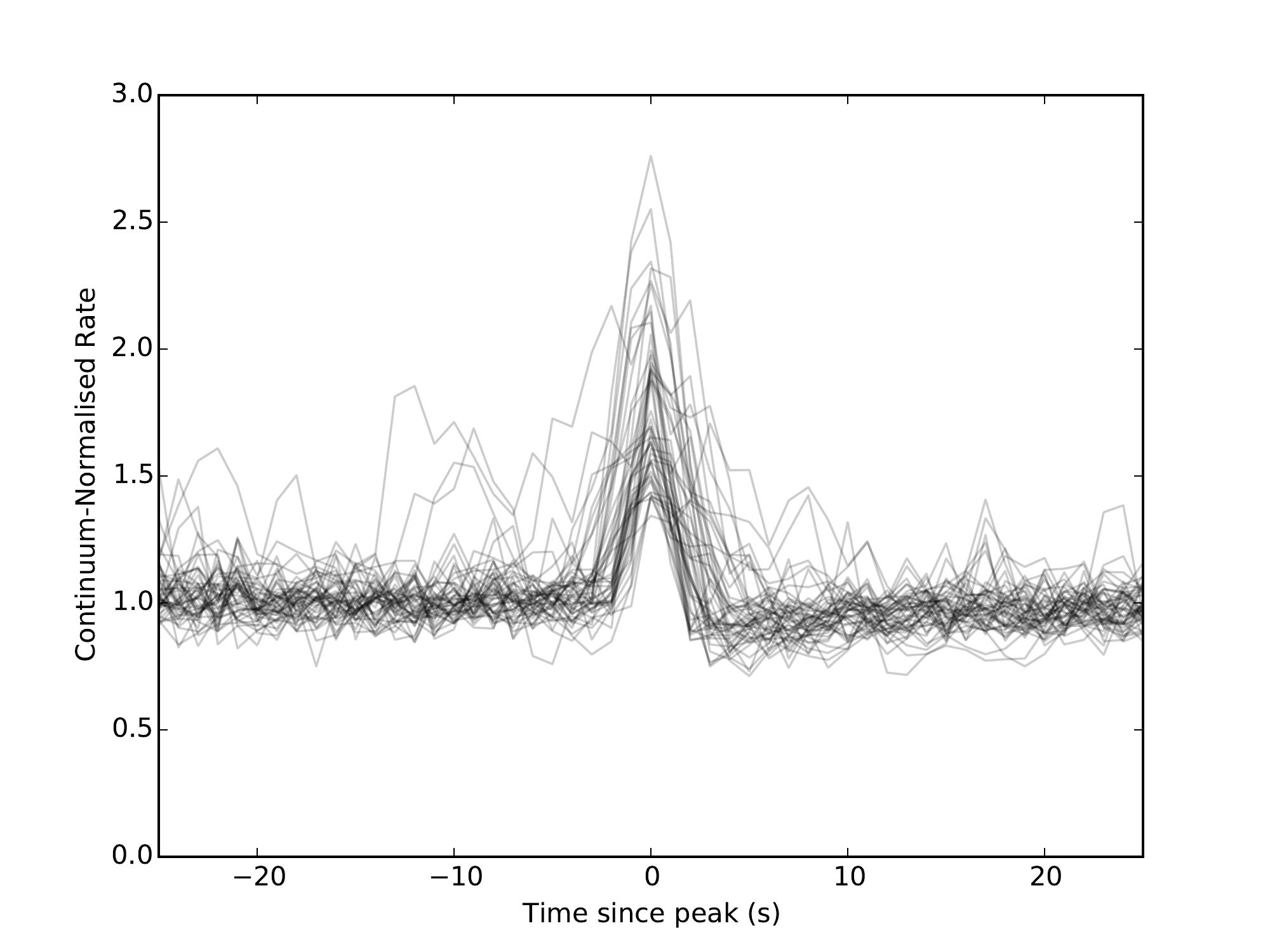}
  \caption{\small  \textbf{Top:} a plot of every Miniburst, centred by the time of its peak, overlaid on top of each other.  \textbf{Bottom:} a plot of every Miniburst in which count rates have been normalised by the persistent emission count rate during the observation from which each burst was observed.}
  \label{fig:mini_over}
\end{figure}

 In Figure \ref{fig:a_mini}, we show a representative Miniburst, and we show all Minibursts overplotted on each other in Figure \ref{fig:mini_over}.  These bursts are roughly Gaussian in shape with a large variation in peak count rate; as can be seen in Figure \ref{fig:mini_over}, however, the persistent-normalised peak count rates of Minibursts are all roughly consistent with 2.


\par Minibursts are all $\sim5$\,s in duration, and some show signs of a `dip' feature similar to those seen in Normal Bursts.  We find that the timescales of these dips are all $\lesssim10$\,s.  We estimate `missing' fluence in each dip by integrating the total persistent-rate-subtracted counts between the end of the burst and a point 10\,s later.  If this `missing fluence' is less than half of the standard deviation in count rate multiplied by 5\,s, which represents the smallest $<10$\,s triangle-shaped dip which would be detectable above noise in a given dataset, we treat the dip in that outburst as not being detected.
\par Due to the relatively short duration and low peak count rates of Minibursts, we are unable to reliably discern whether they contain a single peak or multiple peaks.  For this reason we also do not fit them mathematically.

\subsubsection{Parameters \& Correlations}

\par For each Miniburst, we are able to extract the same parameters that we extracted from Mesobursts (see list in Section \ref{sec:mesostruc}).  The mean and standard deviation of each of these parameters, calculated from \textit{RXTE} data, is presented in Table \ref{tab:mini_param} for Outburst 1, Outburst 2 and the combined population of Minibursts from Outbursts 1 \& 2.  The standard deviations on the fluence and peak rates of Minibursts are very large, suggesting that these parameters are distributed broadly.

\begin{table}
\centering
\begin{tabular}{r c c c c c c}
\hline
\hline
 & \multicolumn{2}{c}{\scriptsize Outburst 1} & \multicolumn{2}{c}{\scriptsize Outburst 2} & \multicolumn{2}{c}{\scriptsize Outbursts 1\&2}  \\
 &Mean&S.D.&Mean&S.D.&Mean&S.D.\\
\hline
\scriptsize Fluence&6792&5776&4474&3307&5422&4627\\
\scriptsize Peak Rate&3501&2851&2473&1664&2902&2293\\
\scriptsize Fluence/$k$&3.67&1.13&3.58&1.47&3.61&1.34\\
\scriptsize Peak Rate/$k$&1.90&0.37&1.76&0.28&1.82&0.32\\
\scriptsize Rise Time&2.33&0.8&2.03&1.1&2.15&1.0\\
\scriptsize Fall Time&2.32&0.9&2.35&1.0&2.32&0.9\\
\scriptsize Tot. Time&4.61&1.0&4.38&01.0&4.47&1.0\\
\hline
\hline
\end{tabular}
\caption{A table showing the mean and standard deviation of 7 parameters of \textit{RXTE}-sampled Minibursts from Outburst 1, Outburst 2 and both outbursts combined.  Fluence is given in cts\,PCU$^{-1}$, peak rate is given in cts\,s$^{-1}$\,PCU$^{-1}$ and rise, fall and total time are given in s.  $k$ is the persistent emission rate during the observation in which a given burst was detected.}
\label{tab:mini_param}
\end{table}

\par Using the Spearman's Rank metric, we find only two correlations above the 5$\,\sigma$ level:
\begin{itemize}
\item Fluence is correlated with peak rate ($7.3\,\sigma$).
\item Fluence divided by persistent rate is correlated with peak rate divided by persistent rate ($7.1\,\sigma$).
\end{itemize}
As in Normal Bursts, a correlation between peak rate and fluence is to be expected.  However, due to the poor statstics associated with Miniburst parameters, it is likely that other parameter pairs are also correlated.

\subsubsection{Colour Evolution}

\par
\begin{figure}
  \centering
  \includegraphics[width=.9\linewidth, trim={0.7cm 1.4cm 0.2cm 1.4cm},clip]{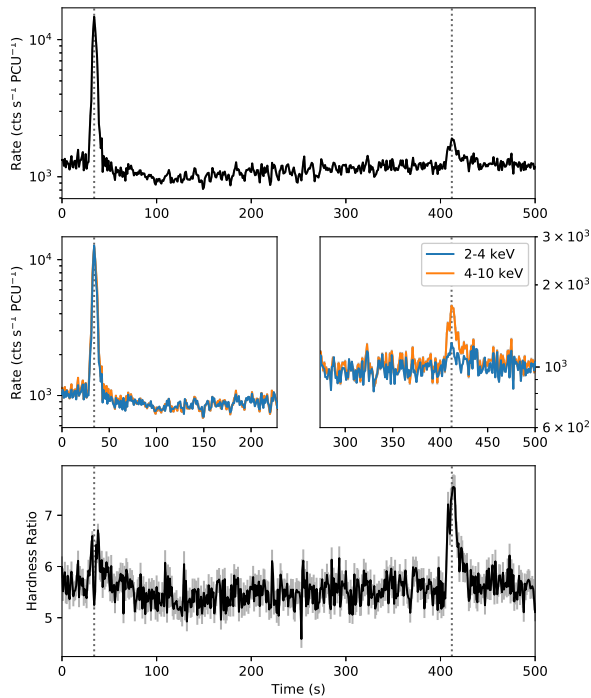}
  \caption{\small A portion of observation 10401-01-16-00, featuring a Normal Burst ($\sim30$\,s) and a Miniburst ($\sim410$\,s).  The top panel shows the total 2--10\,keV lightcurve.  The middle panel shows lightcurves from two different energy bands; the count rates from the soft energy band have been multiplied by 5.4 so they can more easily be compared with the hard energy band.  The bottom panel shows the evolution over time of the ratio between the rates in the two bands.   As can be seen in panels 2 and 3, the Miniburst has a significantly higher fractional amplitude in the 4--10\,keV energy band than in the 2--4\,keV band.}
  \label{fig:minihard}
\end{figure}

 Minibursts show the greatest magnitude of evolution in colour of all the classes of burst.  In Figure \ref{fig:minihard}, we show how the hardness ratio between the 4--10 and 2--4\,keV energy bands changes during an observation containing both a Miniburst and a Normal Burst.  We find that the hardness ratio increases by $\sim50\%$ in a Miniburst, significantly more than the change in hardness during Normal or Mesobursts.  The statistics in minibursts were too poor to check for the presence of hysteresis.

\subsection{Mesobursts}

\par We define Mesobursts as the set of all bursts with a persistent-emission-subtracted peak 1\,s binned \textit{RXTE}/PCA-equivalent count rate below 3000\,cts\,s$^{-1}$\,PCU$^{-1}$ in which the peak of the burst reaches at least $300\%$ of the persistent rate.  Mesobursts account for 43 out of the 190 bursts identified for this study.  They are observed in \textit{RXTE} data from both Outbursts 1 \& 2; in both cases they occur after the main outburst and before or during a rebrightening event.  Mesobursts occurred between MJDs 50238 and 50248 in Outburst 1, and between 50562 and 50577 in Outburst 2; during these intervals, \textit{RXTE} observed the source for a total of 44\,ks.  As no soft X-ray instrument monitored the Bursting Pulsar during the latter stages of Outburst 3, it is unclear whether Mesobursts occurred during this outburst.  The one pointed observation of \textit{NuSTAR} made during this time did not detect any Mesobursts.

\subsubsection{Recurrence Time}

\par Only 6 \textit{RXTE} observations in Outburst 1, and 4 in Outburst 2, contain multiple Mesobursts.  From our limited sample we find minimum and maximum recurrence times of $\sim230$ and $\sim1550$\,s in Outburst 1 and minimum and maximum recurrence times of $\sim310$ and $\sim2280$\,s in Outburst 2.

\subsubsection{Structure}

\par
\begin{figure}
  \centering
  \includegraphics[width=.9\linewidth, trim={0.4cm 0 1.1cm 0},clip]{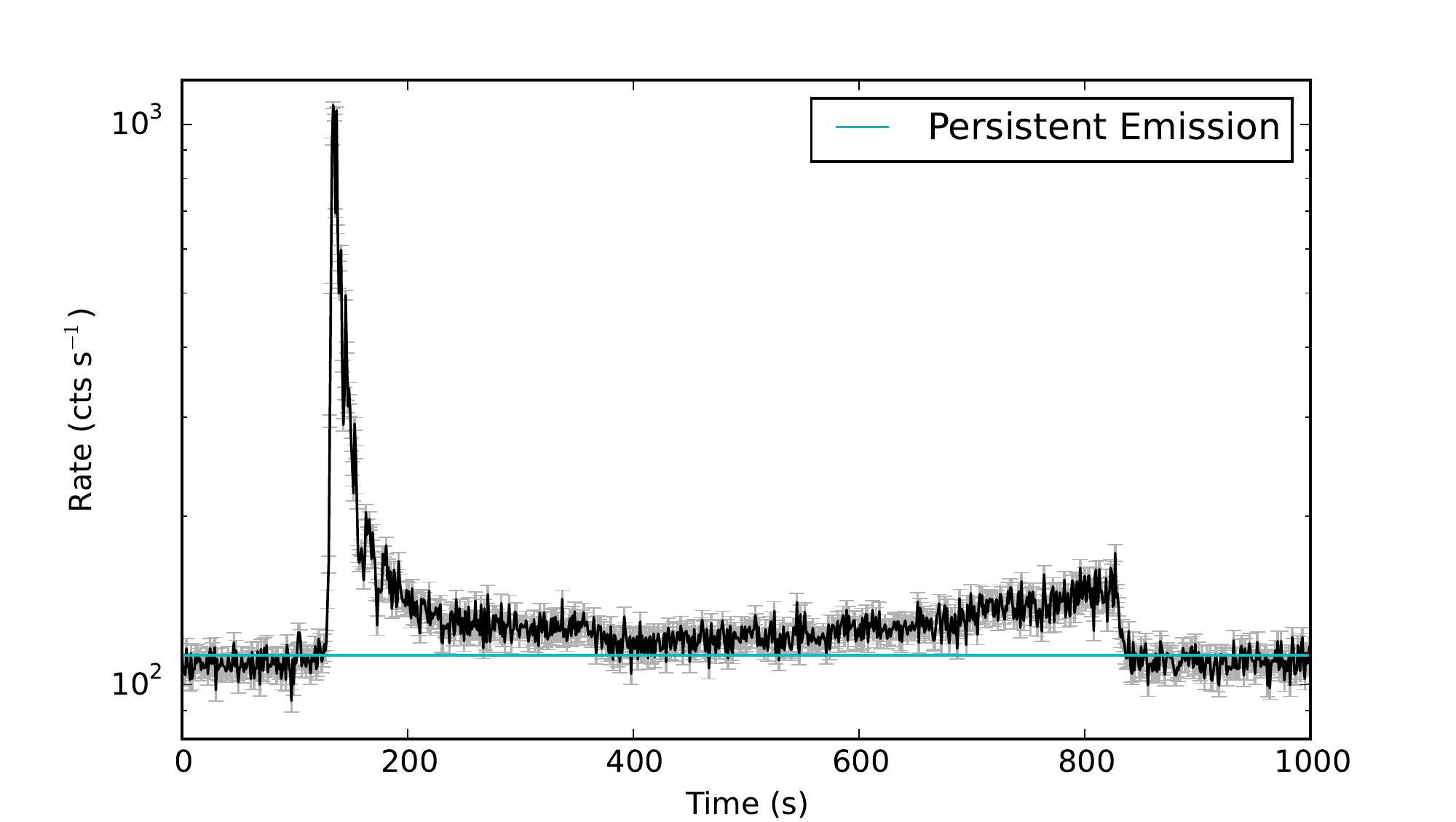}
  \caption{\small A lightcurve from \textit{RXTE} observation 20078-01-17-00 from Outburst 2, showing an apparent `plateau' feature after a Mesoburst.}
  \label{fig:mesoplateau}
\end{figure}

 The structure of the main part of a Mesoburst is significantly more complex than in Normal Bursts, consisting of a large number of secondary peaks near the main peak of the burst.  Mesobursts never show the post-burst `dip' feature that we see in Normal Bursts, but they can show `plateaus'.  In Figure \ref{fig:mesoplateau} we show an example of a Mesoburst with a plateau similar to those seen after Normal Bursts, suggesting a connection between the two classes.

\par
\begin{figure}
  \centering
  \includegraphics[width=.9\linewidth, trim={0.4cm 0 1.1cm 0},clip]{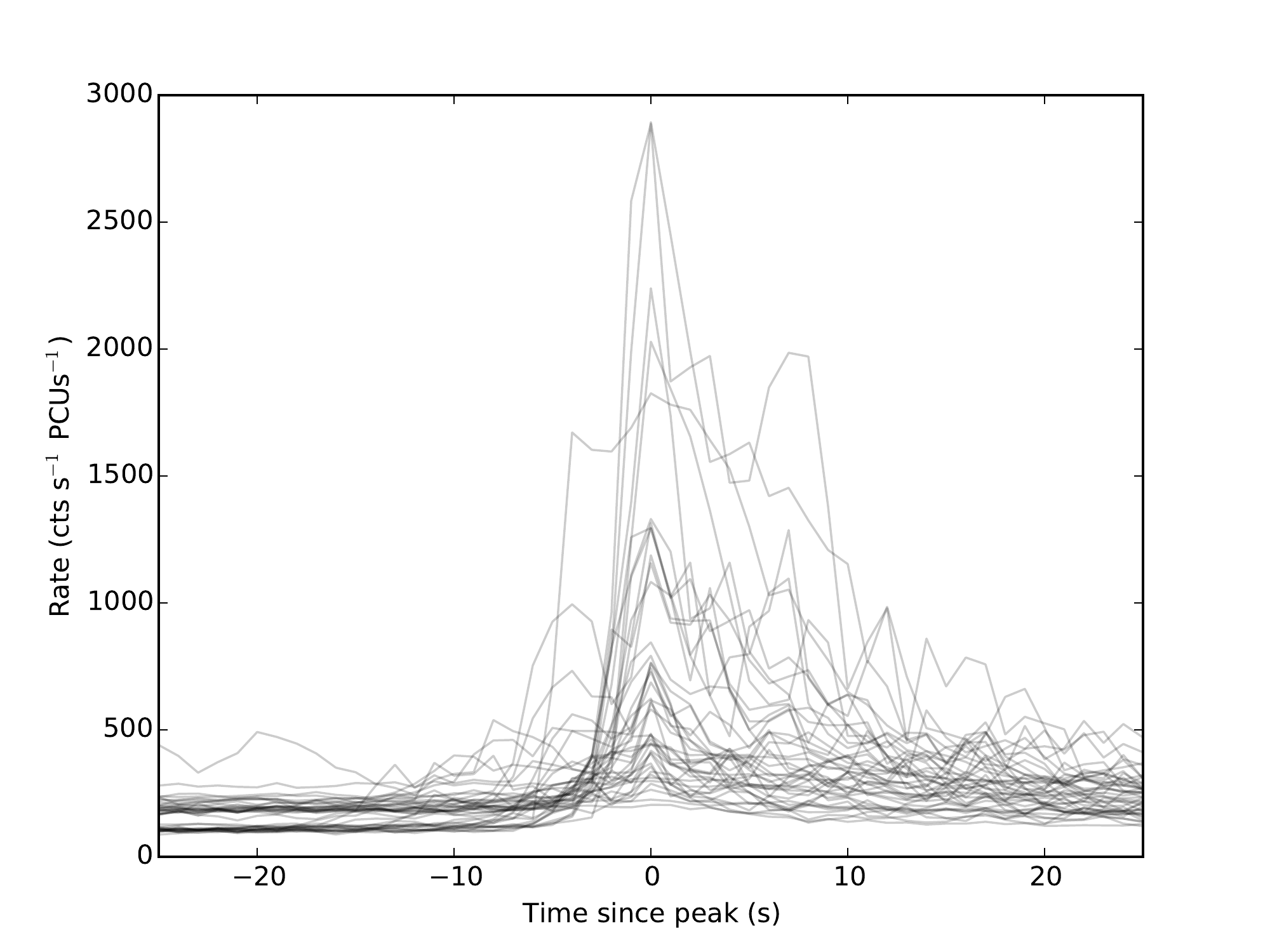}
  \includegraphics[width=.9\linewidth, trim={0.4cm 0 1.1cm 0},clip]{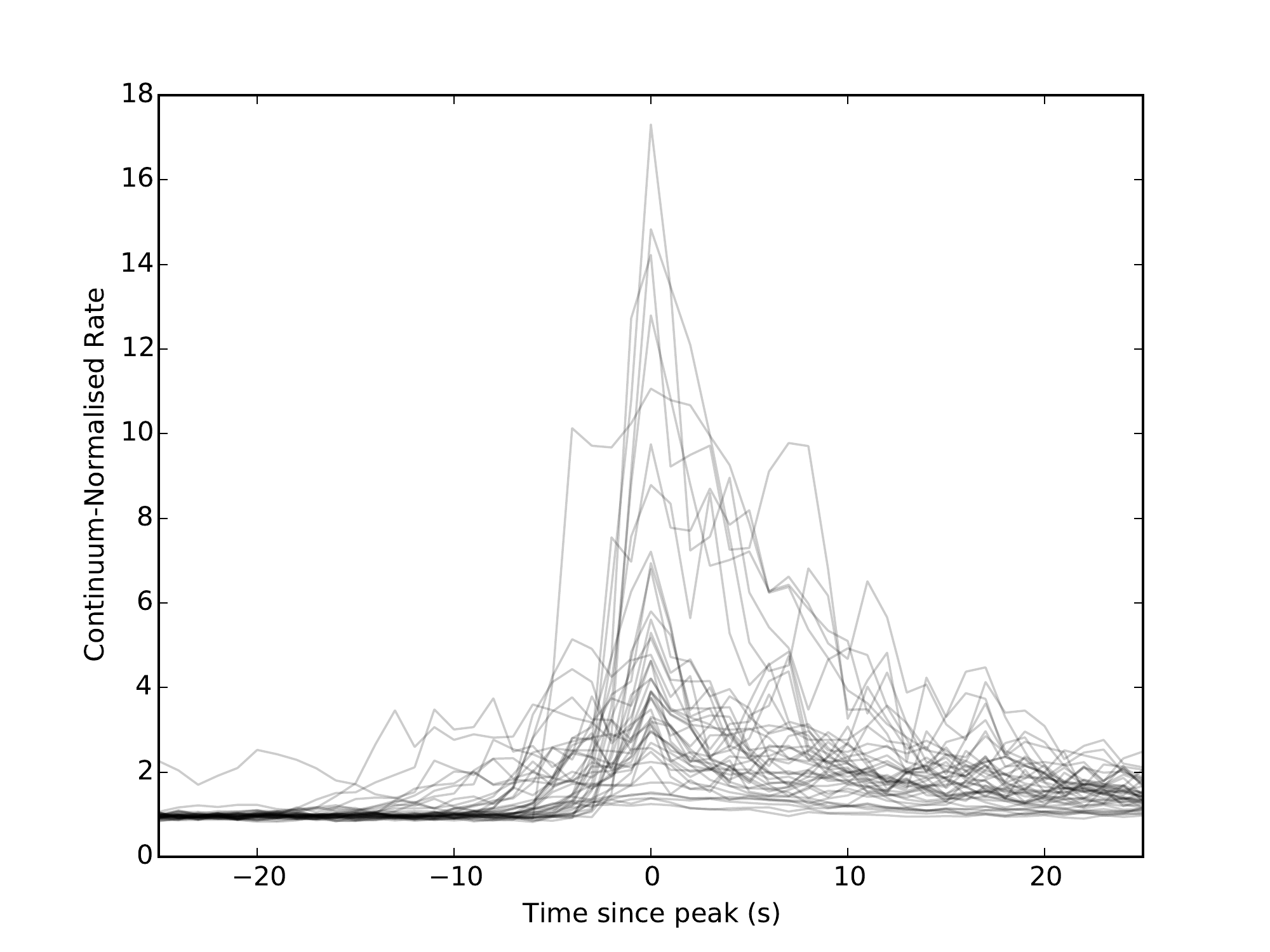}
  \caption{\small  \textbf{Top:} a plot of every Mesoburst, centred by the time of its peak, overlaid on top of each other.  \textbf{Bottom:} a plot of every Mesoburst in which count rates have been normalised by the persistent emission count rate during the observation from which each burst was observed.}
  \label{fig:meso_over}
\end{figure}

 In Figure \ref{fig:meso_over} we show the plot of all Mesobursts observed by \textit{RXTE} overlayed on top of each other before (top panel) and after (bottom panel) being renormalised by persistent emission rate.  It can be seen that the intensity and structure of these bursts is much more variable than in Normal Bursts (see Figure \ref{fig:norm_overlay}).  However, each Mesoburst has a fast rise followed by a slow decay, and they occur over similar timescales of $\sim10$--$30$\,s.

\subsubsection{Parameters \& Correlations}

\label{sec:mesostruc}

\par Due to the complexity structure of Mesobursts, we do not fit them mathematically as we did for Normal Bursts.  Instead we define a number of different parameters for each Mesoburst, listed below:
\begin{itemize}
\item Total burst fluence and burst fluence divided by persistent emission.
\item Peak 1\,s binned rate and peak rate divided by persistent emission.
\item Rise time, fall time and total time.
\end{itemize}
\par The mean and standard deviation of each of these parameters, calculated from \textit{RXTE} data, is presented in Table \ref{tab:meso_param}.  Due to the relative low number of Mesobursts compared to Normal Bursts, we only present the results from the combined set of bursts in both Outbursts 1 \& 2.  In general, Mesobursts are longer in duration than Normal Bursts, and have significantly smaller amplitudes and fluences (compare e.g. Table \ref{tab:params_perob}).

\begin{table}
\centering
\begin{tabular}{l c c}
\hline
\hline
&Mean&Standard Deviation\\
\hline
Fluence \scriptsize(cts\,PCU$^{-1}$)&6067&6707\\
Peak Rate \scriptsize(cts\,s$^{-1}$\,PCU$^{-1}$)&665.4&658.4\\
Fluence/$k$&48.6&32.8\\
Peak Rate/$k$&5.32&4.0\\
Rise Time \scriptsize(s)&6.95&4.9\\
Fall Time \scriptsize(s)&18.28&10.8\\
Total Time \scriptsize(s)&25.88&13.3\\
\hline
\hline
\end{tabular}
\caption{A table showing the mean and standard deviation of 7 burst parameters of \textit{RXTE}-sampled Mesobursts from Outbursts 1 \& 2.  $k$ is the persistent emission rate during the observation in which a given burst was detected.}
\label{tab:meso_param}
\end{table}

\par Using the Spearman's Rank metric, we find a number correlations above the 5$\,\sigma$ level:
\begin{itemize}
\item Fluence is correlated with peak rate ($>10\,\sigma$), peak rate divided by persistent rate ($6.7\,\sigma$), fall time ($6.8\,\sigma$), total time ($6.0\,\sigma$).
\item Fluence divided by persistent rate is correlated with peak rate divided by persistent rate ($7.3\,\sigma$).
\item Peak rate is also correlated with peak rate divided by persistent rate ($7.4\,\sigma$), fall time ($5.8\,\sigma$) and persistent level ($6.2\,\sigma$).
\item Rise time correlates with total time ($5.4\,\sigma$).
\item Fall time correlates with total time ($>10\,\sigma$).
\end{itemize}

Again, the correlation between fluence and peak rate is expected, as is the correlation between peak rate and peak rate divided by persistent rate.

\subsubsection{Colour Evolution}

\par The hardness ratio of the emission from the source decreases significantly during Mesobursts, with the PCA 8--60/2--7\,keV colour decreases from $\sim0.6$ between bursts to $\sim0.2$ at the peak of a burst.  Due to the poor statistics of these features compared with Normal Bursts, we were unable to check for evidence of hardness-intensity hysteresis.

\subsection{Structured ``Bursts''}

\par We define Structured Burst observations as observations in which the recurrence time between bursts is less than, or approximately the same as, the duration of a single burst.  Structured Bursts constitute the most complex behaviour we find in our dataset.  Unlike the other classes of burst we identify, Structured Bursts are not easily described as discrete phenomena.  We find Structured Bursts in 54 observations which are listed in Appendix \ref{app:obs}.
\par
\begin{figure}
  \centering
  \includegraphics[width=.9\linewidth, trim={0.4cm 0 1.1cm 0},clip]{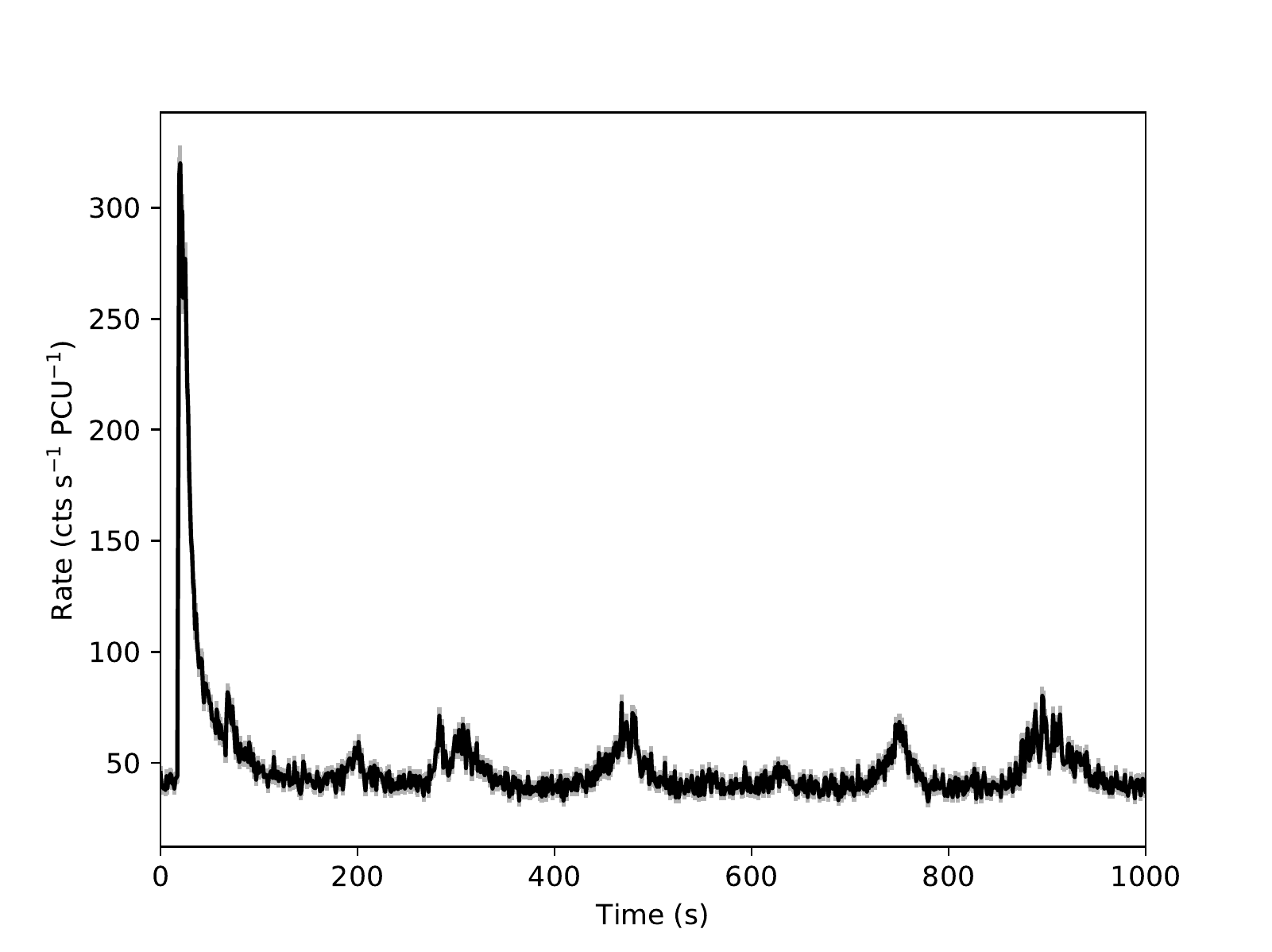}
  \caption{\small A lightcurve from \textit{RXTE}/PCA observation 10401-01-57-03, showing a Mesoburst occuring during a period of Structured Bursting.}
  \label{fig:meso_in_struc}
\end{figure}

 In both outbursts covered by \textit{RXTE}, Structured Bursts occur in the time between the end of the main outburst and the start of a rebrightening event.  In both cases these periods of structured outbursts are preceded by a period populated by Mesobursts.  Mesobursts occurred between MJDs 50248 and 50261 in Outburst 1, and between 50577 and 50618 in Outburst 2; during these intervals, \textit{RXTE} observed the source for a total of 81\,ks.  Notably, as we show in Figure \ref{fig:meso_in_struc}, one Outburst 1 \textit{RXTE} lightcurve containing Structured Bursting also contains a bright Mesoburst.

\par In both outbursts, the amplitude of Structured Bursting behaviour decreases as the outburst approaches the peak of the rebrightening event.  This amplitude continues to decrease as the Structured Burst behaviour evolves into the low-amplitude noisy behaviour associated with the source's evolution towards the hard state.

\subsubsection{Colour Evolution}

\par
\begin{figure}
  \centering
  \includegraphics[width=.9\linewidth, trim={0.4cm 0 1.cm 0},clip]{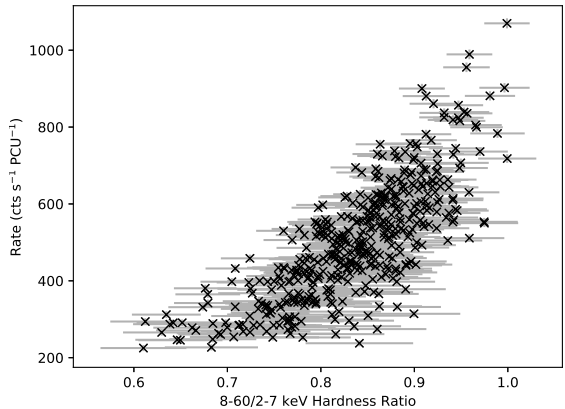}
  \caption{\small A 1\,s-binned hardness-intensity diagram from \textit{RXTE} observation 20078-01-23-00, showing that hardness tends to correlate with intensity during Structured Bursting.  Data are binned to 8\,s, and background has been estimated by subtracting mean count rates in the relevant energy bands from \textit{RXTE} OBSID 30075-01-26-00.}
  \label{fig:struc_hard}
\end{figure}

 We produce hardness-intensity diagrams for a number of Structured Burst observations; we show a representative example in Figure \ref{fig:struc_hard}.  We find that hardness is strongly correlated with count rate during this class of bursting, but that the magnitude of the change in hardness is no greater than $\sim30\%$.  This is less than the change in hardness that we see during Normal or Minibursts.  We also find no evidence of hysteretic hardness-intensity loops from Structured Bursts.

\subsubsection{Types of Structured Bursting}
\label{sec:struc_var}

\par
\begin{figure}
  \centering
  \includegraphics[width=.9\linewidth, trim={0.8cm 0 1.5cm 0},clip]{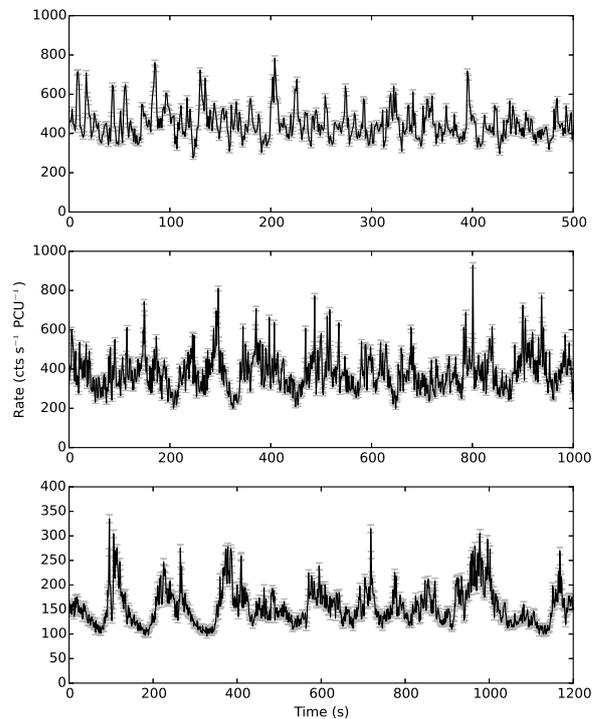}
  \caption{\small A selection of \textit{RXTE} lightcurves from Structured Bursting observations of the Bursting Pulsar.  \textbf{Top:} a lightcurve from Outburst 1 showing flaring on timescales of $\sim10$\,s.  \textbf{Middle:} a lightcurve from Outburst 1 showing the same flaring behaviour with an additional slower modulation over $\sim50$\,s.  \textbf{Bottom:} a lightcurve from Outburst 2 showing a regular sequence of flat-bottomed dips and multi-peaked flaring.  These show the wide variety of variability patterns that we classify as `Structured Bursting'.}
  \label{fig:Types_Struc}
\end{figure}

 In Figure \ref{fig:Types_Struc}, we present a selection of lightcurves which show the different types of variability that can be seen during periods of Structured Bursting.  These consist of a variety of patterns of peaks and flat-bottomed dips, and both \textit{RXTE}-observed outbursts show several of these different patterns of Structured Bursting.  As all types of Structured Bursting have similar amplitudes and occur in the same part of each outburst, we consider them to be generated by the same physical process.  We do not seperate these patterns into separate subclasses in this paper.

\section{Discussion}

\par We analyse all available X-ray data from the first 3 outbursts of the Bursting Pulsar.  The bursting behaviour evolves in a similar way during these outbursts, strongly associating them with the Bursting Pulsar and suggesting an underlying connection between the classes of burst.  We also find that both Outbursts 1 \& 2 showed `rebrightening events' similar to those seen in a number of other LMXBs as well as in dwarf novae (e.g. \citealp{Wijnands_1808,Patruno_Reflares2})
\par We find that the Type II X-ray bursts from these data can be best described as belonging to four phenomenological classes: Normal Bursts, Minibursts, Mesobursts and Structured Bursts.  For each of these four classes, we collect a number of statistics to shed light on the physical mechanisms that generate these lightcurve features.
\par Normal Bursts and Minibursts both represent the ``Type II'' bursting behaviour which is observed most commonly from this source.   Mesobursts occur much later on in the outburst and show fast-rise slow-decay profiles; they are generally much fainter and more structured than Normal Bursts.  Finally, Structured Bursts form continuous highly structured regions of bursting over timescales of days.  All Normal Bursts and some Minibursts show count rate `dips' after the main burst, while Mesobursts and Structured Bursts do not.  In addition to this, some Normal and Mesobursts show count rate `plateaus'; regions of roughly stable count rate above the persistent level which last for $\sim10$s of seconds.  These features are also sometimes seen in Mesobursts, while Minibursts and Structured Bursts never show these structures.
\par Here we discuss these results in the context of models proposed to explain Type II bursting.  We also compare our results with those of previous studies on bursting in both the Bursting Pulsar and the Rapid Burster.

\subsection{Evolution of Outburst and Bursting Behaviour}

\par In general, Outburst 1 was brighter than Outburst 2, with the former having a peak 2--60\,keV intensity a factor of $\sim1.7$ greater than the latter.  However, in Figure \ref{fig:global_ob} we show that both outbursts evolve in a similar way.  In both outbursts, the intensity of the Bursting Pulsar reaches a peak of order $\sim1$\,Crab before decreasing over the next $\sim100$ days to a level of a few tens of mCrab.  A few 10s of days after reaching this level, the lightcurves of both outbursts show a pronounced `rebrightening' event, during which the intensity increases to $\sim100$\,mCrab for $\sim10$ days.  Outburst 1 shows a second rebrightening event $\sim50$ days after the first.  It is unclear whether any rebrightening events occurred in Outburst 3 due to a lack of late-time observations with soft X-ray telescopes.  X-ray `rebrightening' events have been seen after the outbursts of a number of other LMXBs with both neutron star and black hole primaries: including SAX J1808.4-3658 \citep{Wijnands_1808}, XTE J1650-500 \citep{Tomsick_MiniOutbursts} and IGR J17091-3625 \citep{Court_IGRClasses}.
\par As we have shown in Figures \ref{fig:ob_evo1} \& \ref{fig:ob_evo2}, the nature of bursts from the Bursting Pulsar evolves in a similar way in both Outbursts 1 \& 2.  Starting from around the peak of each outburst, both Normal and Minibursts are observed.  The fluence of these bursts decrease over time as the X-ray intensity of the source decreases, before bursting shuts off entirely when the 2--16\,keV flux falls below $\sim0.1$\,Crab.  After a few 10\,s of days with no bursts, bursting switches back on in the form of Mesobursts; this occurs during the tail of a rebrightening event in Outburst 1, but in the tail of the main outburst in Outburst 2.  Mesobursting continues until the 2--16\,keV source flux falls below $\sim0.03$\,Crab, at which point we observe the onset of Structured Bursting.  In both Outbursts, Structured Bursting stops being visible a few 10s of days later during the start of a rebrightening event.  Because this evolution is common to both of the outbursts observed by \textit{RXTE}, this strongly indicates that the nature of bursting in the Bursting Pulsar is connected with the evolution of its outbursts.  Additionally, with the exceptions of Normal and Minibursts, we show that each class of burst is mostly found in a distinct part of the outburst corresponding to a different level of persistent emission.
\par
\begin{figure*}
  \centering
  \includegraphics[width=.9\linewidth, trim={3.7cm 0cm 4.2cm 0cm},clip]{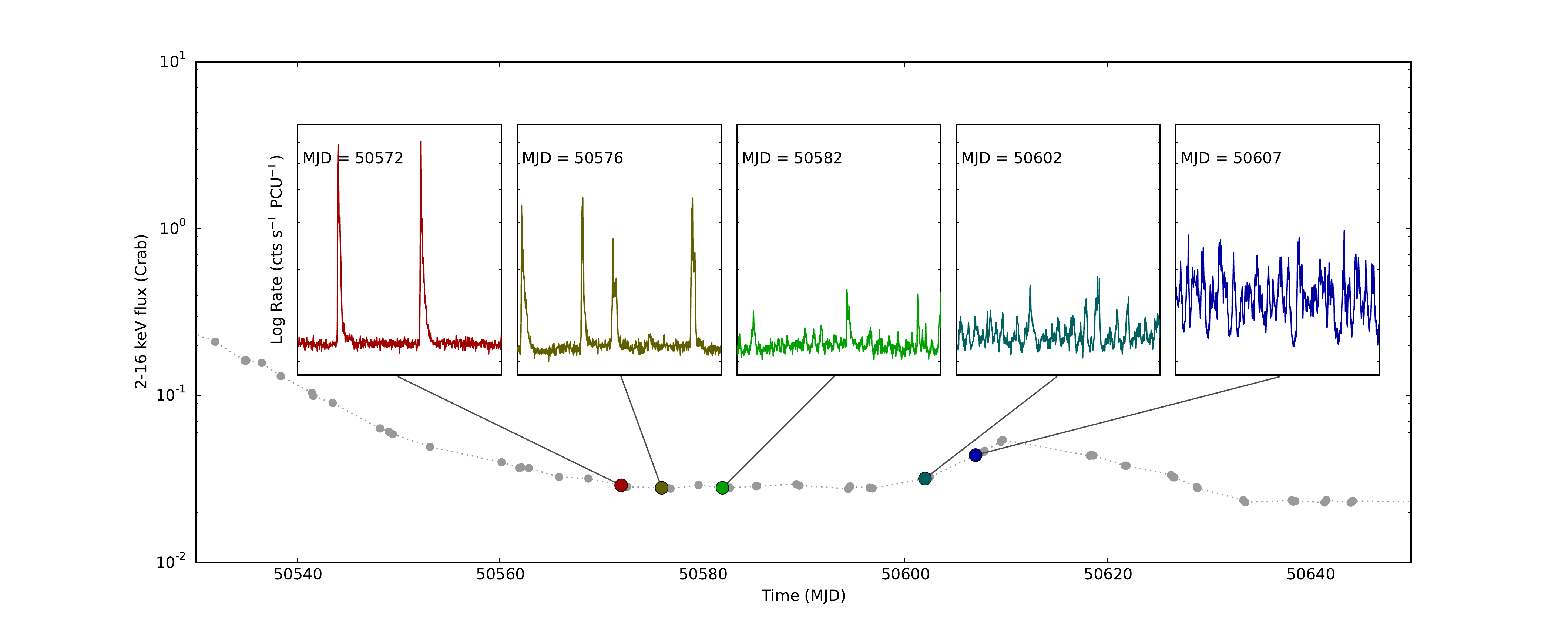}
  \caption{\small A series of lightcurves from \textit{RXTE}/PCA observations of Outburst 2, showing a gradual evolution from Mesobursts to Structured Bursting over a period of $\sim30$ days.  Each inset lightcurve is plotted with the same $y$-scaling, and each corresponds to 2\,ks of data.}
  \label{fig:meso_to_struc}
\end{figure*}

 In Figure \ref{fig:meso_to_struc}, we show lightcurves from Outburst 2 taken a few days before and after the transition from Mesobursts to Structured Bursting.  We can see that, as the system approaches this transition, Mesobursts become more frequent and decrease in amplitude.  Additionally in Figure \ref{fig:meso_in_struc} we show a lightcurve which contains both a Mesoburst and Structured Bursting.  We find that, instead of a well-defined transition between these bursting classes, there is a more gradual change as Mesobursting evolves into Structured Bursting.  This suggests that the same mechanism is likely to be responsible for both of these types of burst.
\par The transition between Normal Bursts and Mesobursts, however, is not smooth; in both outbursts these two classes of bursting are separated by $\sim10$ day gaps in which no bursts of any kind were observed at all.  If all our classes of burst are caused by the same or similar processes, any model to explain them will also have to explain these periods with no bursts.

\subsection{Parameter Correlations}

\par We extracted a number of phenomenological parameters from each Normal Burst, Miniburst and Mesoburst.  For Normal Bursts, we extracted a large number of parameters by fitting a phenomenological model described in Section \ref{sec:struc}.  For Minibursts and Mesobursts we extracted recurrence times and persistent emission-subtracted peak rates; we also calculated burst fluences by integrating the persistent emission-subtracted rate over the duration of the burst.  We do not extract similar parameters for Structured Bursts due to their complex nature.
\par  In all three of the classes of burst we consider, we found that fluence and peak rate correlate strongly with persistent emission.  For each type of burst case, the slope of these correlations is consistent with being equal during Outbursts 1 \& 2.
\par We also compared the Normal Bursts in Outburst 1 with the Normal Bursts in Outburst 2.  The only significant statistical differences we found between these two populations were in the burst peak rate and the burst fluence; both of these parameters are generally higher for Normal Bursts in Outburst 1.  As both of these parameters strongly depend on the persistent emission, both of these differences can be attributed to the fact that Outburst 1 was significantly brighter at peak than Outburst 2.
\par For Normal Bursts, we found additional correlations.  Of particular note, we found that both the fall time and the recovery timescale of a `dip' is proportional to its amplitude, which has implications for the possible mechanism behind these features.  We discuss this further in Section \ref{sec:mod}.
\par These findings strongly suggest that the properties of Normal, Mini and Mesobursts depend on the persistent luminosity of the Bursting Pulsar.  Assuming that this persistent luminosity is proportional to $\dot{M}$, this suggests that all classes of bursting are sensitive to the accretion rate of the system.  Additionally, with the exceptions of Normal and Minibursts, we find that each class of burst is mostly found in a distinct part of the outburst corresponding to a different level of persistent emission.  We suggest that Normal, Meso and Structured Bursts may in fact be manifestations of the same physical instability but at different accretion rates.  This is supported by the observation of a Mesoburst during a period of Structured Bursting, which we show in the lightcurve in Figure \ref{fig:meso_in_struc}.  This shows that the conditions for both Meso and Structured Bursting can be met at the same time.

\subsection{Comparison with Previous Studies}

\par In their study of bursts in the Bursting Pulsar, \citet{Giles_BP} found evidence for three distinct classes of Type II bursts in the Bursting Pulsar:

\begin{itemize}
\item ``Bursts'' (hereafter G$_1$ bursts to avoid confusion), the common Type II bursts seen from the source.
\item ``Minibursts'' (hereafter G$_2$  Bursts), with smaller amplitudes up to $\sim2$ times the persistent emission level.
\item ``Microbursts'' (hereafter G$_3$  Bursts), second-scale bursts with amplitudes of $\sim50$--$100\%$ of the persistent level.
\end{itemize}

We find that Giles et al's G$_1$ category contains the bursts that we identify as Normal Bursts, while our Miniburst category contains the same bursts as Giles' G$_2$ category.  \citet{Giles_BP} only consider bursts up to MJD 50204 in their classification, and they could not classify any bursts that we identify as Mesobursts; under their framework, we find that Mesobursts would also be categorised as G$_1$.  We present the full mapping between Giles classes and our classes in a schematic way in Table \ref{tab:classcomp}.

\begin{table}
\centering
\begin{tabular}{c c}
\hline
\hline
 \scriptsize Our Class & \scriptsize Giles et al. Class  \\
\hline
Normal Bursts & G$_1$ \\
Mesobursts & G$_1$ \\
Minibursts & G$_2$ \\
Structured Bursts & - \\
 - & G$_3$ \\
\hline
\hline
\end{tabular}
\caption{A table showing how our burst classes map to those described in \citet{Giles_BP}.  Giles et al. do not consider the times during the outburst when Structured Bursts appear, and we consider G$_3$ bursts described by Giles et al. to be consistent with flicker noise.}
\label{tab:classcomp}
\end{table}

\par \citet{Giles_BP} note the presence of both dips and plateaus in Normal Bursts.  To calculate the fluence of each main burst and its associated dip, Giles et al. integrate the total persistent-emission-subtracted counts in each feature.  They calculate that ratio between burst fluence and `missing' dip fluence ($\phi_{B}/\phi_{d}$) is between 0.26 and 0.56 in Outburst 1 before correcting for dead-time effects.  Using bursts in which our mathematical fit gave well-constrained ($>5\,\sigma$) values for both burst and dip fluence, we find that $\phi_{B}/\phi_{d}$ is between 1.3 and 2.0 in Outburst 1 and between 1.3 and 2.9 in Outburst 2.  Our values differ significantly from those reported from Giles et al.; this is likely due to differing definitions of the persistent emission level and the start and end times of each dip, as Giles et al. do not report how they define these features.
\par Our values for the ratios between burst and dip fluences, as well as those of Giles et al., are affected by dead-time.  These effects cause the fluence of bursts to be under-reported, as can be inferred from Figure \ref{fig:minidips}, but the integrated counts in dips are not significantly affected \citep{Giles_BP}.  Therefore correcting for dead-time can only increase the value of $\phi_{B}/\phi_{d}$, and our result shows that the fluence of a burst is always greater than the fluence `missing' from a dip.
\par We find evidence of significant colour evolution during both Normal Bursts and Minibursts, which is strongly indicative of a spectral evolution (see also e.g. \citealp{Woods_OB2}).  Further work on the time-resolved spectra of this source will likely allow us to better understand the underlying physics of its behaviour.
\par Using data from the KONUS experiments aboard the GGS-Wind and Kosmos-2326 satellites, \citet{Aptekar_Recur} have previously found that the recurrence times between consecutive bursts in Outburst 1 are distributed with a constant mean of $\sim1776$\,s.  This is substantially longer than our value of 1209\,s that we find for Outburst 1, but our value is likely an underestimate due to a selection bias caused by the relatively short pointings of \textit{RXTE}.
\par Using \textit{Chandra} and \textit{XMM-Newton} data, we find a mean recurrence time for Outburst 3 of 1986\,s; as pointings with these instruments are significantly longer than the burst recurrence timescale, windowing effects are negligible.  As this value is close to the value that \citet{Aptekar_Recur} find for mean recurrence time, our result is consistent with the burst rate in all three outbursts being approximately the same.
\par Previous studies with \textit{CGRO}/BATSE have found that the burst rate during the first few days of Outbursts 1 \& 2 was significantly higher than during the rest of each outburst \citep{Kouveliotou_BP,Woods_OB2}.  As \textit{RXTE} did not observe either of these times, we are unable to test this result.

\subsection{Comparison with other objects}

\par In \citet{Court_BPTMSP} we discuss the possibility that some of the behaviour in the Bursting Pulsar could be due to fluctuations in the magnetospheric radius of the system close to the corotation radius.  This behaviour (e.g. \citealp{Bogdanov_TMSPVar,Ferrigno_TMSPVar}) is also seen in `Transitional Millisecond Pulsars' (TMSPs): objects which alternate between appearing as X-ray pulsars and radio pulsars (see e.g. \citealp{Archibald_Link,Papitto_Swings}).
\par Another natural comparison to the Bursting Pulsar is the Rapid Burster \citep{Lewin_RBDiscovery}, a neutron star LMXB in the globular cluster Liller I.  This object is the only LMXB other than the Bursting Pulsar known to unambiguously exhibit Type II bursting behaviour during outbursts.  \citet{Rappaport_BPHistory} have previously proposed that the Bursting Pulsar, the Rapid Burster and other neutron star LMXBs form a continuum of objects with different magnetic field strengths.
\par We compare our study of bursts in the Bursting Pulsar with studies of Type II bursts in the Rapid Burster, particularly the detailed population study performed by \citet{Bagnoli_PopStudy}.  \citet{Bagnoli_PopStudy} found that Type II bursting begins during the decay of an outburst in the Rapid Burster.  This is the same as what we see in the Bursting Pulsar, where we find Normal Bursting behaviour starts during the outburst decay.  \citet{Bagnoli_PopStudy} found that all bursting in the Rapid Burster shuts off above an Eddington Fraction of $\gtrsim0.05$, whereas we find bursting in the Bursting Pulsar shuts off \textit{below} a 2--16\,keV flux of Eddington fraction of $\sim0.1\,Crab$: assuming that the peak persistent luminosity of the Bursting Pulsar was approximately Eddington Limited (e.g. \citealp{Sazonov_BPGranat}), this value corresponds to an Eddington fraction of order $\sim0.1$.  This suggests that Type II bursting in these two objects happen in very different accretion regimes.
\par Bagnoli et al. showed that bursting behaviour in the Rapid Burster falls into a number of `bursting modes', defined by the morphology of individual Type II bursts.  In particular, they find that Type II bursts in the Rapid Burster fall into two classes (see also \citealp{Marshall_2types}), lightcurves of which we reproduce in Figure \ref{fig:bagnoli_lcs}:

\begin{figure}
  \centering
  \includegraphics[width=.9\linewidth, trim={0.8cm 0 1.4cm 0},clip]{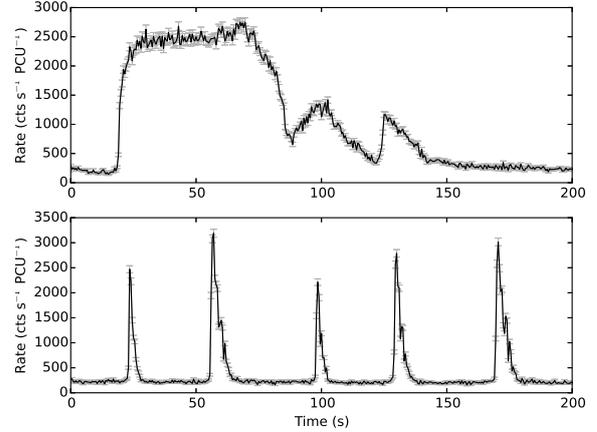}
  \caption{\small  \textit{RXTE} lightcurves of representative Long (top) and Short (bottom) bursts from the Rapid Burster.  These bursts were identified and classified by \citet{Bagnoli_PopStudy}.}
  \label{fig:bagnoli_lcs}
\end{figure}

\begin{itemize}
\item Short near-symmetric Bursts with timescales of $\sim10s$ of seconds and peak rates near the Eddington Limit.
\item Long bursts with a fast rise, a long $\sim100$\,s plateau at peak rate followed by a fast decay.  The level of the plateau is generally at or near the Eddington Limit.
\end{itemize}

\par Short bursts are very similar in shape to Normal Bursts in the Bursting Pulsar, but we find no analogue of long bursts in our study.  \citet{Bagnoli_PopStudy} suggests that the `flat-top' profile of long bursts could be due to the effects of near-Eddington accretion, and they show that the intensity at the `flat top' of these bursts is close to Eddington limit.  Previous works have shown that the persistent emission of the Bursting Pulsar is Eddington-limited at peak, and therefore bursts from the Bursting Pulsar are significantly super-Eddington \citep{Sazonov_BPGranat}.   We suggest, therefore, that Long Bursts cannot occur in systems with a persistent rate approaching the Eddington Limit.  This could explain why Long Bursts are not seen during periods of Normal Bursting in the Bursting Pulsar (during which the persistent emission is $\gtrsim{20}$\% of Eddington), but it remains unclear why these features are not seen later in each outburst when the Bursting Pulsar is fainter.  Alternatively, all the differences we see between bursts produced by the Rapid Burster and the Bursting Pulsar could be explained if the physical mechanisms behind these bursts are indeed different between the objects.
\par \citet{Bagnoli_PopStudy} also find a number of correlations between burst parameters in the Rapid Burster, which we can compare with our results for the Bursting Pulsar.  We find a number of similarities between the two objects:

\begin{itemize}
\item The fluence of a burst correlates with its amplitude.
\item The duration of a burst does not correlate\footnote{We state two parameters do not correlate if their Spearman Rank score corresponds to a significance $<3\sigma$.} with the persistent emission.
\item The recurrence time between consecutive bursts does not depend on the persistent emission.
 \end{itemize}

\par There are also a number of differences between the set of correlations between burst parameters in these two systems:

\begin{itemize}
\item Burst duration is correlated with burst fluence in the Rapid Burster, but these have not been seen to correlate in the Bursting Pulsar.
\item Burst duration, peak rate and burst fluence are all correlated with burst recurrence time in the Rapid Burster.  We have not found any of these parameters to correlate with burst recurrence time in the Bursting Pulsar.
\item Peak rate and burst fluence correlate with persistent emission in the Bursting Pulsar, but this is not true for bursts of a given type in the Rapid Burster.
\end{itemize}

\par As the neither the fluence nor the class of a burst in the Rapid Burster depend strongly on persistent emission, which can be used as a proxy for $\dot{M}$, this suggests that the process that triggers Type-II bursts in this source is not strongly dependent on the global accretion rate.  However the strong correlations between persistent emission and burst peak and fluence we find in the Bursting Pulsar show that the energetics of individual bursts strongly depend global accretion rate at that time.
\par It has previously been noted that consecutive Normal Bursts in the Bursting pulsar do not show a strong correlation between recurrence time and fluence (\citealp{Taam_Evo,Lewin_BP}, however see \citealp{Aptekar_OscRel}).  This correlation would be expected if the instability took the form of a relaxation oscillator, as it does in the Rapid Burster \citep{Lewin_TypeII}.  However, we also find that the arrival times of Normal Bursts from the Bursting Pulsar are not consistent with a Poisson distribution with constant mean.  This implies either that bursts are also not independent events in the Bursting Pulsar, or that the frequency of these bursts is not constant throughout an outburst as reported by \citet{Aptekar_Recur}.

\subsection{Comparison with Models of Type II Bursts}

\label{sec:mod}

\par To our knowledge no models have been proposed which can fully explain Type II bursting behaviour, but several models have been proposed in the context of Type II bursting from the Rapid Burster MXB 1730-33.  A number of models invoke viscous instabilities in the inner disk as the source of cyclical bursting (e.g. \citealp{Taam_Evo,Hayakawa_Type2Mod}), but these fail to explain why the majority of Neutron Star LMXBs do not show this behaviour.
\par \citet{Spruit_Type2Mod} use a different approach.  They show that, in some circumstances, the interaction between an accretion disk and a rapidly rotating magnetospheric boundary can naturally set up a cycle of discrete accretion events rather than a continuous flow  (see also \citealp{Dangelo_Episodic1,Dangelo_Episodic2,vandenEijnden_RB,Scaringi_Gating}).  \citet{Walker_Type2Mod} suggests that, for a neutron star with a radius less than its ISCO, a similar cycle of accretion can be set up when considering the effects of a high radiative torque.  All of these models suggest that Type II bursts are caused by sporadic accretion events onto the neutron star, which in turn are caused by instabilities that originate in the inner part of the accretion disk.  For a more detailed review of these models, see \citet{Lewin_Bursts}.
\par
\begin{figure}
  \centering
  \includegraphics[width=.9\linewidth, trim={0cm 0 0cm 0},clip]{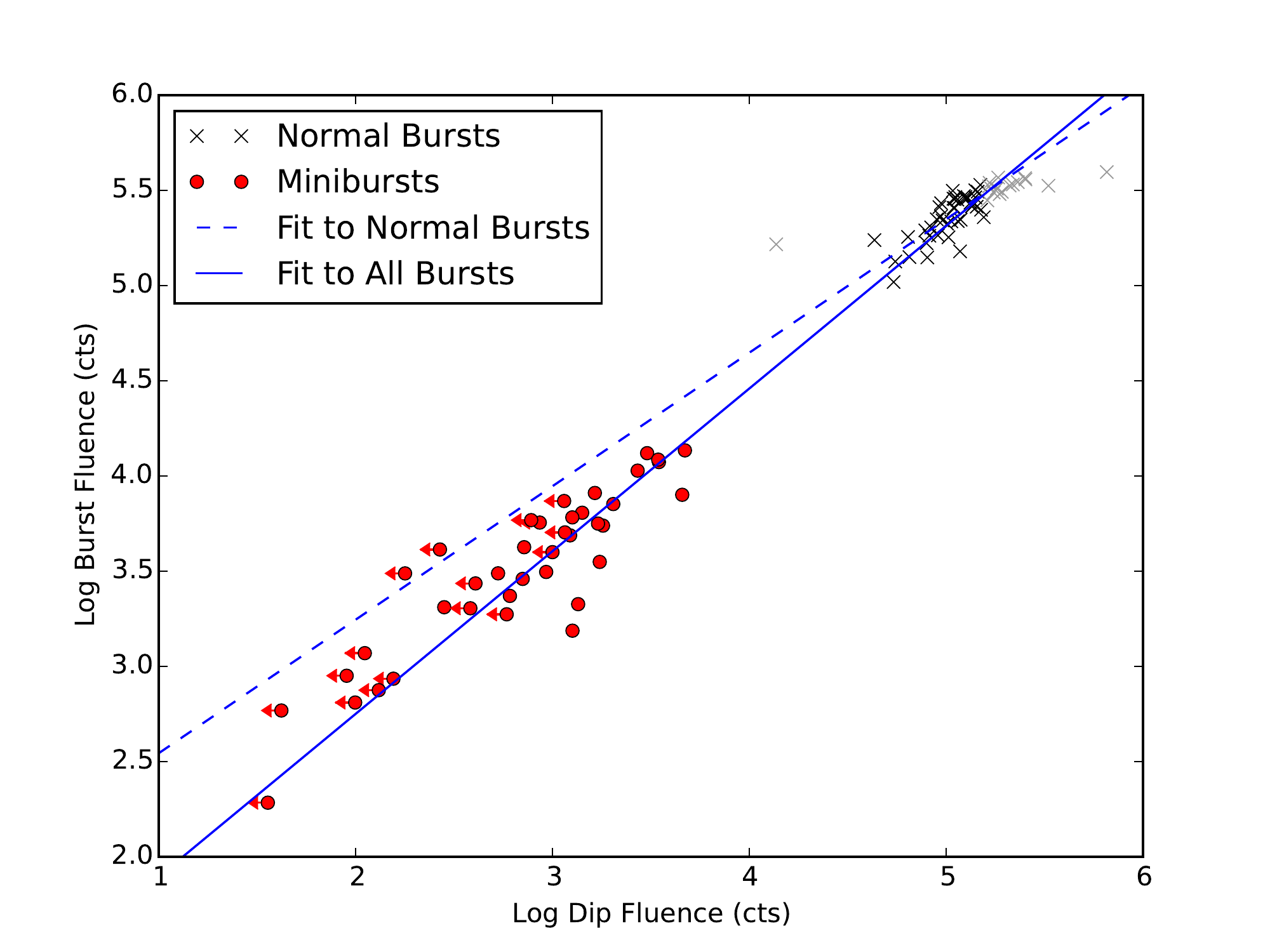}
  \caption{\small A scatter plot showing the relationship between burst fluence and `missing' dip fluence for Normal Bursts (black) and Minibursts (Red), with the best fit power law plotted in solid blue.  A power law fit to just the Normal Bursts (blue dashed line) also approaches the Minibursts.  Note that the Normal Bursts plotted in grey were not used to calculate this latter fit, as the effects of instrumental dead-time cause high burst fluences to be under-reported.  Upper limits on Miniburst dip fluences are shown with arrows.}
  \label{fig:minidips}
\end{figure}

 All of the models discussed above are able to reproduce some of the features we see from bursts in the Bursting Pulsar.  In particular, the `dip' we see after Normal Bursts has previously been interpreted as being caused by the inner disk refilling after a sudden accretion event (e.g. \citealp{Younes_Expo}).  As these dips are also seen after some Minibursts, we could also interpret Minibursts as being caused by a similar cycle.  To test this idea, in Figure \ref{fig:minidips} we present a scatter plot of the burst and dip fluences for all Normal Bursts and Minibursts.  In both classes of burst, there is a strong correlation between these two parameters.  We find that a power law fit to the Normal Bursts in this parameter space also describes the Minibursts.  This suggests that the same relationship between burst fluence and missing dip fluence holds for both types of burst, although the two populations are not continuous.  This suggests that Minibursts are energetically consistent with being significantly fainter versions of Normal Bursts.

\par The models of \citet{Spruit_Type2Mod} and \citet{Walker_Type2Mod} also have shortcomings when used to describe the Bursting Pulsar.  \citet{Walker_Type2Mod} state that their model only produces Type II bursts for a very specific set of criteria on the system parameters.   One of these criteria is an essentially non-magnetic ($B=0$) neutron star.  This is inconsistent with observations of cyclotron lines from the Bursting Pulsar and the presence of a persistent pulsar, which suggest a surface field strength of order 10$^{11}$\,G \citep{Doroshenko_NBFlash}.
\par Unlike models based on viscous instability, the model of \citet{Spruit_Type2Mod} does not impose a correlation between burst fluence and burst recurrence time (see e.g. the evaluation of this model in the context of the Rapid Burster performed by \citealp{Bagnoli_PopStudy}).  However, it does predict a strong correlation between burst recurrence time and mean accretion rate, which is not consistent with our results for the Bursting Pulsar.
\par In general, we find that models established to explain bursting in the Rapid Burster are poor at explaining bursting in the Bursting Pulsar.  Any model which can produce Type II bursting in both systems fails to explain why other systems do not also show this behaviour.  Our results suggest that Type II bursts in the Rapid Burster and the Bursting Pulsar may require two separate models to be explained.

\subsubsection{Evidence of Thermonuclear Burning}

\par We also consider the possibility that some of our observations could be explained by thermonuclear burning in the Bursting Pulsar.  A thermonuclear origin for the main part of Normal Type II X-ray bursts has been ruled out by previous authors (e.g. \citealp{Lewin_BP}), but it is less clear that associated features could not be explained by this process.
\par It has been shown that, above a certain accretion rate, thermonuclear burning on the surface of a neutron star should be stable; below this rate, thermonuclear burning takes place in the form of Type I bursts (e.g. \citealp{Fujimoto_Shellstab,Bildsten_Regimes}).  \citet{Bildsten_Nuclear} have previously studied which form thermonuclear burning on the Bursting Pulsar would take.  They find that the presence and profile of a thermonuclear burning event on the Bursting Pulsar would be strongly dependent on both the accretion rate $\dot{M}$ and the magnetic field strength $B$.  They predict that, for $B\gtrsim3\times10^{10}$\,G, burning events would take the form of a slowly propagating burning front which would result in a low-amplitude X-ray burst with a timescale of several minutes.  Measurements of the Bursting Pulsar taken during Outburst 3 suggest a surface field strength of $>10^{11}$\,G, in turn suggesting that the Bursting Pulsar exists in the regime in which this burning behaviour is possible.
\par The `plateau' events after Normal Bursts are consistent with the slow burning predicted by \citet{Bildsten_Nuclear}.  This picture is consistent with models for Type II X-ray bursts involving spasmodic accretion events (e.g. \citealp{Spruit_Type2Mod,Walker_Type2Mod}), as plateaus always occur after a Type II burst has deposited a large amount of ignitable material onto the neutron star surface.  However in this picture it would be unclear why many Normal Bursts do not show this plateau feature.  Mesobursts can also exhibit plateaus, and are therefore may also be products of spasmodic accretion onto the neutron star.
\par However, the interpretation of Mesobursts as being caused by discrete accretion events is difficult to reconcile with the fact that these features never show dips.  \citet{Bildsten_Nuclear} show that, at smaller values of $\dot{M}$, nuclear burning on the Bursting Pulsar could become unstable.  Mesobursts are only seen during the latter stages of Outbursts 1 \& 2, when the accretion rate is well below 0.1 Eddington.  An interesting alternative possibility is that Mesobursts are a hybrid event, consisting of a flash of unstable thermonuclear X-ray burning followed by a slower quasi-stable burning of residual material in the form of a propagating burning front.
\par This picture would also be able to explain why Mesobursts are only seen during the latter parts of each outburst.  As the accretion rate onto the Bursting Pulsar approaches Eddington during the peak of its outbursts, it is likely that the accretion rate is high enough that only stable burning is permitted.  During the smaller rebrightening events after the main part of each outburst, the accretion rate is $\sim1$--2 orders of magnitude lower, and hence the system may then be back in the regime in which Type I burning is possible.  Additional studies of the spectral evolution of Mesobursts will be required to further explore this possibility.
\par Previous authors have discussed the possibility of a marginally stable burning regime on the surface of neutron stars (not to be confused with the previously mentioned quasi-stable burning).  In this regime, which occurs close to the boundary between stable and unstable burning, \citet{Heger_MargStab} showed that an oscillatory mode of burning may occur.  They associated this mode of burning with the mHz QPOs which have been observed in a number of neutron star LMXBs (e.g. \citealp{Revnivtsev_MargStab,Altamirano_MargStab}).  These QPOs only occur over a narrow range of source luminosities, show a strong decrease in amplitude at higher energies, and they disappear after a Type I burst (e.g. \citealp{Altamirano_MargStab}).
\par Lightcurves of objects undergoing marginally stable burning qualitatively resemble those of Structured Bursting in the Bursting Pulsar, raising the possibility of a thermonuclear explanation for Structured Bursting.  However, as we show in Figure \ref{fig:ob_evo1}, Structured Bursting during Outburst 1 occurred during a period of time in which the Bursting Pulsar's luminosity changed by $\sim1$ order of magnitude.  In addition to this, in Figure \ref{fig:meso_in_struc} we show an example of a Mesoburst during a period of Structured Bursting.  If Mesobursts can be associated with Type I bursts, any marginally stable burning on the surface of the Bursting Pulsar should have stopped after this event.  Due to these inconsistencies with observations of marginally stable burning on other sources, it is unlikely that Structured Bursting is a manifestation of marginally stable burning on the Bursting Pulsar.
\par \citet{Linares_MargStab} observed yet another mode of thermonuclear burning during the 2010 outburst of the LMXB Terzan 5 X-2.  They observed a smooth evolution from discrete Type I bursts into a period of quasi-periodic oscillations resembling Structured Bursting.  This behaviour resembles the evolution we observe between Mesobursts and Structured Bursting in Outbursts 1 \& 2 of the Bursting Pulsar (as shown in Figure \ref{fig:meso_to_struc}; compare with Figure 1 in \citealp{Linares_MargStab}).  However there are a number of differences between the evolutions seen in both objects.  In Terzan 5 X-2 the recurrence timescale of Type I bursts during the evolution is strongly related to the accretion rate of the source at the time, whereas there is no such strong relation between the two in Mesobursts from the Bursting Pulsar.  Additionally, the quasi-periodic oscillations in Terzan X-2 evolved smoothly back into Type I bursts later in the outburst, whereas Structured Bursting does not evolve back into Mesobursts in the Bursting Pulsar.  As such, it is unclear that Mesobursts and Structured Bursting can be associated with the unusual burning mode seen on Terzan 5 X-2.

\section{Conclusions}

\par We analyse all X-ray bursts from the Rapid Burster seen by \textit{RXTE}/PCA during its first and second outbursts, as well as bursts seen by other missions during the third outburst of the source.  We conclude that these bursts are best described as belonging to four separate classes of burst: Normal Bursts, Mesobursts, Minibursts and Structured Bursts.  We find that the bursting behaviour in these four classes evolves in a similar way throughout the first two outbursts of the Bursting Pulsar.  We present a new semi-mathematical model to fit to the Normal Bursts in this object.  Using this new framework, we will be able better quantify Bursting-Pulsar-like X-ray bursts when they are observed in other objects in the future.
\par We find the bursts in the Rapid Burster and the Bursting Pulsar to be different in burst profile, peak Eddington ratio, and durations.  While the fluence of Type II bursts in the Bursting Pulsar depend strongly on the persistent emission at the time, this is not the case in the Rapid Burster.  Additionally the waiting time between bursts in the Rapid Burster depend heavily on the fluence of the preceding burst, but we do not find this in the Bursting Pulsar.  Therefore, it would be reasonable to conclude that the bursting in these two objects is generated by two different mechanisms.
\par However, it is also important to note a number of similarities between the Bursting Pulsar and the Rapid Burster.  Bursting behaviour in both objects depends on the global accretion rate of the system and the evolution of its outbursts.  For example, the recurrence times of bursts does not depend on persistent emission in either object, and nor does the duration of an individual burst.  Notably while Type II bursts in the Rapid Burster only occur at luminosities $L\lesssim0.05L_{Edd}$, we find that Normal bursts in the Bursting Pulsar only occur at $L\gtrsim0.1L_{Edd}$.  There is no overlap between the luminosity regimes, in terms of the Eddington Luminosity, at which bursting is observed in the two objects.  This leads to the alternative hypothesis that bursts in the two systems may be caused by similar processes, but that these processes take place in very different physical regimes.

\section*{Acknowledgements}

\par J. C. thanks the Science \& Technology Facilities Council and the Royal Astronomical Society for their financial support.  D.A. thanks the Royal Society, and the International Space Science Institute (\textit{ISSI}) for its support.  N.D. is supported by a Vidi grant from the Dutch Organisation for Scientific Research (NWO).  T.B. acknowledges financial contribution from the agreement ASI-INAF n.2017-14-H.0.
\par The authors acknowledge the use of AstroPy \citep{Astropy}, NumPy \citep{NumPy} and MatPlotLib \citep{Hunter_MatPlotLib} libraries for Python.  We also acknowledge the use of public data from the \textit{RXTE} \citep{Bradt_RXTE} archive, as well as \texttt{FTOOLS} \citep{Blackburn_FTools} for data manipulation.  Results provided by the ASM/\textit{RXTE} teams at MIT and at the RXTE SOF and GOF at NASA's GSFC.





\bibliographystyle{mnras}
\bibliography{/home/jamie/Documents/Bibliographies/refs}

\newcommand{\noop}[1]{}
\begin{thebibliography}{}
\makeatletter
\relax
\def\mn@urlcharsother{\let\do\@makeother \do\$\do\&\do\#\do\^\do\_\do\%\do\~}
\def\mn@doi{\begingroup\mn@urlcharsother \@ifnextchar [ {\mn@doi@}
  {\mn@doi@[]}}
\def\mn@doi@[#1]#2{\def\@tempa{#1}\ifx\@tempa\@empty \href
  {http://dx.doi.org/#2} {doi:#2}\else \href {http://dx.doi.org/#2} {#1}\fi
  \endgroup}
\def\mn@eprint#1#2{\mn@eprint@#1:#2::\@nil}
\def\mn@eprint@arXiv#1{\href {http://arxiv.org/abs/#1} {{\tt arXiv:#1}}}
\def\mn@eprint@dblp#1{\href {http://dblp.uni-trier.de/rec/bibtex/#1.xml}
  {dblp:#1}}
\def\mn@eprint@#1:#2:#3:#4\@nil{\def\@tempa {#1}\def\@tempb {#2}\def\@tempc
  {#3}\ifx \@tempc \@empty \let \@tempc \@tempb \let \@tempb \@tempa \fi \ifx
  \@tempb \@empty \def\@tempb {arXiv}\fi \@ifundefined
  {mn@eprint@\@tempb}{\@tempb:\@tempc}{\expandafter \expandafter \csname
  mn@eprint@\@tempb\endcsname \expandafter{\@tempc}}}

\bibitem[\protect\citeauthoryear{{Altamirano}, {van der Klis}, {Wijnands}  \&
  {Cumming}}{{Altamirano} et~al.}{2008a}]{Altamirano_MargStab}
{Altamirano} D.,  {van der Klis} M.,  {Wijnands} R.,   {Cumming} A.,  2008a,
  \mn@doi [\apjl] {10.1086/527355}, \href
  {http://adsabs.harvard.edu/abs/2008ApJ...673L..35A} {673, L35}

\bibitem[\protect\citeauthoryear{{Altamirano}, {van der Klis}, {M{\'e}ndez},
  {Jonker}, {Klein-Wolt}  \& {Lewin}}{{Altamirano}
  et~al.}{2008b}]{Altamirano_CrabNorm}
{Altamirano} D.,  {van der Klis} M.,  {M{\'e}ndez} M.,  {Jonker} P.~G.,
  {Klein-Wolt} M.,   {Lewin} W.~H.~G.,  2008b, \mn@doi [\apj] {10.1086/590897},
  \href {http://adsabs.harvard.edu/abs/2008ApJ...685..436A} {685, 436}

\bibitem[\protect\citeauthoryear{{Angelini}, {White}  \& {Stella}}{{Angelini}
  et~al.}{1991}]{Angelini_SMC}
{Angelini} L.,  {White} N.~E.,   {Stella} L.,  1991, \mn@doi [\apj]
  {10.1086/169895}, \href {http://adsabs.harvard.edu/abs/1991ApJ...371..332A}
  {371, 332}

\bibitem[\protect\citeauthoryear{{Aptekar} et~al.,}{{Aptekar}
  et~al.}{1997}]{Aptekar_OscRel}
{Aptekar} R.~L.,  et~al., 1997, Astronomy Letters, \href
  {http://adsabs.harvard.edu/abs/1997AstL...23..147A} {23, 147}

\bibitem[\protect\citeauthoryear{{Aptekar} et~al.,}{{Aptekar}
  et~al.}{1998}]{Aptekar_Recur}
{Aptekar} R.~L.,  et~al., 1998, \mn@doi [\apj] {10.1086/305100}, \href
  {http://adsabs.harvard.edu/abs/1998ApJ...493..404A} {493, 404}

\bibitem[\protect\citeauthoryear{{Archibald} et~al.,}{{Archibald}
  et~al.}{2009}]{Archibald_Link}
{Archibald} A.~M.,  et~al., 2009, \mn@doi [Science] {10.1126/science.1172740},
  \href {http://adsabs.harvard.edu/abs/2009Sci...324.1411A} {324, 1411}

\bibitem[\protect\citeauthoryear{{Astropy Collaboration} et~al.,}{{Astropy
  Collaboration} et~al.}{2013}]{Astropy}
{Astropy Collaboration} et~al., 2013, \mn@doi [\aap]
  {10.1051/0004-6361/201322068}, \href
  {http://adsabs.harvard.edu/abs/2013A%26A...558A..33A} {558, A33}

\bibitem[\protect\citeauthoryear{Azzalini}{Azzalini}{1985}]{Azzalini_Dist}
Azzalini A.,  1985, Scandinavian Journal of Statistics, 12, 171

\bibitem[\protect\citeauthoryear{{Bagnoli}, {in't Zand}, {D'Angelo}  \&
  {Galloway}}{{Bagnoli} et~al.}{2015}]{Bagnoli_PopStudy}
{Bagnoli} T.,  {in't Zand} J.~J.~M.,  {D'Angelo} C.~R.,   {Galloway} D.~K.,
  2015, \mn@doi [\mnras] {10.1093/mnras/stv330}, \href
  {http://adsabs.harvard.edu/abs/2015MNRAS.449..268B} {449, 268}

\bibitem[\protect\citeauthoryear{{Bildsten}}{{Bildsten}}{1995}]{Bildsten_Regimes}
{Bildsten} L.,  1995, \mn@doi [\apj] {10.1086/175128}, \href
  {http://adsabs.harvard.edu/abs/1995ApJ...438..852B} {438, 852}

\bibitem[\protect\citeauthoryear{{Bildsten} \& {Brown}}{{Bildsten} \&
  {Brown}}{1997}]{Bildsten_Nuclear}
{Bildsten} L.,  {Brown} E.~F.,  1997, \mn@doi [\apj] {10.1086/303752}, \href
  {http://adsabs.harvard.edu/abs/1997ApJ...477..897B} {477, 897}

\bibitem[\protect\citeauthoryear{{Bildsten} et~al.,}{{Bildsten}
  et~al.}{1997}]{Bildsten_Rev}
{Bildsten} L.,  et~al., 1997, \mn@doi [\apjs] {10.1086/313060}, \href
  {http://adsabs.harvard.edu/abs/1997ApJS..113..367B} {113, 367}

\bibitem[\protect\citeauthoryear{{Blackburn}}{{Blackburn}}{1995}]{Blackburn_FTools}
{Blackburn} J.~K.,  1995, in {Shaw} R.~A.,  {Payne} H.~E.,   {Hayes} J.~J.~E.,
  eds,  Astronomical Society of the Pacific Conference Series Vol. 77,
  Astronomical Data Analysis Software and Systems IV. p.~367

\bibitem[\protect\citeauthoryear{{Bogdanov} et~al.,}{{Bogdanov}
  et~al.}{2015}]{Bogdanov_TMSPVar}
{Bogdanov} S.,  et~al., 2015, \mn@doi [\apj] {10.1088/0004-637X/806/2/148},
  \href {http://adsabs.harvard.edu/abs/2015ApJ...806..148B} {806, 148}

\bibitem[\protect\citeauthoryear{{Bradt}, {Rothschild}  \& {Swank}}{{Bradt}
  et~al.}{1993}]{Bradt_RXTE}
{Bradt} H.~V.,  {Rothschild} R.~E.,   {Swank} J.~H.,  1993, \aaps, \href
  {http://adsabs.harvard.edu/abs/1993A%26AS...97..355B} {97, 355}

\bibitem[\protect\citeauthoryear{{Burrows} et~al.,}{{Burrows}
  et~al.}{2003}]{Burrows_XRT}
{Burrows} D.~N.,  et~al., 2003, in {Truemper} J.~E.,  {Tananbaum} H.~D.,  eds,
  \procspie Vol. 4851, X-Ray and Gamma-Ray Telescopes and Instruments for
  Astronomy.. pp 1320--1325, \mn@doi{10.1117/12.461279}

\bibitem[\protect\citeauthoryear{Court}{Court}{2017}]{Court_PANTHEON}
Court J.,  2017, jmcourt/PANTHEON: Zenodo Release,
  \mn@doi{10.5281/zenodo.1040704}, \url
  {https://doi.org/10.5281/zenodo.1040704}

\bibitem[\protect\citeauthoryear{{Court}, {Altamirano}, {Pereyra}, {Boon},
  {Yamaoka}, {Belloni}, {Wijnands}  \& {Pahari}}{{Court}
  et~al.}{2017}]{Court_IGRClasses}
{Court} J.~M.~C.,  {Altamirano} D.,  {Pereyra} M.,  {Boon} C.~M.,  {Yamaoka}
  K.,  {Belloni} T.,  {Wijnands} R.,   {Pahari} M.,  2017, \mn@doi [\mnras]
  {10.1093/mnras/stx773}, \href
  {http://adsabs.harvard.edu/abs/2017MNRAS.468.4748C} {468, 4748}

\bibitem[\protect\citeauthoryear{{Court}, {Altamirano}  \& {Sanna}}{{Court}
  et~al.}{2018}]{Court_BPTMSP}
{Court} J.~M.~C.,  {Altamirano} D.,   {Sanna} A.,  2018, \mn@doi [\mnras]
  {10.1093/mnrasl/sly056}, \href
  {http://adsabs.harvard.edu/abs/2018MNRAS.477L.106C} {477, L106}

\bibitem[\protect\citeauthoryear{{D'A{\`i}} et~al.,}{{D'A{\`i}}
  et~al.}{2015}]{Dai_OB3}
{D'A{\`i}} A.,  et~al., 2015, \mn@doi [\mnras] {10.1093/mnras/stv531}, \href
  {http://adsabs.harvard.edu/abs/2015MNRAS.449.4288D} {449, 4288}

\bibitem[\protect\citeauthoryear{{D'A{\`i}}, {Burderi}, {Di Salvo}, {Iaria},
  {Pintore}, {Riggio}  \& {Sanna}}{{D'A{\`i}} et~al.}{2016}]{Dai_Hlags}
{D'A{\`i}} A.,  {Burderi} L.,  {Di Salvo} T.,  {Iaria} R.,  {Pintore} F.,
  {Riggio} A.,   {Sanna} A.,  2016, \mn@doi [\mnras] {10.1093/mnrasl/slw112},
  \href {http://adsabs.harvard.edu/abs/2016MNRAS.463L..84D} {463, L84}

\bibitem[\protect\citeauthoryear{{D'Angelo} \& {Spruit}}{{D'Angelo} \&
  {Spruit}}{2010}]{Dangelo_Episodic1}
{D'Angelo} C.~R.,  {Spruit} H.~C.,  2010, \mn@doi [\mnras]
  {10.1111/j.1365-2966.2010.16749.x}, \href
  {http://adsabs.harvard.edu/abs/2010MNRAS.406.1208D} {406, 1208}

\bibitem[\protect\citeauthoryear{{D'Angelo} \& {Spruit}}{{D'Angelo} \&
  {Spruit}}{2012}]{Dangelo_Episodic2}
{D'Angelo} C.~R.,  {Spruit} H.~C.,  2012, \mn@doi [\mnras]
  {10.1111/j.1365-2966.2011.20046.x}, \href
  {http://adsabs.harvard.edu/abs/2012MNRAS.420..416D} {420, 416}

\bibitem[\protect\citeauthoryear{{Daigne}, {Goldoni}, {Ferrando}, {Goldwurm},
  {Decourchelle}  \& {Warwick}}{{Daigne} et~al.}{2002}]{Daigne_BPQ}
{Daigne} F.,  {Goldoni} P.,  {Ferrando} P.,  {Goldwurm} A.,  {Decourchelle} A.,
    {Warwick} R.~S.,  2002, \mn@doi [\aap] {10.1051/0004-6361:20020223}, \href
  {http://adsabs.harvard.edu/abs/2002A%26A...386..531D} {386, 531}

\bibitem[\protect\citeauthoryear{{Degenaar}, {Wijnands}, {Cackett}, {Homan},
  {in't Zand}, {Kuulkers}, {Maccarone}  \& {van der Klis}}{{Degenaar}
  et~al.}{2012}]{Degenaar_BPQuiescence}
{Degenaar} N.,  {Wijnands} R.,  {Cackett} E.~M.,  {Homan} J.,  {in't Zand}
  J.~J.~M.,  {Kuulkers} E.,  {Maccarone} T.~J.,   {van der Klis} M.,  2012,
  \mn@doi [\aap] {10.1051/0004-6361/201219470}, \href
  {http://cdsads.u-strasbg.fr/abs/2012A%26A...545A..49D} {545, A49}

\bibitem[\protect\citeauthoryear{{Degenaar}, {Miller}, {Harrison}, {Kennea},
  {Kouveliotou}  \& {Younes}}{{Degenaar} et~al.}{2014}]{Degenaar_BPSpec}
{Degenaar} N.,  {Miller} J.~M.,  {Harrison} F.~A.,  {Kennea} J.~A.,
  {Kouveliotou} C.,   {Younes} G.,  2014, \mn@doi [\apjl]
  {10.1088/2041-8205/796/1/L9}, \href
  {http://adsabs.harvard.edu/abs/2014ApJ...796L...9D} {796, L9}

\bibitem[\protect\citeauthoryear{{Doroshenko}, {Santangelo}, {Doroshenko},
  {Suleimanov}  \& {Piraino}}{{Doroshenko} et~al.}{2015}]{Doroshenko_NBFlash}
{Doroshenko} R.,  {Santangelo} A.,  {Doroshenko} V.,  {Suleimanov} V.,
  {Piraino} S.,  2015, \mn@doi [\mnras] {10.1093/mnras/stv1418}, \href
  {http://adsabs.harvard.edu/abs/2015MNRAS.452.2490D} {452, 2490}

\bibitem[\protect\citeauthoryear{{Evans} et~al.,}{{Evans}
  et~al.}{2007}]{Evans_Swift1}
{Evans} P.~A.,  et~al., 2007, \mn@doi [\aap] {10.1051/0004-6361:20077530},
  \href {http://adsabs.harvard.edu/abs/2007A%26A...469..379E} {469, 379}

\bibitem[\protect\citeauthoryear{{Ferrigno} et~al.,}{{Ferrigno}
  et~al.}{2014}]{Ferrigno_TMSPVar}
{Ferrigno} C.,  et~al., 2014, \mn@doi [\aap] {10.1051/0004-6361/201322904},
  \href {http://adsabs.harvard.edu/abs/2014A%26A...567A..77F} {567, A77}

\bibitem[\protect\citeauthoryear{{Finger}, {Koh}, {Nelson}, {Prince}, {Vaughan}
   \& {Wilson}}{{Finger} et~al.}{1996}]{Finger_Pulse}
{Finger} M.~H.,  {Koh} D.~T.,  {Nelson} R.~W.,  {Prince} T.~A.,  {Vaughan}
  B.~A.,   {Wilson} R.~B.,  1996, \mn@doi [\nat] {10.1038/381291a0}, \href
  {http://adsabs.harvard.edu/abs/1996Natur.381..291F} {381, 291}

\bibitem[\protect\citeauthoryear{{Fishman}, {Kouveliotou}, {van Paradijs},
  {Harmon}, {Paciesas}, {Briggs}, {Kommers}  \& {Lewin}}{{Fishman}
  et~al.}{1995}]{Fishman_Discovery}
{Fishman} G.~J.,  {Kouveliotou} C.,  {van Paradijs} J.,  {Harmon} B.~A.,
  {Paciesas} W.~S.,  {Briggs} M.~S.,  {Kommers} J.,   {Lewin} W.~H.~G.,  1995,
  \iaucirc, \href {http://adsabs.harvard.edu/abs/1995IAUC.6272....1F} {6272}

\bibitem[\protect\citeauthoryear{{Fruscione} et~al.,}{{Fruscione}
  et~al.}{2006}]{Fruscione_Ciao}
{Fruscione} A.,  et~al., 2006, in Society of Photo-Optical Instrumentation
  Engineers (SPIE) Conference Series. p. 62701V, \mn@doi{10.1117/12.671760}

\bibitem[\protect\citeauthoryear{{Fujimoto}, {Hanawa}  \& {Miyaji}}{{Fujimoto}
  et~al.}{1981}]{Fujimoto_Shellstab}
{Fujimoto} M.~Y.,  {Hanawa} T.,   {Miyaji} S.,  1981, \mn@doi [\apj]
  {10.1086/159034}, \href {http://adsabs.harvard.edu/abs/1981ApJ...247..267F}
  {247, 267}

\bibitem[\protect\citeauthoryear{{Galloway}, {Muno}, {Hartman}, {Psaltis}  \&
  {Chakrabarty}}{{Galloway} et~al.}{2008}]{Galloway_TypeI}
{Galloway} D.~K.,  {Muno} M.~P.,  {Hartman} J.~M.,  {Psaltis} D.,
  {Chakrabarty} D.,  2008, \mn@doi [\apjs] {10.1086/592044}, \href
  {http://adsabs.harvard.edu/abs/2008ApJS..179..360G} {179, 360}

\bibitem[\protect\citeauthoryear{{Garmire}, {Bautz}, {Ford}, {Nousek}  \&
  {Ricker}}{{Garmire} et~al.}{2003}]{Garmire_ACIS}
{Garmire} G.~P.,  {Bautz} M.~W.,  {Ford} P.~G.,  {Nousek} J.~A.,   {Ricker} Jr.
  G.~R.,  2003, in {Truemper} J.~E.,  {Tananbaum} H.~D.,  eds,  \procspie Vol.
  4851, X-Ray and Gamma-Ray Telescopes and Instruments for Astronomy.. pp
  28--44, \mn@doi{10.1117/12.461599}

\bibitem[\protect\citeauthoryear{{Gehrels}}{{Gehrels}}{2004}]{Gehrels_Swift}
{Gehrels} N.,  2004, in {Schoenfelder} V.,  {Lichti} G.,   {Winkler} C.,  eds,
  ESA Special Publication Vol. 552, 5th INTEGRAL Workshop on the INTEGRAL
  Universe. p.~777

\bibitem[\protect\citeauthoryear{{Gehrels}, {Chipman}  \& {Kniffen}}{{Gehrels}
  et~al.}{1994}]{Gehrels_CGRO}
{Gehrels} N.,  {Chipman} E.,   {Kniffen} D.,  1994, \mn@doi [\apjs]
  {10.1086/191978}, \href {http://adsabs.harvard.edu/abs/1994ApJS...92..351G}
  {92, 351}

\bibitem[\protect\citeauthoryear{{Giles}, {Swank}, {Jahoda}, {Zhang},
  {Strohmayer}, {Stark}  \& {Morgan}}{{Giles} et~al.}{1996}]{Giles_BP}
{Giles} A.~B.,  {Swank} J.~H.,  {Jahoda} K.,  {Zhang} W.,  {Strohmayer} T.,
  {Stark} M.~J.,   {Morgan} E.~H.,  1996, \mn@doi [\apjl] {10.1086/310262},
  \href {http://adsabs.harvard.edu/abs/1996ApJ...469L..25G} {469, L25}

\bibitem[\protect\citeauthoryear{{Gosling}, {Bandyopadhyay}, {Miller-Jones}  \&
  {Farrell}}{{Gosling} et~al.}{2007}]{Gosling_BPCompanion}
{Gosling} A.~J.,  {Bandyopadhyay} R.~M.,  {Miller-Jones} J.~C.~A.,   {Farrell}
  S.~A.,  2007, \mn@doi [\mnras] {10.1111/j.1365-2966.2007.12152.x}, \href
  {http://adsabs.harvard.edu/abs/2007MNRAS.380.1511G} {380, 1511}

\bibitem[\protect\citeauthoryear{{Harrison} et~al.,}{{Harrison}
  et~al.}{2013}]{Harrison_NuSTAR}
{Harrison} F.~A.,  et~al., 2013, \mn@doi [\apj] {10.1088/0004-637X/770/2/103},
  \href {http://adsabs.harvard.edu/abs/2013ApJ...770..103H} {770, 103}

\bibitem[\protect\citeauthoryear{{Hayakawa}}{{Hayakawa}}{1985}]{Hayakawa_Type2Mod}
{Hayakawa} S.,  1985, \mn@doi [\physrep] {10.1016/0370-1573(85)90053-5}, \href
  {http://adsabs.harvard.edu/abs/1985PhR...121..317H} {121, 317}

\bibitem[\protect\citeauthoryear{{Heger}, {Cumming}  \& {Woosley}}{{Heger}
  et~al.}{2007}]{Heger_MargStab}
{Heger} A.,  {Cumming} A.,   {Woosley} S.~E.,  2007, \mn@doi [\apj]
  {10.1086/517491}, \href {http://adsabs.harvard.edu/abs/2007ApJ...665.1311H}
  {665, 1311}

\bibitem[\protect\citeauthoryear{{Hoffman}, {Marshall}  \& {Lewin}}{{Hoffman}
  et~al.}{1978}]{Hoffman_RB}
{Hoffman} J.~A.,  {Marshall} H.~L.,   {Lewin} W.~H.~G.,  1978, \mn@doi [\nat]
  {10.1038/271630a0}, \href {http://adsabs.harvard.edu/abs/1978Natur.271..630H}
  {271, 630}

\bibitem[\protect\citeauthoryear{{Hunter}}{{Hunter}}{2007}]{Hunter_MatPlotLib}
{Hunter} J.~D.,  2007, Computing In Science \& Engineering, 9, 90

\bibitem[\protect\citeauthoryear{{Jahoda}, {Swank}, {Giles}, {Stark},
  {Strohmayer}, {Zhang}  \& {Morgan}}{{Jahoda} et~al.}{1996}]{Jahoda_PCA}
{Jahoda} K.,  {Swank} J.~H.,  {Giles} A.~B.,  {Stark} M.~J.,  {Strohmayer} T.,
  {Zhang} W.,   {Morgan} E.~H.,  1996, in {Siegmund} O.~H.,  {Gummin} M.~A.,
  eds,  \procspie Vol. 2808, EUV, X-Ray, and Gamma-Ray Instrumentation for
  Astronomy VII. pp 59--70

\bibitem[\protect\citeauthoryear{{Jahoda}, {Markwardt}, {Radeva}, {Rots},
  {Stark}, {Swank}, {Strohmayer}  \& {Zhang}}{{Jahoda}
  et~al.}{2006}]{Jahoda_Calibrate}
{Jahoda} K.,  {Markwardt} C.~B.,  {Radeva} Y.,  {Rots} A.~H.,  {Stark} M.~J.,
  {Swank} J.~H.,  {Strohmayer} T.~E.,   {Zhang} W.,  2006, \mn@doi [\apjs]
  {10.1086/500659}, \href {http://adsabs.harvard.edu/abs/2006ApJS..163..401J}
  {163, 401}

\bibitem[\protect\citeauthoryear{{Jones}, {Oliphant}, {Peterson}
  et~al.}{{Jones} et~al.}{2001}]{NumPy}
{Jones} E.,  {Oliphant} T.,  {Peterson} P.,   et~al., 2001, {{SciPy}: Open
  source scientific tools for {Python}}, \url {http://www.scipy.org/}

\bibitem[\protect\citeauthoryear{{Kennea}, {Kouveliotou}  \& {Younes}}{{Kennea}
  et~al.}{2014}]{Kennea_BPOutburst}
{Kennea} J.~A.,  {Kouveliotou} C.,   {Younes} G.,  2014, The Astronomer's
  Telegram, \href {http://adsabs.harvard.edu/abs/2014ATel.5845....1K} {5845}

\bibitem[\protect\citeauthoryear{{Kouveliotou}, {van Paradijs}, {Fishman},
  {Briggs}, {Kommers}, {Harmon}, {Meegan}  \& {Lewin}}{{Kouveliotou}
  et~al.}{1996}]{Kouveliotou_BP}
{Kouveliotou} C.,  {van Paradijs} J.,  {Fishman} G.~J.,  {Briggs} M.~S.,
  {Kommers} J.,  {Harmon} B.~A.,  {Meegan} C.~A.,   {Lewin} W.~H.~G.,  1996,
  \mn@doi [\nat] {10.1038/379799a0}, \href
  {http://adsabs.harvard.edu/abs/1996Natur.379..799K} {379, 799}

\bibitem[\protect\citeauthoryear{{Koyama} et~al.,}{{Koyama}
  et~al.}{2007}]{Koyama_XIS}
{Koyama} K.,  et~al., 2007, \mn@doi [\pasj] {10.1093/pasj/59.sp1.S23}, \href
  {http://adsabs.harvard.edu/abs/2007PASJ...59S..23K} {59, 23}

\bibitem[\protect\citeauthoryear{{Krimm} et~al.,}{{Krimm}
  et~al.}{2013}]{Krimm_BAT}
{Krimm} H.~A.,  et~al., 2013, \mn@doi [\apjs] {10.1088/0067-0049/209/1/14},
  \href {http://adsabs.harvard.edu/abs/2013ApJS..209...14K} {209, 14}

\bibitem[\protect\citeauthoryear{{Lamb}, {Miller}  \& {Taam}}{{Lamb}
  et~al.}{1996}]{Lamb_TypeIBP}
{Lamb} D.~Q.,  {Miller} M.~C.,   {Taam} R.~E.,  1996, ArXiv Astrophysics
  e-prints, \href {http://adsabs.harvard.edu/abs/1996astro.ph..4089L} {}

\bibitem[\protect\citeauthoryear{{Levine}, {Bradt}, {Cui}, {Jernigan},
  {Morgan}, {Remillard}, {Shirey}  \& {Smith}}{{Levine}
  et~al.}{1996}]{Levine_ASM}
{Levine} A.~M.,  {Bradt} H.,  {Cui} W.,  {Jernigan} J.~G.,  {Morgan} E.~H.,
  {Remillard} R.,  {Shirey} R.~E.,   {Smith} D.~A.,  1996, \mn@doi [\apjl]
  {10.1086/310260}, \href {http://adsabs.harvard.edu/abs/1996ApJ...469L..33L}
  {469, L33}

\bibitem[\protect\citeauthoryear{{Lewin} et~al.,}{{Lewin}
  et~al.}{1976a}]{Lewin_TypeII}
{Lewin} W.~H.~G.,  et~al., 1976a, \mn@doi [\apjl] {10.1086/182188}, \href
  {http://adsabs.harvard.edu/abs/1976ApJ...207L..95L} {207, L95}

\bibitem[\protect\citeauthoryear{{Lewin}, {Clark}  \& {Doty}}{{Lewin}
  et~al.}{1976b}]{Lewin_RBDiscovery}
{Lewin} W.~H.~G.,  {Clark} G.,   {Doty} J.,  1976b, \iaucirc, \href
  {http://adsabs.harvard.edu/abs/1976IAUC.2922....1L} {2922}

\bibitem[\protect\citeauthoryear{{Lewin}, {van Paradijs}  \& {Taam}}{{Lewin}
  et~al.}{1993}]{Lewin_Bursts}
{Lewin} W.~H.~G.,  {van Paradijs} J.,   {Taam} R.~E.,  1993, \mn@doi [\ssr]
  {10.1007/BF00196124}, \href
  {http://adsabs.harvard.edu/abs/1993SSRv...62..223L} {62, 223}

\bibitem[\protect\citeauthoryear{{Lewin}, {Rutledge}, {Kommers}, {van Paradijs}
   \& {Kouveliotou}}{{Lewin} et~al.}{1996}]{Lewin_BP}
{Lewin} W.~H.~G.,  {Rutledge} R.~E.,  {Kommers} J.~M.,  {van Paradijs} J.,
  {Kouveliotou} C.,  1996, \mn@doi [\apjl] {10.1086/310022}, \href
  {http://adsabs.harvard.edu/abs/1996ApJ...462L..39L} {462, L39}

\bibitem[\protect\citeauthoryear{{Linares}, {Altamirano}, {Chakrabarty},
  {Cumming}  \& {Keek}}{{Linares} et~al.}{2012}]{Linares_MargStab}
{Linares} M.,  {Altamirano} D.,  {Chakrabarty} D.,  {Cumming} A.,   {Keek} L.,
  2012, \mn@doi [\apj] {10.1088/0004-637X/748/2/82}, \href
  {http://adsabs.harvard.edu/abs/2012ApJ...748...82L} {748, 82}

\bibitem[\protect\citeauthoryear{{Linares}, {Kennea}, {Krimm}  \&
  {Kouveliotou}}{{Linares} et~al.}{2014}]{Linares_NewBurst}
{Linares} M.,  {Kennea} J.,  {Krimm} H.,   {Kouveliotou} C.,  2014, The
  Astronomer's Telegram, \href
  {http://adsabs.harvard.edu/abs/2014ATel.5883....1L} {5883}

\bibitem[\protect\citeauthoryear{{Lubi{\'n}ski}}{{Lubi{\'n}ski}}{2009}]{Lubinski_Heavens}
{Lubi{\'n}ski} P.,  2009, \mn@doi [\aap] {10.1051/0004-6361:200810897}, \href
  {http://adsabs.harvard.edu/abs/2009A%26A...496..557L} {496, 557}

\bibitem[\protect\citeauthoryear{{Marshall}, {Grindlay}  \&
  {Weisskopf}}{{Marshall} et~al.}{1979}]{Marshall_2types}
{Marshall} H.,  {Grindlay} J.,   {Weisskopf} M.,  1979, in Bulletin of the
  American Astronomical Society. p.~788

\bibitem[\protect\citeauthoryear{{Masetti}, {D'Avanzo}, {Blagorodnova}  \&
  {Palazzi}}{{Masetti} et~al.}{2014}]{Masetti_BPCompanion}
{Masetti} N.,  {D'Avanzo} P.,  {Blagorodnova} N.,   {Palazzi} E.,  2014, The
  Astronomer's Telegram, \href
  {http://adsabs.harvard.edu/abs/2014ATel.5999....1M} {5999}

\bibitem[\protect\citeauthoryear{{Mitsuda} et~al.,}{{Mitsuda}
  et~al.}{2007}]{Mitsuda_Suzaku}
{Mitsuda} K.,  et~al., 2007, \mn@doi [\pasj] {10.1093/pasj/59.sp1.S1}, \href
  {http://adsabs.harvard.edu/abs/2007PASJ...59S...1M} {59, 1}

\bibitem[\protect\citeauthoryear{{Negoro} et~al.,}{{Negoro}
  et~al.}{2014}]{Negoro_OB3}
{Negoro} H.,  et~al., 2014, The Astronomer's Telegram, \href
  {http://adsabs.harvard.edu/abs/2014ATel.5790....1N} {5790}

\bibitem[\protect\citeauthoryear{{Paciesas}, {Harmon}, {Fishman}, {Zhang}  \&
  {Robinson}}{{Paciesas} et~al.}{1996}]{Paciesas_BPDiscovery}
{Paciesas} W.~S.,  {Harmon} B.~A.,  {Fishman} G.~J.,  {Zhang} S.~N.,
  {Robinson} C.~R.,  1996, \iaucirc, \href
  {http://adsabs.harvard.edu/abs/1996IAUC.6284....1P} {6284}

\bibitem[\protect\citeauthoryear{{Papitto} et~al.,}{{Papitto}
  et~al.}{2013}]{Papitto_Swings}
{Papitto} A.,  et~al., 2013, \mn@doi [\nat] {10.1038/nature12470}, \href
  {http://adsabs.harvard.edu/abs/2013Natur.501..517P} {501, 517}

\bibitem[\protect\citeauthoryear{{Patruno}, {Maitra}, {Curran}, {D'Angelo},
  {Fridriksson}, {Russell}, {Middleton}  \& {Wijnands}}{{Patruno}
  et~al.}{2016}]{Patruno_Reflares2}
{Patruno} A.,  {Maitra} D.,  {Curran} P.~A.,  {D'Angelo} C.,  {Fridriksson}
  J.~K.,  {Russell} D.~M.,  {Middleton} M.,   {Wijnands} R.,  2016, \mn@doi
  [\apj] {10.3847/0004-637X/817/2/100}, \href
  {http://adsabs.harvard.edu/abs/2016ApJ...817..100P} {817, 100}

\bibitem[\protect\citeauthoryear{{Poisson}}{{Poisson}}{1837}]{Poisson_Distribution}
{Poisson} S.,  1837.
Mathematics of Statistics, Bachelier, \url
  {https://books.google.co.uk/books?id=uB8OAAAAQAAJ}

\bibitem[\protect\citeauthoryear{{Rappaport} \& {Joss}}{{Rappaport} \&
  {Joss}}{1997}]{Rappaport_BPHistory}
{Rappaport} S.,  {Joss} P.~C.,  1997, \mn@doi [\apj] {10.1086/304506}, \href
  {http://adsabs.harvard.edu/abs/1997ApJ...486..435R} {486, 435}

\bibitem[\protect\citeauthoryear{{Revnivtsev}, {Churazov}, {Gilfanov}  \&
  {Sunyaev}}{{Revnivtsev} et~al.}{2001}]{Revnivtsev_MargStab}
{Revnivtsev} M.,  {Churazov} E.,  {Gilfanov} M.,   {Sunyaev} R.,  2001, \mn@doi
  [\aap] {10.1051/0004-6361:20010434}, \href
  {http://adsabs.harvard.edu/abs/2001A%26A...372..138R} {372, 138}

\bibitem[\protect\citeauthoryear{{Sanna} et~al.,}{{Sanna}
  et~al.}{2017a}]{Sanna_BP}
{Sanna} A.,  et~al., 2017a, \mn@doi [\mnras] {10.1093/mnras/stx635}, \href
  {http://adsabs.harvard.edu/abs/2017MNRAS.469....2S} {469, 2}

\bibitem[\protect\citeauthoryear{{Sanna}, {D'Ai}, {Bozzo}, {Riggio}, {Pintore},
  {Burderi}, {Di Salvo}  \& {Iaria}}{{Sanna} et~al.}{2017b}]{Sanna_BPOutburst}
{Sanna} A.,  {D'Ai} A.,  {Bozzo} E.,  {Riggio} A.,  {Pintore} F.,  {Burderi}
  L.,  {Di Salvo} T.,   {Iaria} R.,  2017b, The Astronomer's Telegram, \href
  {http://adsabs.harvard.edu/abs/2017ATel.10079...1S} {10079}

\bibitem[\protect\citeauthoryear{{Sazonov}, {Sunyaev}  \& {Lund}}{{Sazonov}
  et~al.}{1997}]{Sazonov_BPGranat}
{Sazonov} S.~Y.,  {Sunyaev} R.~A.,   {Lund} N.,  1997, Astronomy Letters, \href
  {http://adsabs.harvard.edu/abs/1997AstL...23..286S} {23, 286}

\bibitem[\protect\citeauthoryear{{Scaringi}, {Maccarone}, {D'Angelo}, {Knigge}
  \& {Groot}}{{Scaringi} et~al.}{2017}]{Scaringi_Gating}
{Scaringi} S.,  {Maccarone} T.~J.,  {D'Angelo} C.,  {Knigge} C.,   {Groot}
  P.~J.,  2017, \mn@doi [\nat] {10.1038/nature24653}, \href
  {http://adsabs.harvard.edu/abs/2017Natur.552..210S} {552, 210}

\bibitem[\protect\citeauthoryear{{Spruit} \& {Taam}}{{Spruit} \&
  {Taam}}{1993}]{Spruit_Type2Mod}
{Spruit} H.~C.,  {Taam} R.~E.,  1993, \mn@doi [\apj] {10.1086/172162}, \href
  {http://adsabs.harvard.edu/abs/1993ApJ...402..593S} {402, 593}

\bibitem[\protect\citeauthoryear{{Strohmayer} \& {Bildsten}}{{Strohmayer} \&
  {Bildsten}}{2006}]{Strohmayer_TypeI}
{Strohmayer} T.,  {Bildsten} L.,  2006, {New views of thermonuclear bursts}.
pp 113--156

\bibitem[\protect\citeauthoryear{{Strohmayer}, {Jahoda}, {Giles}  \&
  {Lee}}{{Strohmayer} et~al.}{1997}]{Strohmayer_BPFieldTypeI}
{Strohmayer} T.~E.,  {Jahoda} K.,  {Giles} A.~B.,   {Lee} U.,  1997, \mn@doi
  [\apj] {10.1086/304522}, \href
  {http://adsabs.harvard.edu/abs/1997ApJ...486..355S} {486, 355}

\bibitem[\protect\citeauthoryear{{Sturner} \& {Dermer}}{{Sturner} \&
  {Dermer}}{1996}]{Sturner_BPNature}
{Sturner} S.~J.,  {Dermer} C.~D.,  1996, \mn@doi [\apjl] {10.1086/310126},
  \href {http://adsabs.harvard.edu/abs/1996ApJ...465L..31S} {465, L31}

\bibitem[\protect\citeauthoryear{{Sturner} \& {Shrader}}{{Sturner} \&
  {Shrader}}{2005}]{Sturner_Failed}
{Sturner} S.~J.,  {Shrader} C.~R.,  2005, \mn@doi [\apj] {10.1086/429815},
  \href {http://adsabs.harvard.edu/abs/2005ApJ...625..923S} {625, 923}

\bibitem[\protect\citeauthoryear{{Taam} \& {Lin}}{{Taam} \&
  {Lin}}{1984}]{Taam_Evo}
{Taam} R.~E.,  {Lin} D.~N.~C.,  1984, \mn@doi [\apj] {10.1086/162734}, \href
  {http://adsabs.harvard.edu/abs/1984ApJ...287..761T} {287, 761}

\bibitem[\protect\citeauthoryear{{Tan}, {Lewin}, {Lubin}, {van Paradijs},
  {Penninx}, {van der Klis}, {Damen}  \& {Stella}}{{Tan}
  et~al.}{1991}]{Tan_RBBursts}
{Tan} J.,  {Lewin} W.~H.~G.,  {Lubin} L.~M.,  {van Paradijs} J.,  {Penninx} W.,
   {van der Klis} M.,  {Damen} E.,   {Stella} L.,  1991, \mn@doi [\mnras]
  {10.1093/mnras/251.1.1}, \href
  {http://adsabs.harvard.edu/abs/1991MNRAS.251....1T} {251, 1}

\bibitem[\protect\citeauthoryear{{Tomsick}, {Kalemci}, {Corbel}  \&
  {Kaaret}}{{Tomsick} et~al.}{2003}]{Tomsick_MiniOutbursts}
{Tomsick} J.~A.,  {Kalemci} E.,  {Corbel} S.,   {Kaaret} P.,  2003, \mn@doi
  [\apj] {10.1086/375811}, \href
  {http://adsabs.harvard.edu/abs/2003ApJ...592.1100T} {592, 1100}

\bibitem[\protect\citeauthoryear{{Uchiyama} et~al.,}{{Uchiyama}
  et~al.}{2008}]{Uchiyama_SuzPSF}
{Uchiyama} Y.,  et~al., 2008, \mn@doi [\pasj] {10.1093/pasj/60.sp1.S35}, \href
  {http://adsabs.harvard.edu/abs/2008PASJ...60S..35U} {60, S35}

\bibitem[\protect\citeauthoryear{{Walker}}{{Walker}}{1992}]{Walker_Type2Mod}
{Walker} M.~A.,  1992, \mn@doi [\apj] {10.1086/170970}, \href
  {http://adsabs.harvard.edu/abs/1992ApJ...385..651W} {385, 651}

\bibitem[\protect\citeauthoryear{{Weisskopf}}{{Weisskopf}}{1999}]{Weisskopf_Chandra}
{Weisskopf} M.~C.,  1999, ArXiv Astrophysics e-prints, \href
  {http://adsabs.harvard.edu/abs/1999astro.ph.12097W} {}

\bibitem[\protect\citeauthoryear{{Wijnands} \& {Wang}}{{Wijnands} \&
  {Wang}}{2002}]{Wijnands_BPQ}
{Wijnands} R.,  {Wang} Q.~D.,  2002, \mn@doi [\apjl] {10.1086/340332}, \href
  {http://adsabs.harvard.edu/abs/2002ApJ...568L..93W} {568, L93}

\bibitem[\protect\citeauthoryear{{Wijnands}, {M{\'e}ndez}, {Markwardt}, {van
  der Klis}, {Chakrabarty}  \& {Morgan}}{{Wijnands}
  et~al.}{2001}]{Wijnands_1808}
{Wijnands} R.,  {M{\'e}ndez} M.,  {Markwardt} C.,  {van der Klis} M.,
  {Chakrabarty} D.,   {Morgan} E.,  2001, \mn@doi [\apj] {10.1086/323073},
  \href {http://adsabs.harvard.edu/abs/2001ApJ...560..892W} {560, 892}

\bibitem[\protect\citeauthoryear{{Winkler} et~al.,}{{Winkler}
  et~al.}{2003}]{Winkler_IBIS}
{Winkler} C.,  et~al., 2003, \mn@doi [\aap] {10.1051/0004-6361:20031288}, \href
  {http://adsabs.harvard.edu/abs/2003A%26A...411L...1W} {411, L1}

\bibitem[\protect\citeauthoryear{{Woods} et~al.,}{{Woods}
  et~al.}{1999}]{Woods_OB2}
{Woods} P.~M.,  et~al., 1999, \mn@doi [\apj] {10.1086/307171}, \href
  {http://adsabs.harvard.edu/abs/1999ApJ...517..431W} {517, 431}

\bibitem[\protect\citeauthoryear{{Woods}, {Kouveliotou}, {van Paradijs},
  {Koshut}, {Finger}, {Briggs}, {Fishman}  \& {Lewin}}{{Woods}
  et~al.}{2000}]{Woods_PulseBursts}
{Woods} P.~M.,  {Kouveliotou} C.,  {van Paradijs} J.,  {Koshut} T.~M.,
  {Finger} M.~H.,  {Briggs} M.~S.,  {Fishman} G.~J.,   {Lewin} W.~H.~G.,  2000,
  \mn@doi [\apj] {10.1086/309367}, \href
  {http://adsabs.harvard.edu/abs/2000ApJ...540.1062W} {540, 1062}

\bibitem[\protect\citeauthoryear{{Younes} et~al.,}{{Younes}
  et~al.}{2015}]{Younes_Expo}
{Younes} G.,  et~al., 2015, \mn@doi [\apj] {10.1088/0004-637X/804/1/43}, \href
  {http://adsabs.harvard.edu/abs/2015ApJ...804...43Y} {804, 43}

\bibitem[\protect\citeauthoryear{{in't Zand}, {Cumming}, {Triemstra},
  {Mateijsen}  \& {Bagnoli}}{{in't Zand} et~al.}{2014}]{intZand_Decay}
{in't Zand} J.~J.~M.,  {Cumming} A.,  {Triemstra} T.~L.,  {Mateijsen}
  R.~A.~D.~A.,   {Bagnoli} T.,  2014, \mn@doi [\aap]
  {10.1051/0004-6361/201322913}, \href
  {http://adsabs.harvard.edu/abs/2014A%26A...562A..16I} {562, A16}

\bibitem[\protect\citeauthoryear{{van den Eijnden}, {Bagnoli}, {Degenaar},
  {Lohfink}, {Parker}, {in 't Zand}  \& {Fabian}}{{van den Eijnden}
  et~al.}{2017}]{vandenEijnden_RB}
{van den Eijnden} J.,  {Bagnoli} T.,  {Degenaar} N.,  {Lohfink} A.~M.,
  {Parker} M.~L.,  {in 't Zand} J.~J.~M.,   {Fabian} A.~C.,  2017, \mn@doi
  [\mnras] {10.1093/mnrasl/slw244}, \href
  {http://adsabs.harvard.edu/abs/2017MNRAS.466L..98V} {466, L98}

\makeatother
\end{thebibliography}



\appendix

\section{List of \textit{RXTE} Observations}
\label{app:obs}

\par In Table \ref{tab:obslist} we present a table of all \textit{RXTE} observations used in this study.  The prefixes \textbf{A}, \textbf{B}, \textbf{C}, \textbf{D} and \textbf{E} correspond to OBSIDs beginning with 10401-01, 20077-01, 20078-01, 20401-01 and 30075-01 respectively.

\begin{table*}
\centering
\begin{tabular}{lllllllllllllll}
\hline
\hline
\scriptsize Obsid&\scriptsize Exp.&\scriptsize Date&\scriptsize Obsid&\scriptsize Exp.&\scriptsize Date&\scriptsize Obsid&\scriptsize Exp.&\scriptsize Date&\scriptsize Obsid&\scriptsize Exp.&\scriptsize Date&\scriptsize Obsid&\scriptsize Exp.&\scriptsize Date\\
\hline
\textbf{A}-01-00&3105&119&\textbf{A}-57-00&2432&250&\textbf{A}-94-00&3216&332&\textbf{C}-16-00&4941&562&\textbf{C}-40-01&3419&730\\
\textbf{A}-02-00&1655&117&\textbf{A}-57-01&894&253&\textbf{A}-95-00&9487&333&\textbf{C}-16-01&671&561&\textbf{C}-40-02&896&764\\
\textbf{A}-03-00&6724&122&\textbf{A}-57-02&1408&253&\textbf{A}-96-00&2627&337&\textbf{C}-16-02&1159&562&\textbf{C}-41-00&5255&735\\
\textbf{A}-03-000&2372&122&\textbf{A}-57-03&1792&253&\textbf{A}-97-00&3341&340&\textbf{C}-17-00&3537&568&\textbf{C}-41-01&2387&735\\
\textbf{A}-03-01&768&122&\textbf{A}-58-00&1024&255&\textbf{A}-98-00&99&343&\textbf{C}-18-00&2981&576&\textbf{C}-41-02&1141&744\\
\textbf{A}-04-00&639&128&\textbf{A}-58-01&1401&255&\textbf{A}-99-00&2783&345&\textbf{C}-18-01&3103&576&\textbf{C}-42-00&1476&744\\
\textbf{A}-05-00&1990&129&\textbf{A}-58-02&1679&255&\textbf{A}-99-01&1001&344&\textbf{C}-19-00&3286&582&\textbf{C}-43-00&5277&764\\
\textbf{A}-06-00&1280&134&\textbf{A}-58-03&1683&255&\textbf{B}-01-00&1664&467&\textbf{C}-19-01&2893&582&\textbf{C}-44-00&6712&769\\
\textbf{A}-08-00&2431&142&\textbf{A}-59-00&1152&257&\textbf{B}-02-00&1920&468&\textbf{C}-19-02&470&582&\textbf{D}-01-00&2688&523\\
\textbf{A}-09-00&640&138&\textbf{A}-59-01&2203&257&\textbf{B}-03-00&2982&469&\textbf{C}-20-00&3460&589&\textbf{D}-02-00&3469&525\\
\textbf{A}-10-00&2470&143&\textbf{A}-59-02&768&257&\textbf{B}-04-00&3530&470&\textbf{C}-20-01&1126&589&\textbf{D}-03-00&3026&528\\
\textbf{A}-11-00&2381&148&\textbf{A}-60-00&1907&260&\textbf{B}-05-00&2025&472&\textbf{C}-21-00&3659&596&\textbf{D}-04-00&3050&531\\
\textbf{A}-12-00&3352&151&\textbf{A}-60-01&3376&260&\textbf{B}-06-00&2677&473&\textbf{C}-21-01&2907&596&\textbf{D}-05-00&3485&536\\
\textbf{A}-13-00&3480&155&\textbf{A}-60-02&1783&260&\textbf{B}-07-00&3365&473&\textbf{C}-21-02&1086&596&\textbf{D}-06-00&1367&538\\
\textbf{A}-14-00&1839&158&\textbf{A}-60-03&1559&260&\textbf{B}-08-00&3113&475&\textbf{C}-22-00&1967&602&\textbf{D}-07-00&3196&543\\
\textbf{A}-15-00&1595&161&\textbf{A}-61-00&3292&262&\textbf{B}-09-00&2868&480&\textbf{C}-22-01&3086&602&\textbf{D}-08-00&2617&548\\
\textbf{A}-16-00&3470&156&\textbf{A}-61-01&3035&262&\textbf{B}-10-00&1009&482&\textbf{C}-22-02&1024&602&\textbf{D}-09-00&2598&553\\
\textbf{A}-17-00&4481&164&\textbf{A}-61-02&2013&262&\textbf{B}-11-00&2864&487&\textbf{C}-23-00&3697&607&\textbf{D}-10-00&4069&560\\
\textbf{A}-18-00&384&171&\textbf{A}-62-00&2390&264&\textbf{B}-12-00&1847&489&\textbf{C}-23-01&3091&607&\textbf{D}-11-00&2686&572\\
\textbf{A}-19-00&128&172&\textbf{A}-62-01&1703&264&\textbf{B}-13-00&2805&497&\textbf{C}-24-00&1152&618&\textbf{D}-12-00&2867&565\\
\textbf{A}-20-00&2087&178&\textbf{A}-62-02&2719&264&\textbf{B}-14-00&3741&499&\textbf{C}-24-01&2300&618&\textbf{D}-13-00&2021&585\\
\textbf{A}-21-00&2711&181&\textbf{A}-63-00&517&266&\textbf{B}-15-00&384&501&\textbf{C}-24-02&1386&618&\textbf{D}-13-01&765&585\\
\textbf{A}-22-00&2816&183&\textbf{A}-63-01&3077&266&\textbf{B}-16-00&768&503&\textbf{C}-25-00&4069&626&\textbf{D}-14-00&2640&594\\
\textbf{A}-22-01&2911&185&\textbf{A}-64-00&2381&268&\textbf{B}-17-00&2399&509&\textbf{C}-25-01&1920&626&\textbf{D}-14-01&1719&594\\
\textbf{A}-23-00&1678&187&\textbf{A}-64-01&3110&268&\textbf{B}-18-00&2306&511&\textbf{C}-25-02&768&626&\textbf{D}-15-00&3226&621\\
\textbf{A}-24-00&2509&189&\textbf{A}-65-00&2003&270&\textbf{B}-19-00&3477&516&\textbf{C}-26-00&2071&633&\textbf{D}-15-01&1373&621\\
\textbf{A}-25-00&2846&192&\textbf{A}-65-01&2744&270&\textbf{B}-20-00&1922&520&\textbf{C}-26-01&4043&633&\textbf{D}-16-00&2432&609\\
\textbf{A}-26-00&768&194&\textbf{A}-65-02&4331&270&\textbf{C}-01-00&8200&389&\textbf{C}-27-00&1792&638&\textbf{D}-16-01&1562&609\\
\textbf{A}-27-00&2923&196&\textbf{A}-66-00&2203&272&\textbf{C}-02-00&1408&400&\textbf{C}-27-01&2495&638&\textbf{D}-17-00&1790&628\\
\textbf{A}-28-00&6839&199&\textbf{A}-66-01&1723&272&\textbf{C}-02-01&896&401&\textbf{C}-27-02&3082&638&\textbf{D}-17-01&1291&628\\
\textbf{A}-29-00&3478&201&\textbf{A}-66-02&2533&272&\textbf{C}-02-02&512&401&\textbf{C}-28-00&3454&644&\textbf{D}-18-00&1959&641\\
\textbf{A}-30-00&5906&203&\textbf{A}-67-00&395&274&\textbf{C}-03-00&3409&465&\textbf{C}-28-01&1359&644&\textbf{D}-18-01&2614&641\\
\textbf{A}-31-00&6170&206&\textbf{A}-67-01&3533&274&\textbf{C}-03-01&2635&466&\textbf{C}-28-02&756&644&\textbf{D}-19-00&3158&650\\
\textbf{A}-32-00&2712&209&\textbf{A}-67-02&3466&274&\textbf{C}-03-02&2645&466&\textbf{C}-29-00&1535&652&\textbf{D}-20-00&751&672\\
\textbf{A}-34-00&1831&213&\textbf{A}-68-00&1841&276&\textbf{C}-04-00&2620&478&\textbf{C}-30-01&3435&658&\textbf{E}-01-00&512&831\\
\textbf{A}-35-00&2563&216&\textbf{A}-69-00&3659&278&\textbf{C}-04-01&2956&477&\textbf{C}-31-00&1920&662&\textbf{E}-02-00&1836&845\\
\textbf{A}-36-00&3683&219&\textbf{A}-70-00&2022&280&\textbf{C}-04-02&2515&476&\textbf{C}-31-01&1152&662&\textbf{E}-03-00&1871&859\\
\textbf{A}-37-00&3446&215&\textbf{A}-71-00&3474&283&\textbf{C}-05-00&1421&484&\textbf{C}-31-02&1012&657&\textbf{E}-04-00&1927&873\\
\textbf{A}-38-00&1536&217&\textbf{A}-72-00&5687&285&\textbf{C}-05-01&1995&484&\textbf{C}-32-00&4646&678&\textbf{E}-05-00&2088&889\\
\textbf{A}-39-00&2317&218&\textbf{A}-73-00&3109&287&\textbf{C}-05-02&2505&485&\textbf{C}-32-01&2803&678&\textbf{E}-06-00&2003&901\\
\textbf{A}-40-00&1239&220&\textbf{A}-74-00&1659&289&\textbf{C}-06-00&2770&492&\textbf{C}-33-00&4334&747&\textbf{E}-07-00&1536&914\\
\textbf{A}-41-00&1363&221&\textbf{A}-75-00&1798&291&\textbf{C}-06-01&2375&492&\textbf{C}-33-01&3534&748&\textbf{E}-08-00&967&935\\
\textbf{A}-42-00&2728&224&\textbf{A}-76-00&1558&293&\textbf{C}-06-02&2203&492&\textbf{C}-33-02&2957&748&\textbf{E}-09-00&1598&949\\
\textbf{A}-43-00&2079&225&\textbf{A}-77-00&1738&295&\textbf{C}-07-00&1258&494&\textbf{C}-34-00&3477&687&\textbf{E}-10-00&1835&961\\
\textbf{A}-44-00&2076&226&\textbf{A}-78-00&463&297&\textbf{C}-08-00&3305&505&\textbf{C}-34-01&1008&687&\textbf{E}-11-00&1741&975\\
\textbf{A}-45-00&2050&228&\textbf{A}-79-00&1024&299&\textbf{C}-08-01&777&505&\textbf{C}-34-02&2831&687&\textbf{E}-12-00&1032&991\\
\textbf{A}-47-00&2687&232&\textbf{A}-80-00&5818&301&\textbf{C}-09-00&1377&513&\textbf{C}-35-00&1497&756&\textbf{E}-13-00&1231&1001\\
\textbf{A}-48-00&2267&234&\textbf{A}-81-00&6898&303&\textbf{C}-09-01&1536&513&\textbf{C}-35-01&1959&755&\textbf{E}-14-00&1608&1016\\
\textbf{A}-49-00&35&236&\textbf{A}-82-00&3537&306&\textbf{C}-10-00&1664&517&\textbf{C}-35-02&2023&755&\textbf{E}-15-00&1712&1030\\
\textbf{A}-50-00&3719&238&\textbf{A}-83-00&512&308&\textbf{C}-10-01&3796&518&\textbf{C}-36-00&2825&702&\textbf{E}-16-00&1440&1045\\
\textbf{A}-51-00&3590&240&\textbf{A}-84-00&6361&310&\textbf{C}-11-00&2330&527&\textbf{C}-36-01&1592&702&\textbf{E}-17-00&1888&1057\\
\textbf{A}-52-00&2518&241&\textbf{A}-85-00&10391&312&\textbf{C}-11-01&290&527&\textbf{C}-37-00&2092&709&\textbf{E}-18-00&1847&1071\\
\textbf{A}-53-00&3063&243&\textbf{A}-86-00&9232&314&\textbf{C}-11-02&2399&527&\textbf{C}-37-01&384&710&\textbf{E}-19-00&1792&1086\\
\textbf{A}-55-00&3328&245&\textbf{A}-87-00&3109&316&\textbf{C}-12-00&3345&534&\textbf{C}-38-00&1752&716&\textbf{E}-20-00&1904&1101\\
\textbf{A}-55-01&3395&245&\textbf{A}-88-00&6630&318&\textbf{C}-12-01&2048&534&\textbf{C}-38-01&1536&716&\textbf{E}-21-00&1921&1115\\
\textbf{A}-55-02&2667&245&\textbf{A}-89-00&2569&320&\textbf{C}-13-00&1735&541&\textbf{C}-38-02&1144&716&\textbf{E}-22-00&1769&1129\\
\textbf{A}-56-00&512&250&\textbf{A}-90-00&2209&323&\textbf{C}-13-01&1691&541&\textbf{C}-38-03&338&717&\textbf{E}-23-00&1892&1135\\
\textbf{A}-56-01&1280&250&\textbf{A}-91-00&2317&325&\textbf{C}-14-00&3579&549&\textbf{C}-39-00&2756&723&\textbf{E}-24-00&1943&1197\\
\textbf{A}-56-02&1664&250&\textbf{A}-92-00&2199&327&\textbf{C}-14-01&2785&549&\textbf{C}-39-01&4690&723&\textbf{E}-25-00&2237&1210\\
\textbf{A}-56-03&1920&250&\textbf{A}-93-00&3720&331&\textbf{C}-15-00&7494&579&\textbf{C}-40-00&3419&730&\textbf{E}-26-00&1396&1224\\
\hline
\hline
\end{tabular}
\caption{A list of all \textit{RXTE} observations of the Bursting Pulsar used in this study.  Exposure is given in seconds, and date is given in days from MJD 50000.  The prefixes \textbf{A}, \textbf{B}, \textbf{C}, \textbf{D} and \textbf{E} correspond to OBSIDs beginning with 10401-01, 20077-01, 20078-01, 20401-01 and 30075-01 respectively.}
\label{tab:obslist}
\end{table*}

\section{Normal Burst Histograms}
\label{app:hists}

\par
\begin{figure}
  \centering
  \includegraphics[width=.9\linewidth, trim={0cm 0 0cm 0},clip]{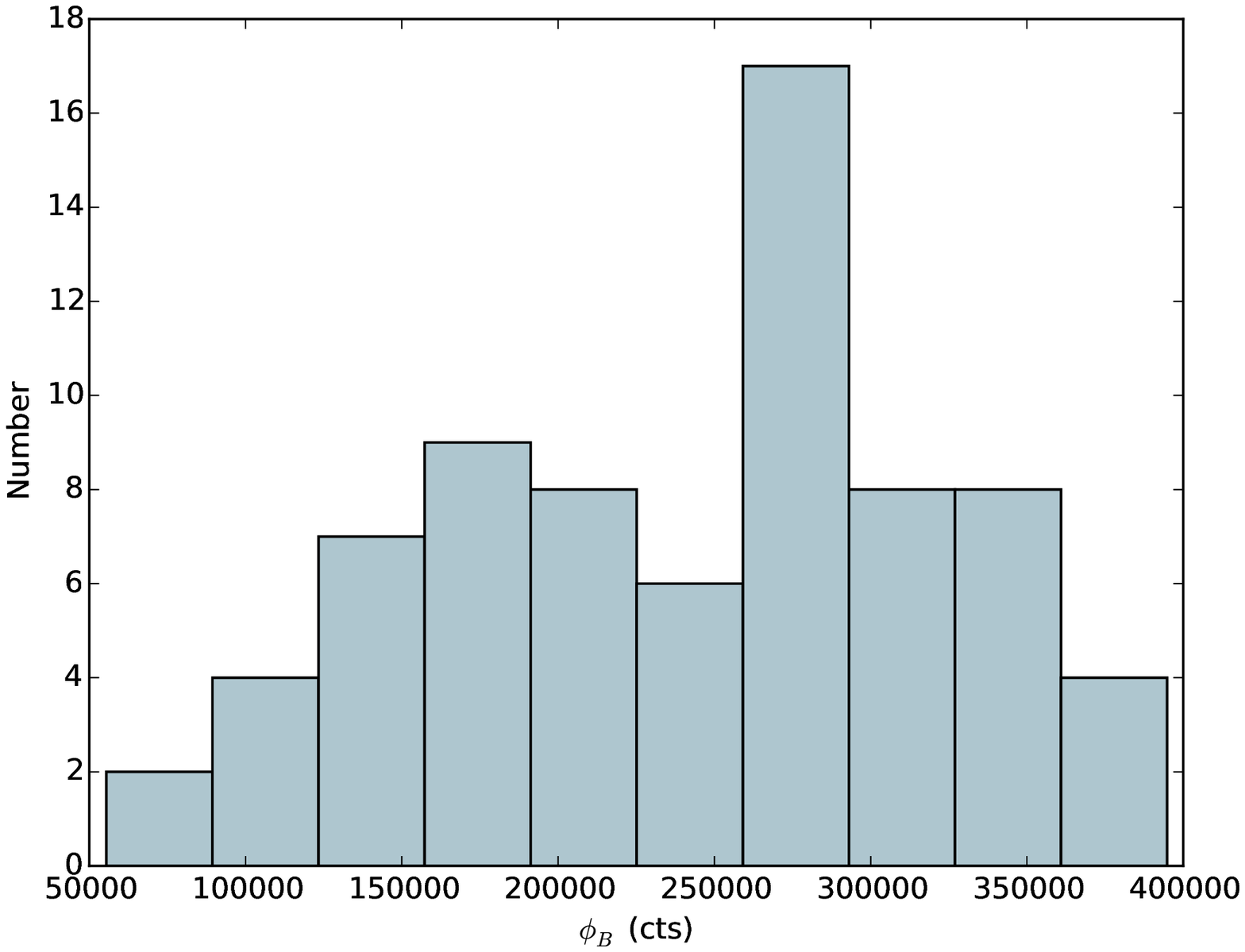}
  \caption{\small A histogram showing the distribution of burst fluence $\phi_B$ amongst our sample of Normal Bursts.}
  \label{fig:app_hist_phib}
\end{figure}

\begin{figure}
  \centering
  \includegraphics[width=.9\linewidth, trim={0cm 0 0cm 0},clip]{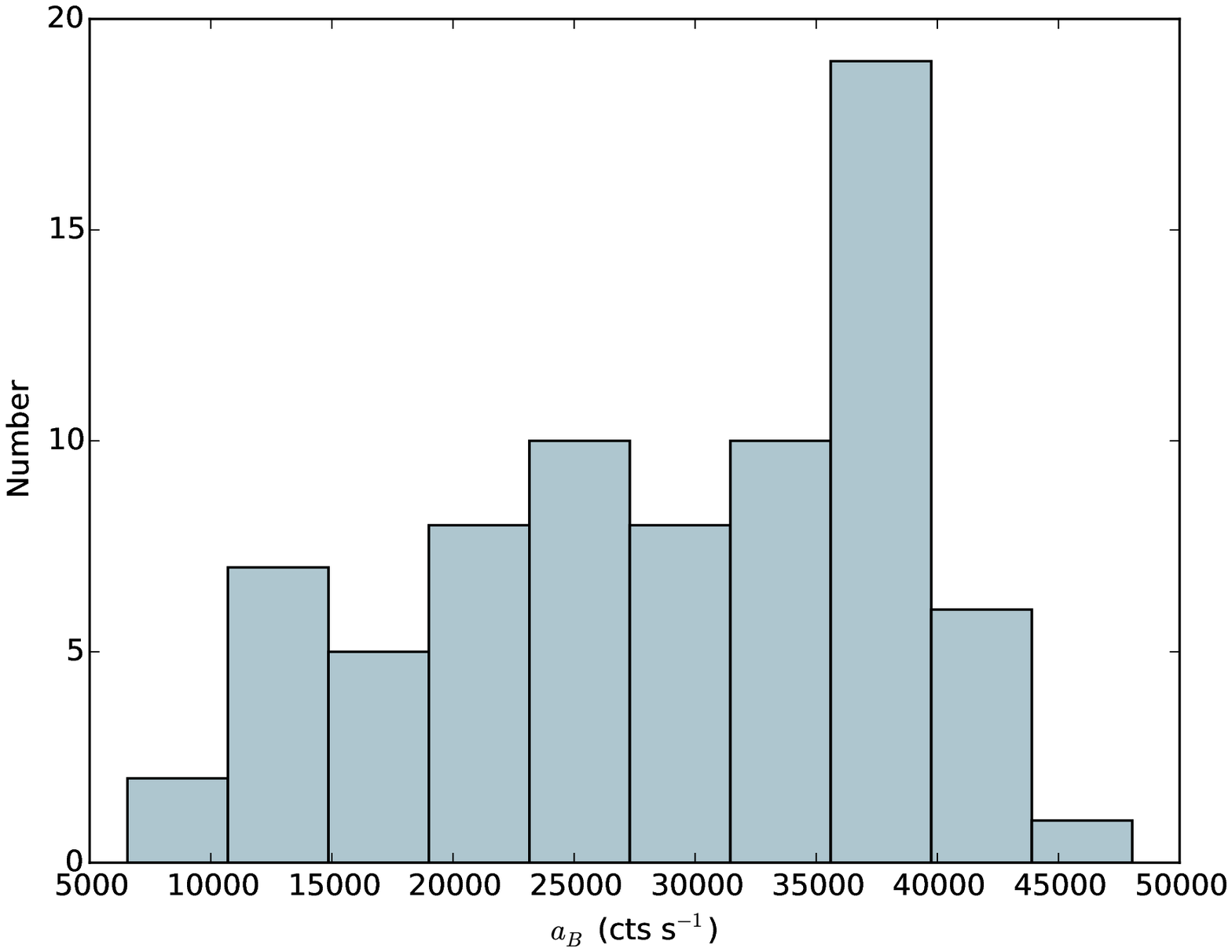}
  \caption{\small A histogram showing the distribution of burst amplitude $a_B$ amongst our sample of Normal Bursts.}
  \label{fig:app_hist_ab}
\end{figure}

\begin{figure}
  \centering
  \includegraphics[width=.9\linewidth, trim={0cm 0 0cm 0},clip]{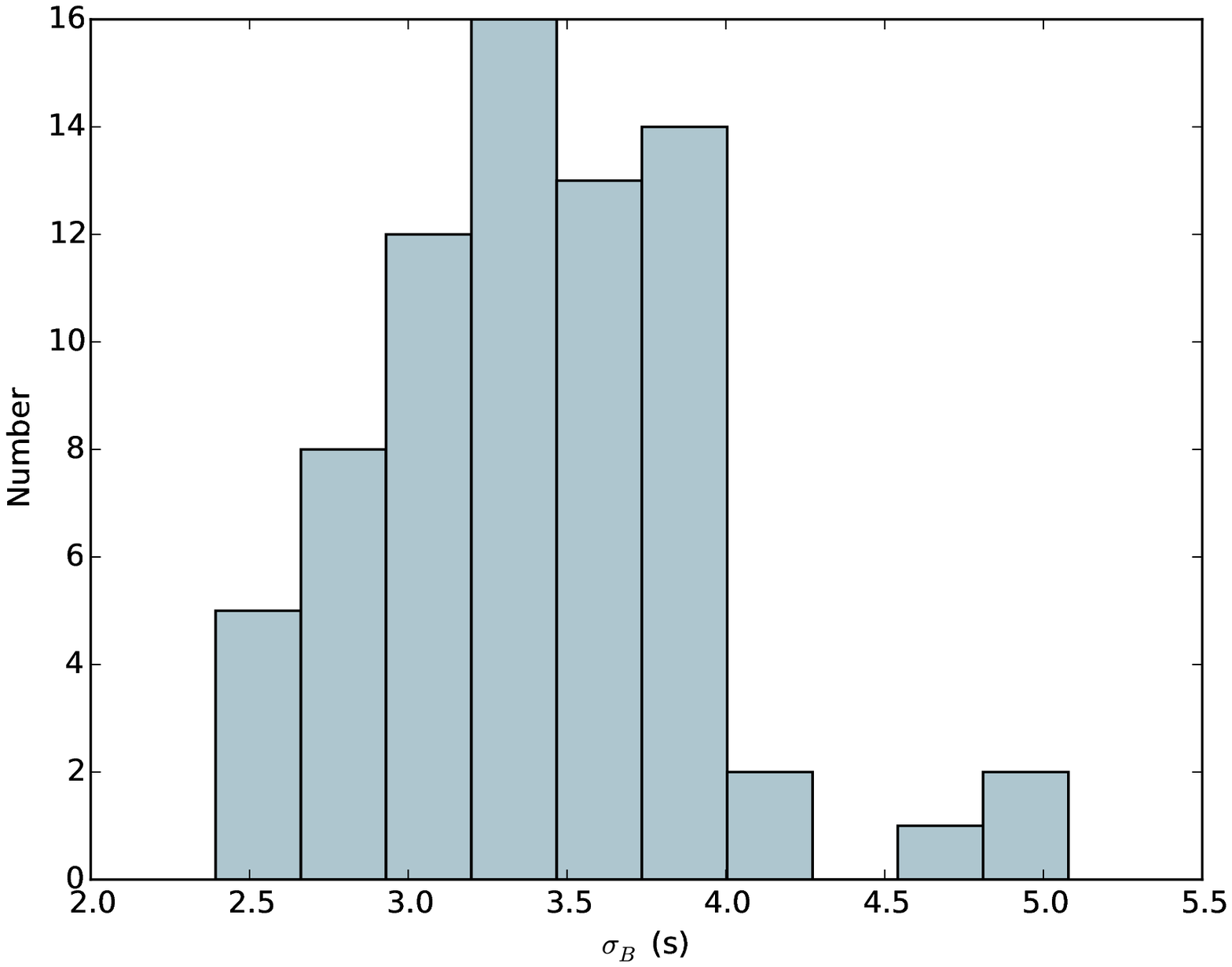}
  \caption{\small A histogram showing the distribution of burst width $\sigma_B$ amongst our sample of Normal Bursts.}
  \label{fig:app_hist_sigb}
\end{figure}

\begin{figure}
  \centering
  \includegraphics[width=.9\linewidth, trim={0cm 0 0cm 0},clip]{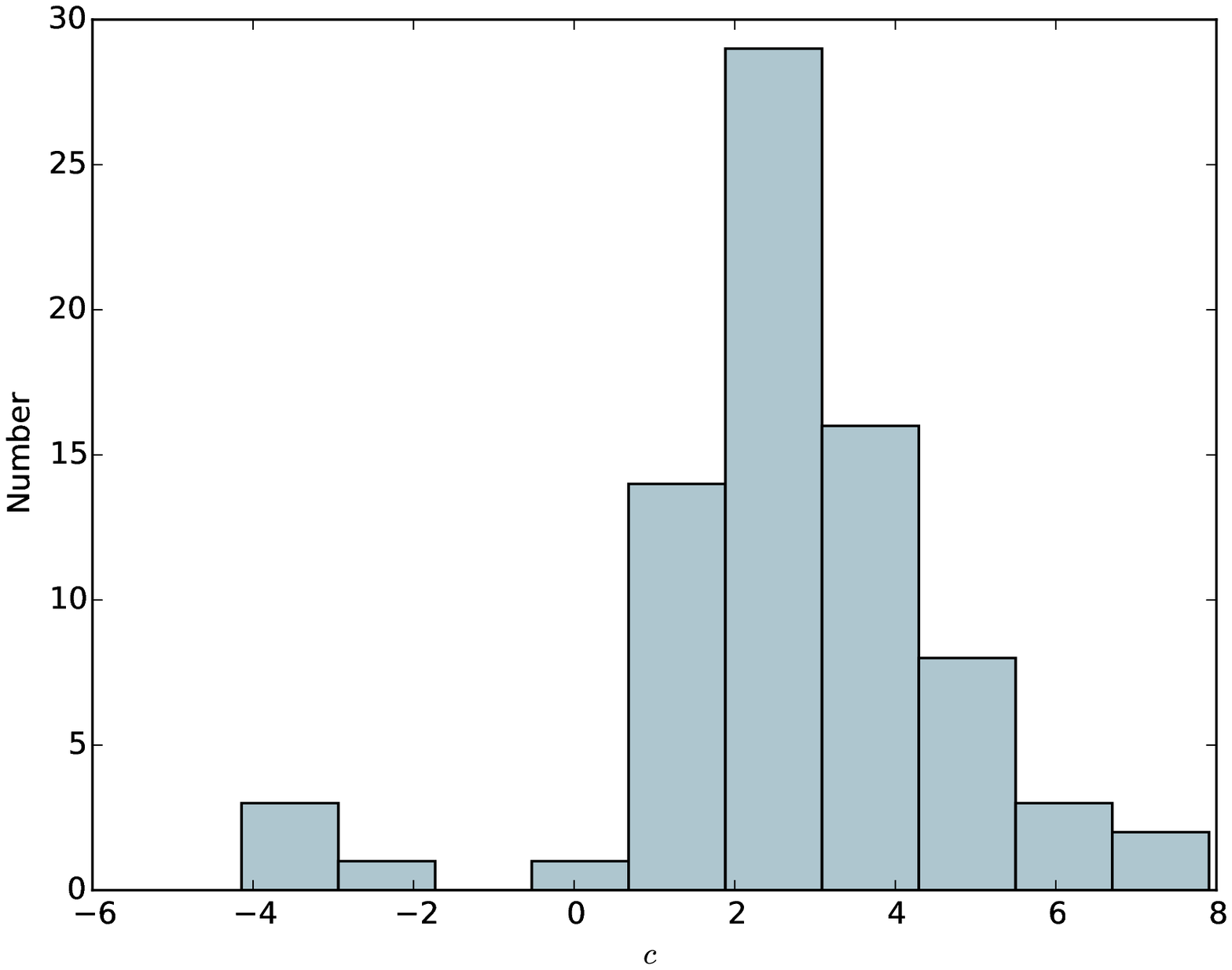}
  \caption{\small A histogram showing the distribution of burst skewness $c$ amongst our sample of Normal Bursts. }
  \label{fig:app_hist_c}
\end{figure}

\begin{figure}
  \centering
  \includegraphics[width=.9\linewidth, trim={0cm 0 0cm 0},clip]{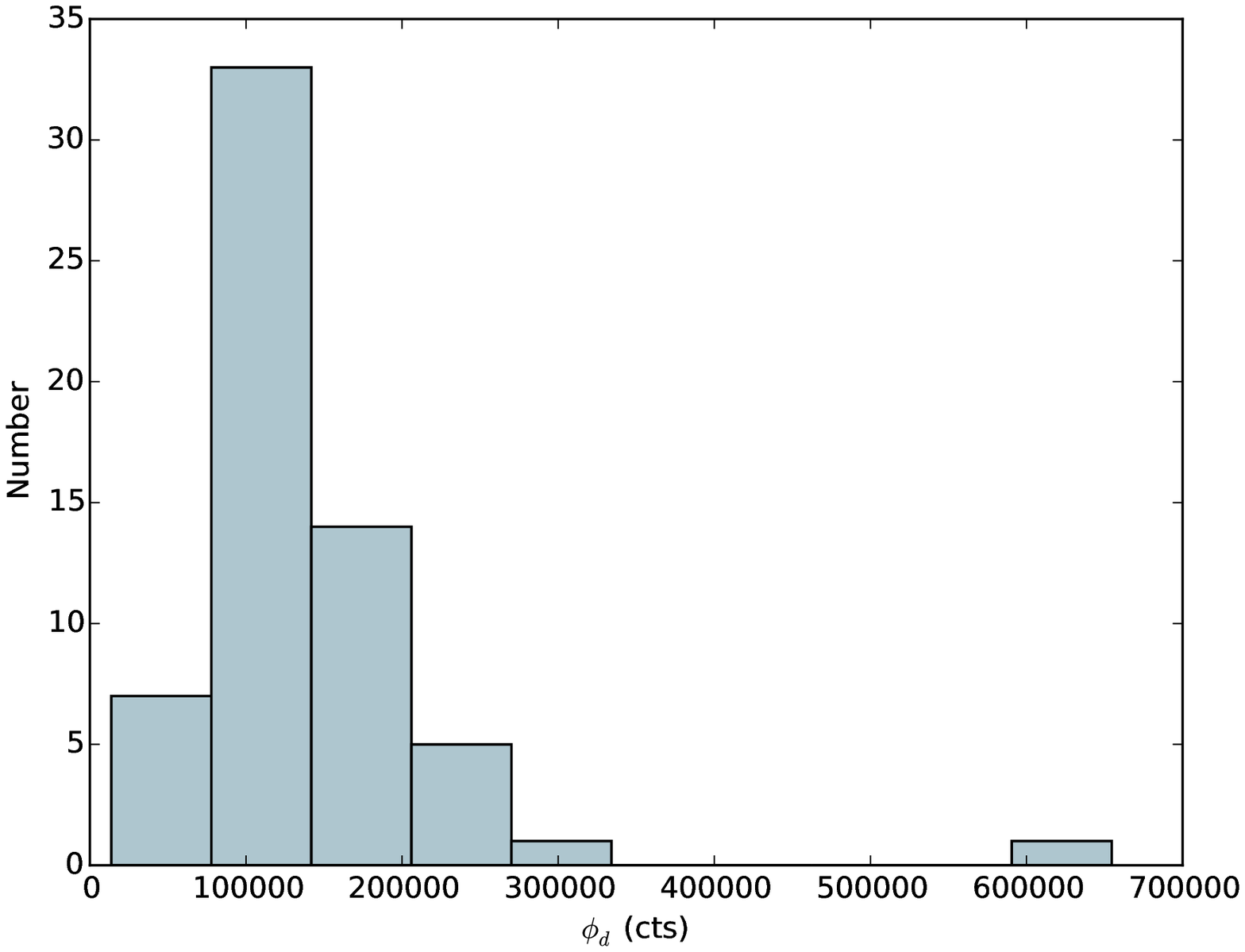}
  \caption{\small A histogram showing the distribution of dip fluence $\phi_d$ amongst our sample of Normal Bursts.}
  \label{fig:app_hist_phid}
\end{figure}

\begin{figure}
  \centering
  \includegraphics[width=.9\linewidth, trim={0cm 0 0cm 0},clip]{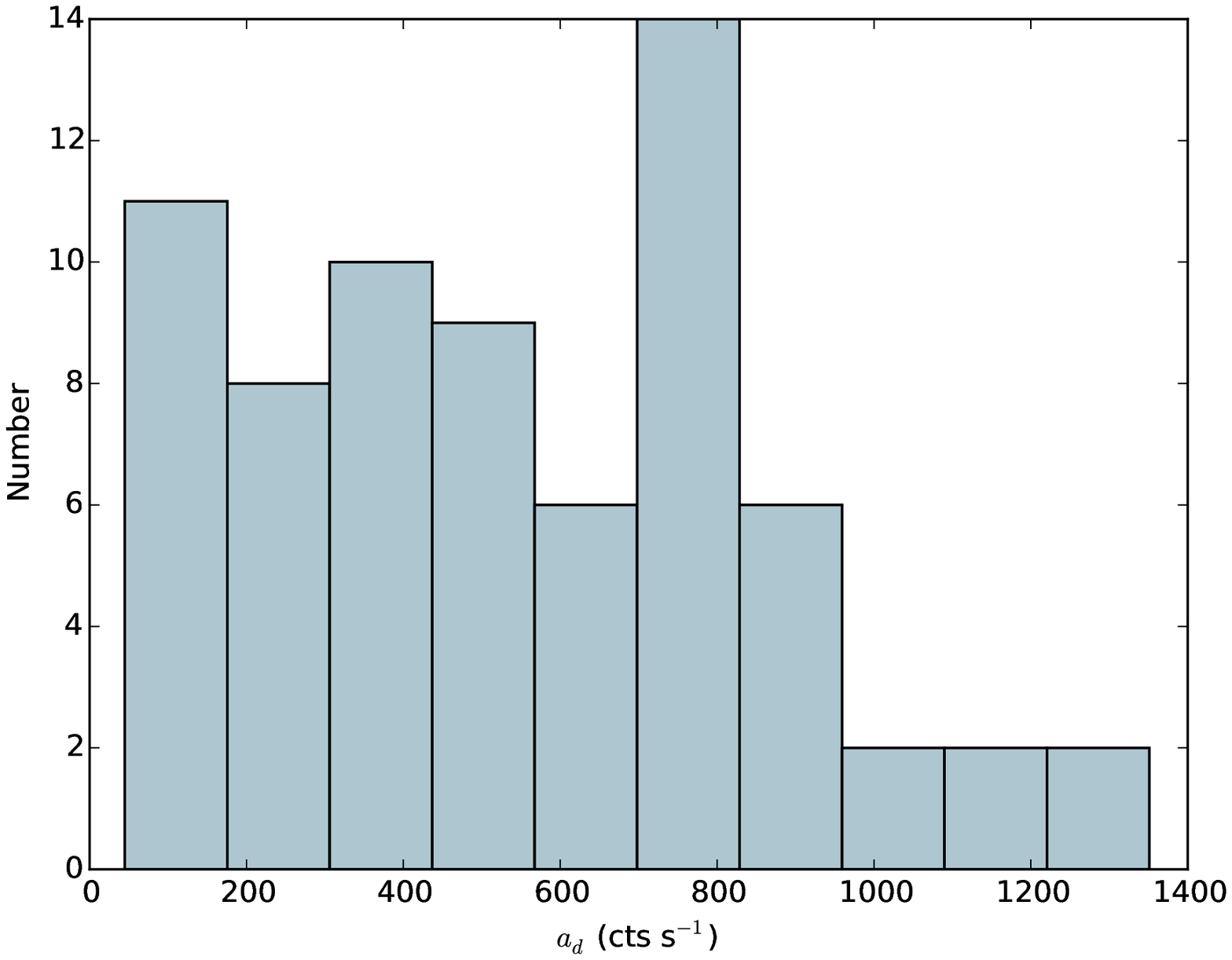}
  \caption{\small A histogram showing the distribution of dip amplitude $a_d$ amongst our sample of Normal Bursts.}
  \label{fig:app_hist_ad}
\end{figure}

\begin{figure}
  \centering
  \includegraphics[width=.9\linewidth, trim={0cm 0 0cm 0},clip]{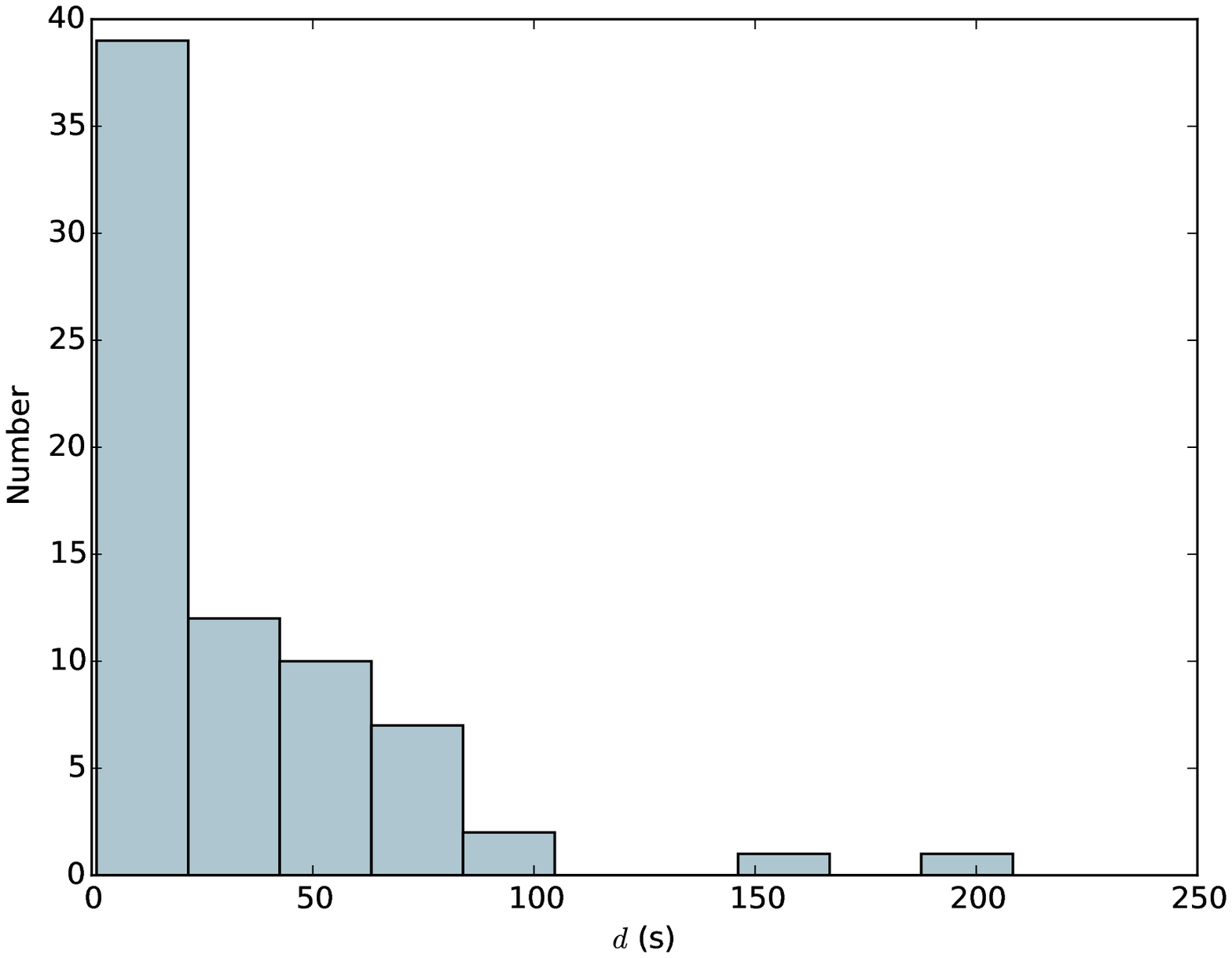}
  \caption{\small A histogram showing the distribution of dip fall-time $d$ amongst our sample of Normal Bursts.}
  \label{fig:app_hist_d}
\end{figure}

\begin{figure}
  \centering
  \includegraphics[width=.9\linewidth, trim={0cm 0 0cm 0},clip]{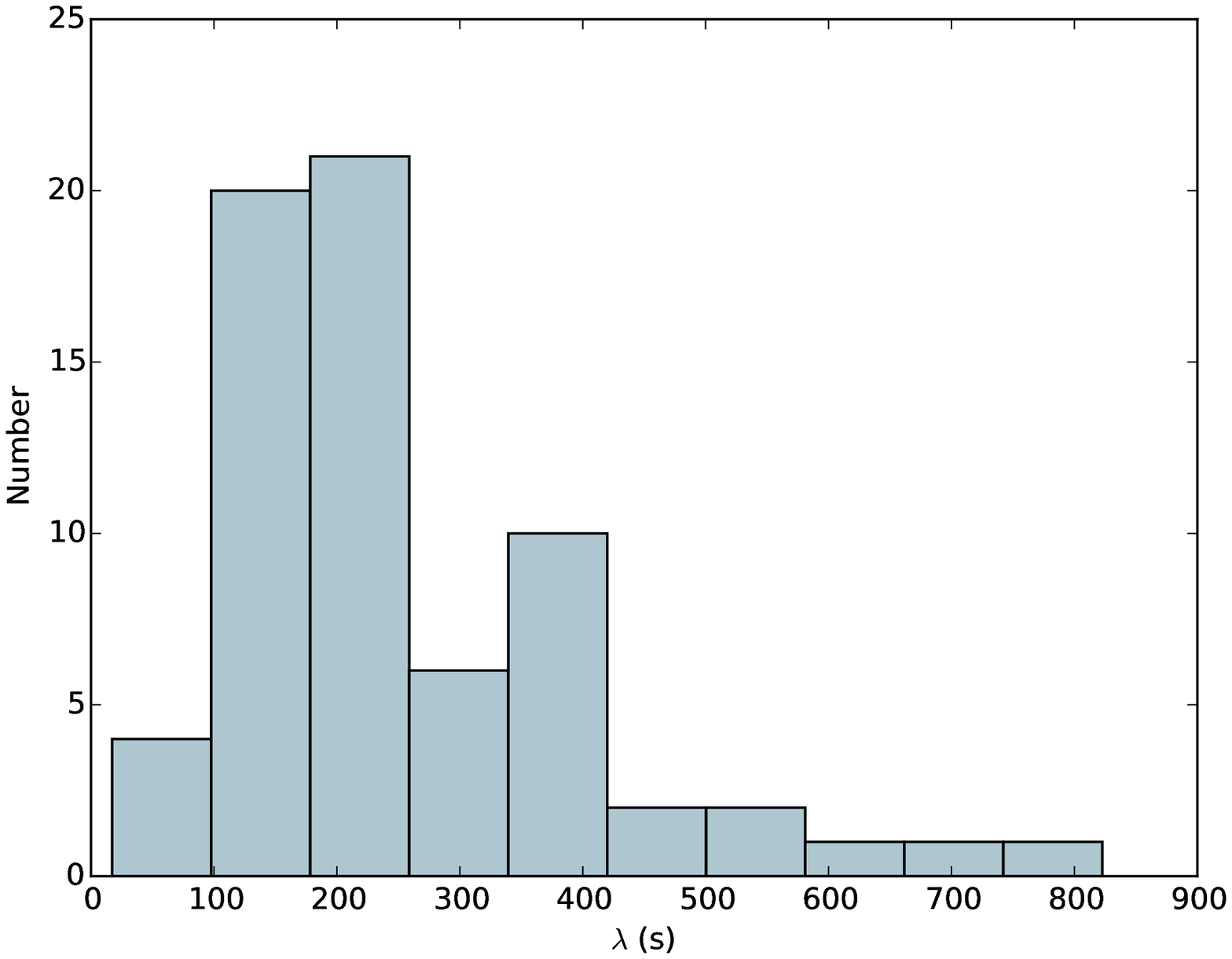}
  \caption{\small A histogram showing the distribution of dip recovery timescale $\lambda$ amongst our sample of Normal Bursts.}
  \label{fig:app_hist_lamb}
\end{figure}

\begin{figure}
  \centering
  \includegraphics[width=.9\linewidth, trim={0cm 0 0cm 0},clip]{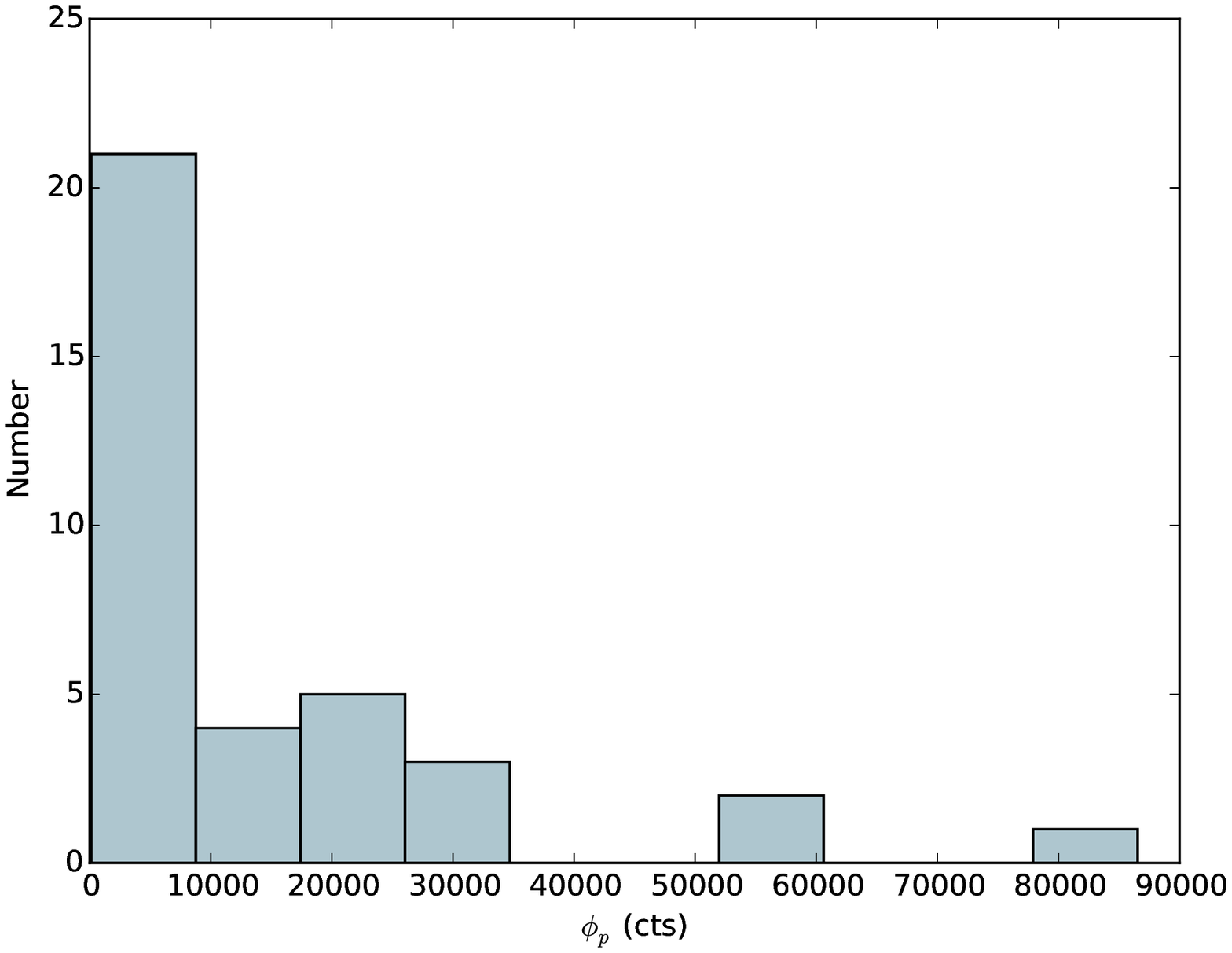}
  \caption{\small A histogram showing the distribution of plateau fluence $\phi_p$ amongst our sample of Normal Bursts.}
  \label{fig:app_hist_phip}
\end{figure}

\begin{figure}
  \centering
  \includegraphics[width=.9\linewidth, trim={0cm 0 0cm 0},clip]{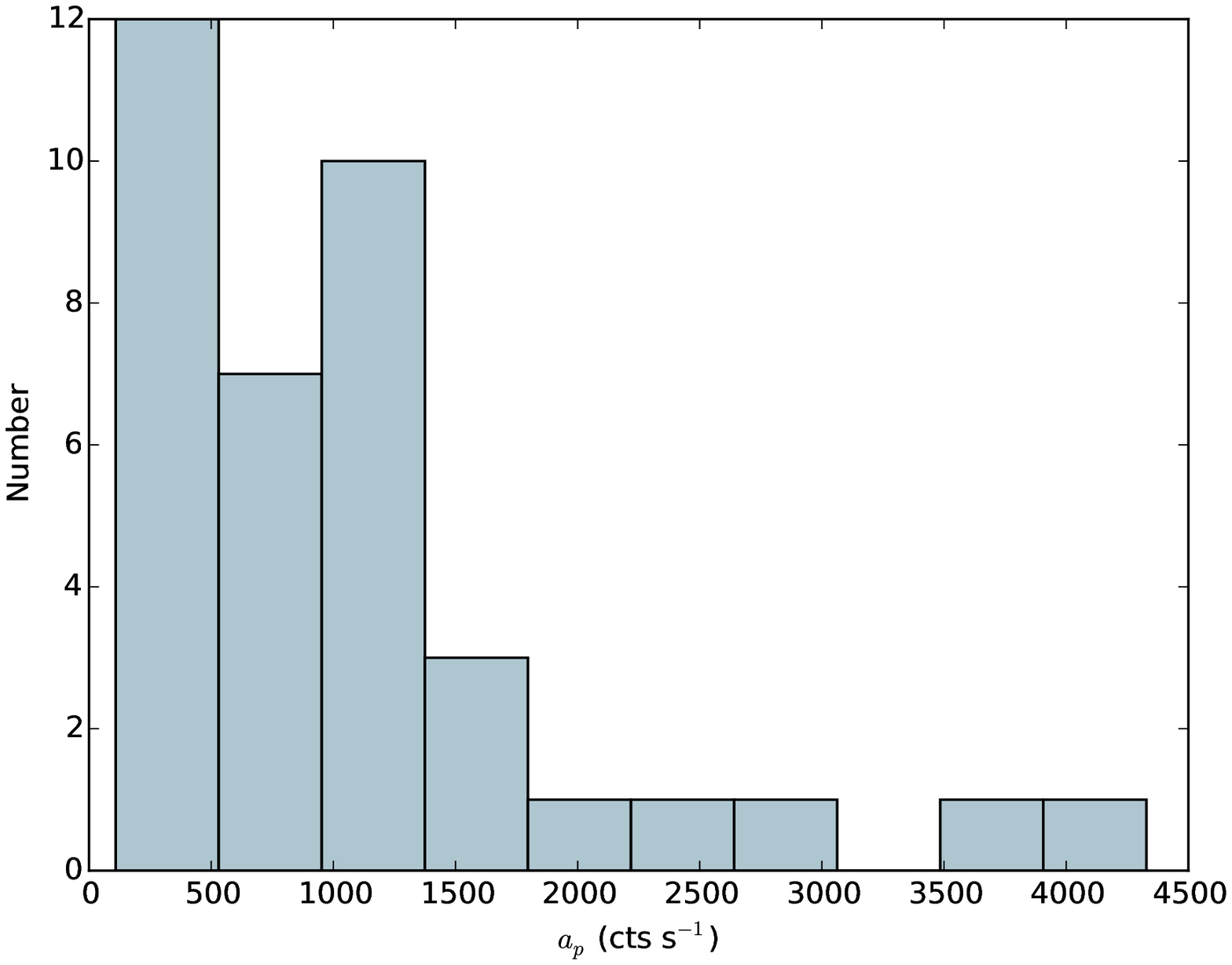}
  \caption{\small A histogram showing the distribution of plateau amplitude $a_p$ amongst our sample of Normal Bursts.}
  \label{fig:app_hist_ap}
\end{figure}

\begin{figure}
  \centering
  \includegraphics[width=.9\linewidth, trim={0cm 0 0cm 0},clip]{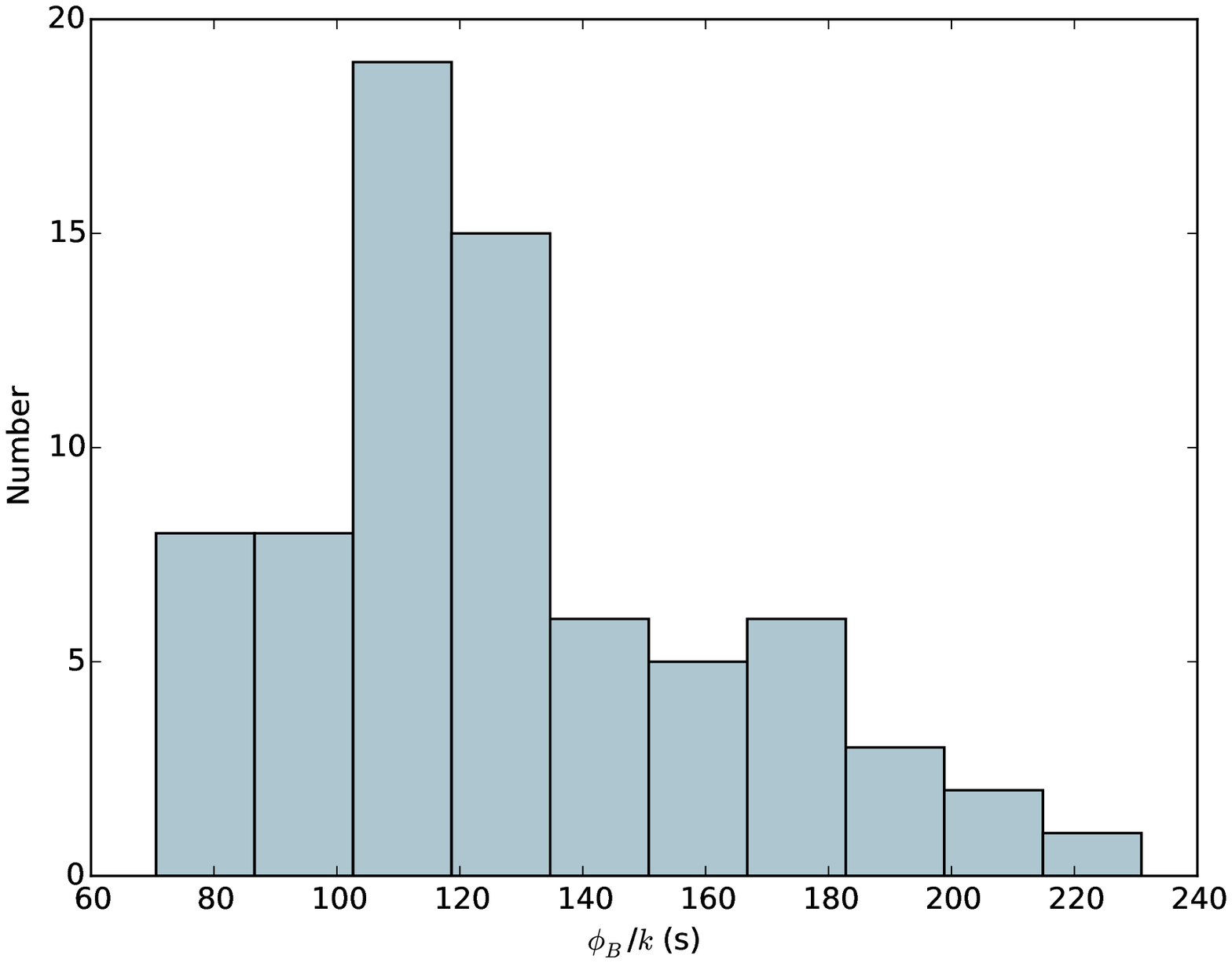}
  \caption{\small A histogram showing the distribution of persistent-emission-normalised burst fluence $\phi_B/k$ amongst our sample of Normal Bursts.}
  \label{fig:app_hist_phib_n}
\end{figure}

\begin{figure}
  \centering
  \includegraphics[width=.9\linewidth, trim={0cm 0 0cm 0},clip]{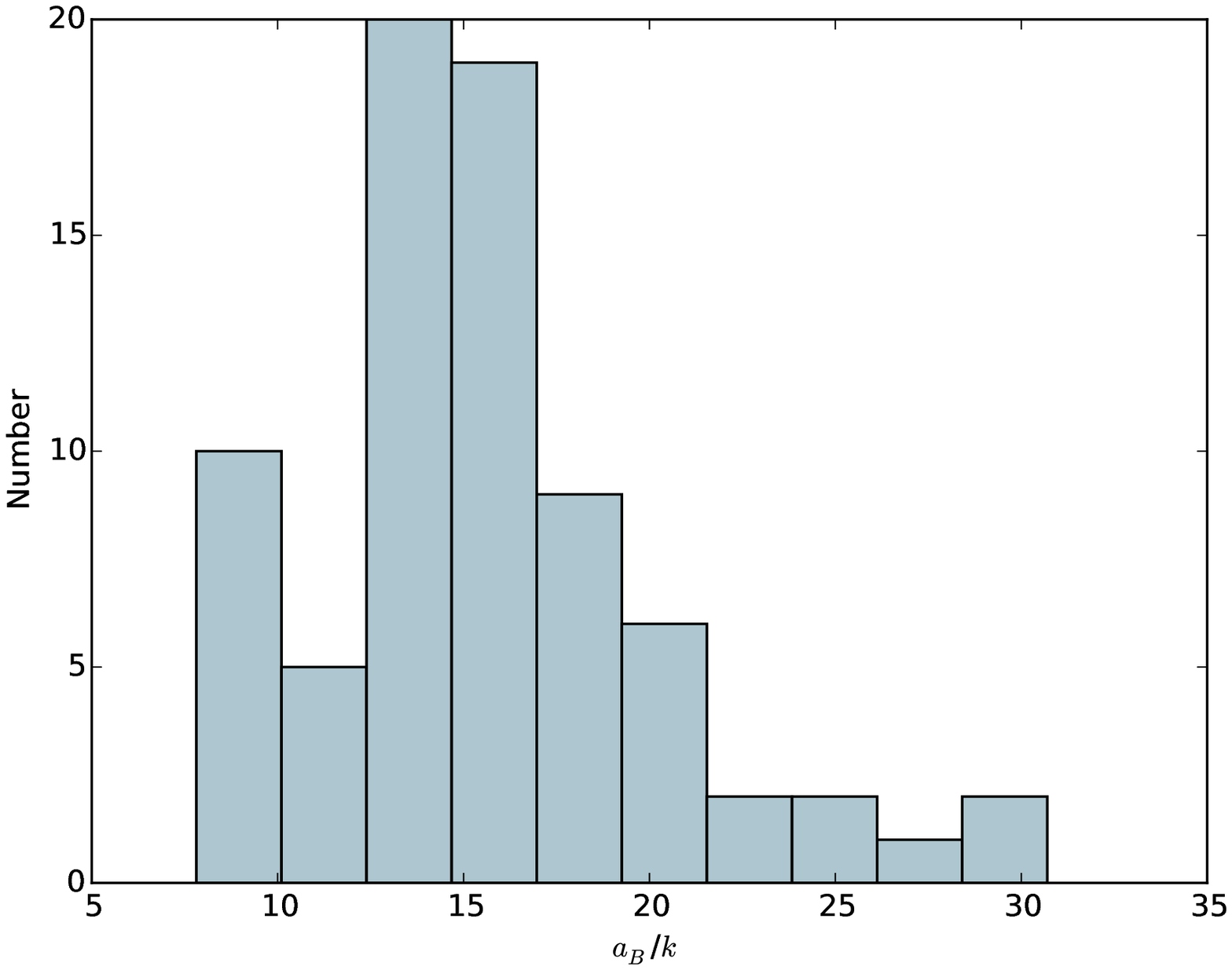}
  \caption{\small A histogram showing the distribution of persistent-emission-normalised burst amplitude $a_B/k$ amongst our sample of Normal Bursts.}
  \label{fig:app_hist_ab_n}
\end{figure}

\begin{figure}
  \centering
  \includegraphics[width=.9\linewidth, trim={0cm 0 0cm 0},clip]{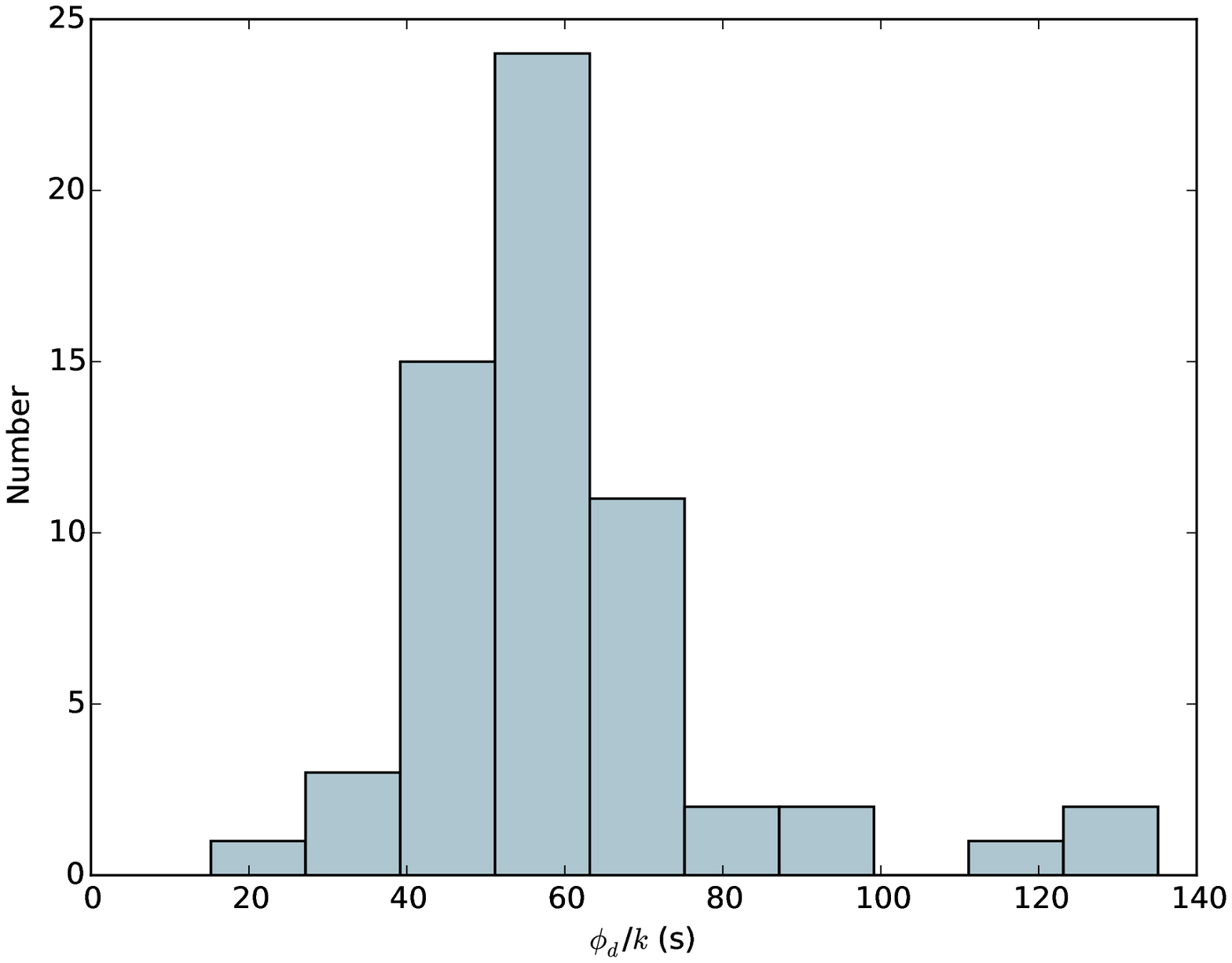}
  \caption{\small A histogram showing the distribution of persistent-emission-normalised dip fluence $\phi_d/k$ amongst our sample of Normal Bursts.}
  \label{fig:app_hist_phid_n}
\end{figure}

\begin{figure}
  \centering
  \includegraphics[width=.9\linewidth, trim={0cm 0 0cm 0},clip]{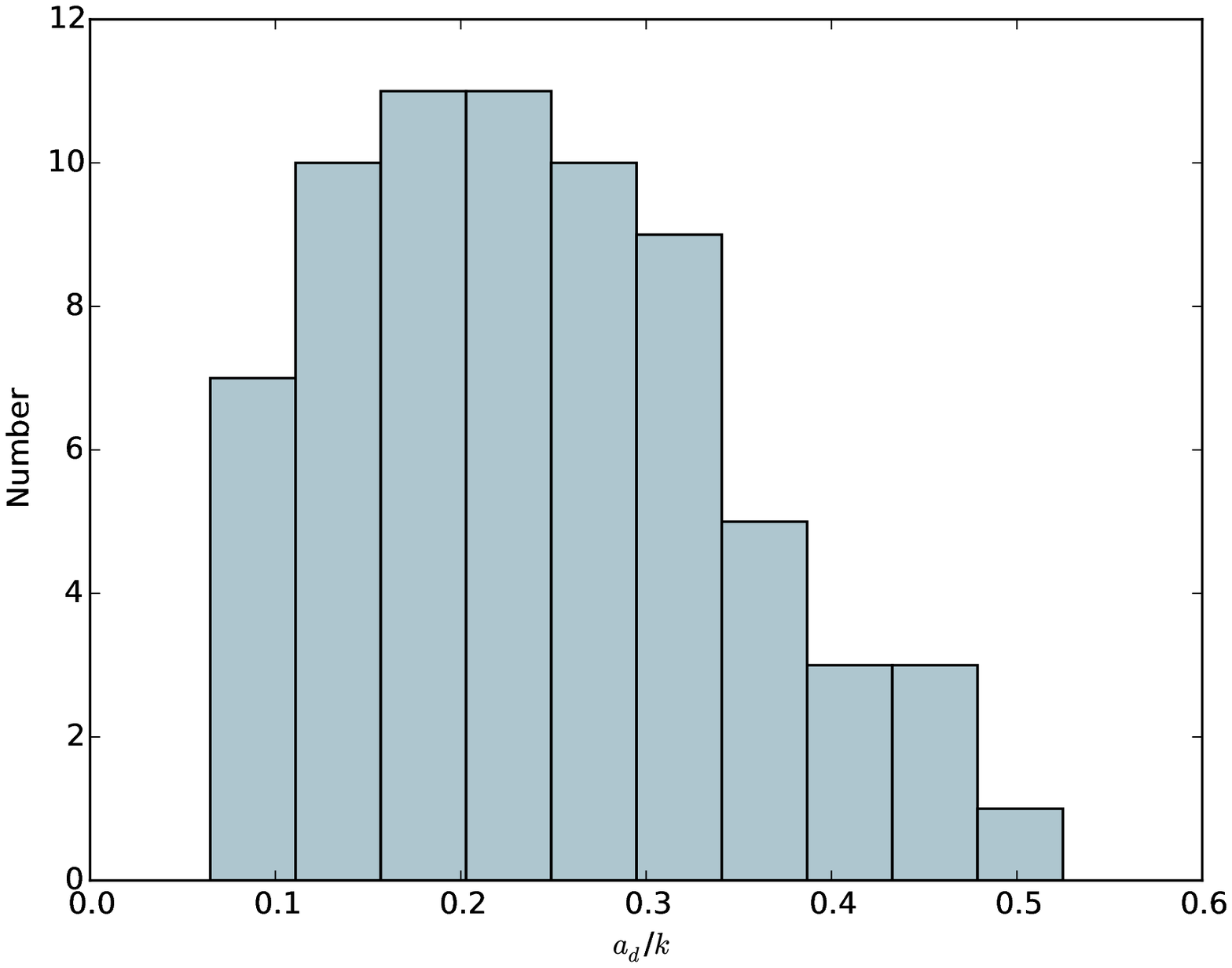}
  \caption{\small A histogram showing the distribution of persistent-emission-normalised dip amplitude $a_d/k$ amongst our sample of Normal Bursts.}
  \label{fig:app_hist_ad_n}
\end{figure}

\begin{figure}
  \centering
  \includegraphics[width=.9\linewidth, trim={0cm 0 0cm 0},clip]{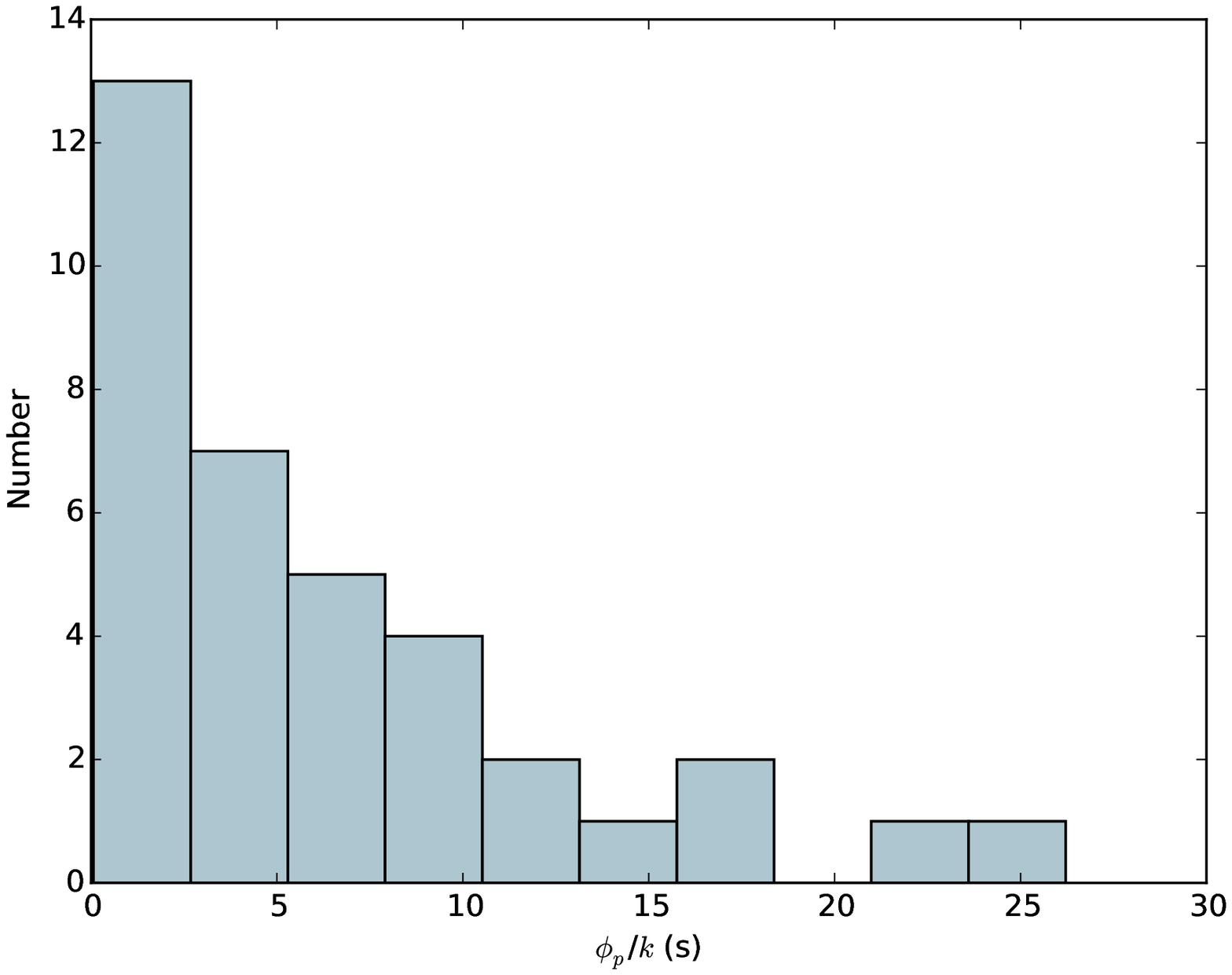}
  \caption{\small A histogram showing the distribution of persistent-emission-normalised plateau fluence $\phi_p/k$ amongst our sample of Normal Bursts.}
  \label{fig:app_hist_phip_n}
\end{figure}

\begin{figure}
  \centering
  \includegraphics[width=.9\linewidth, trim={0cm 0 0cm 0},clip]{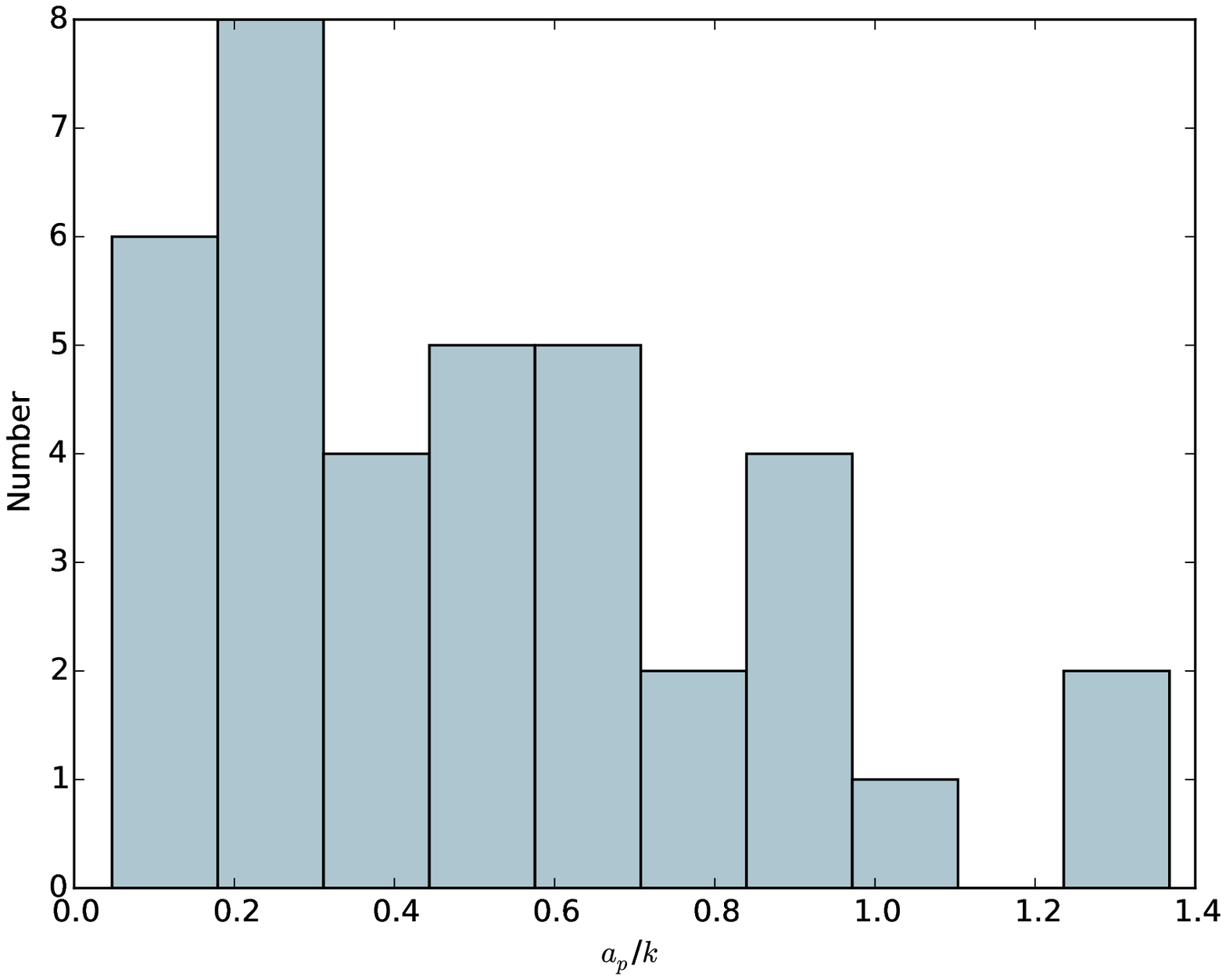}
  \caption{\small A histogram showing the distribution of persistent-emission-normalised plateau amplitude $a_p/k$ amongst our sample of Normal Bursts.}
  \label{fig:app_hist_ap_n}
\end{figure}

 In Figures \ref{fig:app_hist_phib}--\ref{fig:app_hist_ap}, we present histograms showing the distributions of $\phi_B$, $a_B$, $\sigma_B$, $c$, $\phi_d$, $a_d$, $d$, $\lambda$, $\phi_p$ and $a_p$ we find in our population study.  Each of these is a parameter we used to fit the Normal Bursts in our sample: see Section \ref{sec:struc} for a full explanation of these parameters.  In Figures \ref{fig:app_hist_phib_n}--\ref{fig:app_hist_ap_n} we show the distributions of $\phi_B$, $a_B$, $\phi_d$, $a_d$, $\phi_p$ and $a_p$ after being normalised by the persistent emission rate $k$ at the time of each burst.

\section{Parameter Correlations in Normal Bursts}
\label{app:corr}

\par Before normalizing for persistent rate, we find $>5\,\sigma$ correlations between 12 pairs of the parameters we use to describe Normal Bursts:

\begin{itemize}
\item Persistent emission $k$ correlates with burst fluence $\phi_B$ ($>10\,\sigma$), burst amplitude $a_b$ ($>10\,\sigma$), dip fluence $\phi_D$ ($>10\,\sigma$) and dip amplitude $a_d$ ($7.2\,\sigma$).
\item Burst fluence $\phi_B$ also correlates with burst amplitude $a_B$ ($>10\,\sigma$), dip fluence $\phi_D$ ($>10\,\sigma$) and dip amplitude $a_d$ ($7.1\,\sigma$).
\item Burst amplitude $\phi_B$ also correlates with dip fluence $\phi_D$ ($6.2\,\sigma$) and dip amplitude $a_d$ ($5.7\,\sigma$).
\item Burst width $\sigma_B$ correlates with burst skewness $c$ ($5.8\,\sigma$).
\item Dip amplitude $a_d$ anticorrelates with dip recovery timescale $\lambda$ ($5.0\,\sigma$).
\item Plateau fluence $\phi_p$ correlates with plateau amplitude $a_p$ ($6.6\,\sigma$).
\end{itemize}

\begin{figure*}
  \centering
  \includegraphics[width=\linewidth, trim={2.1cm 2cm 3.5cm 3cm},clip]{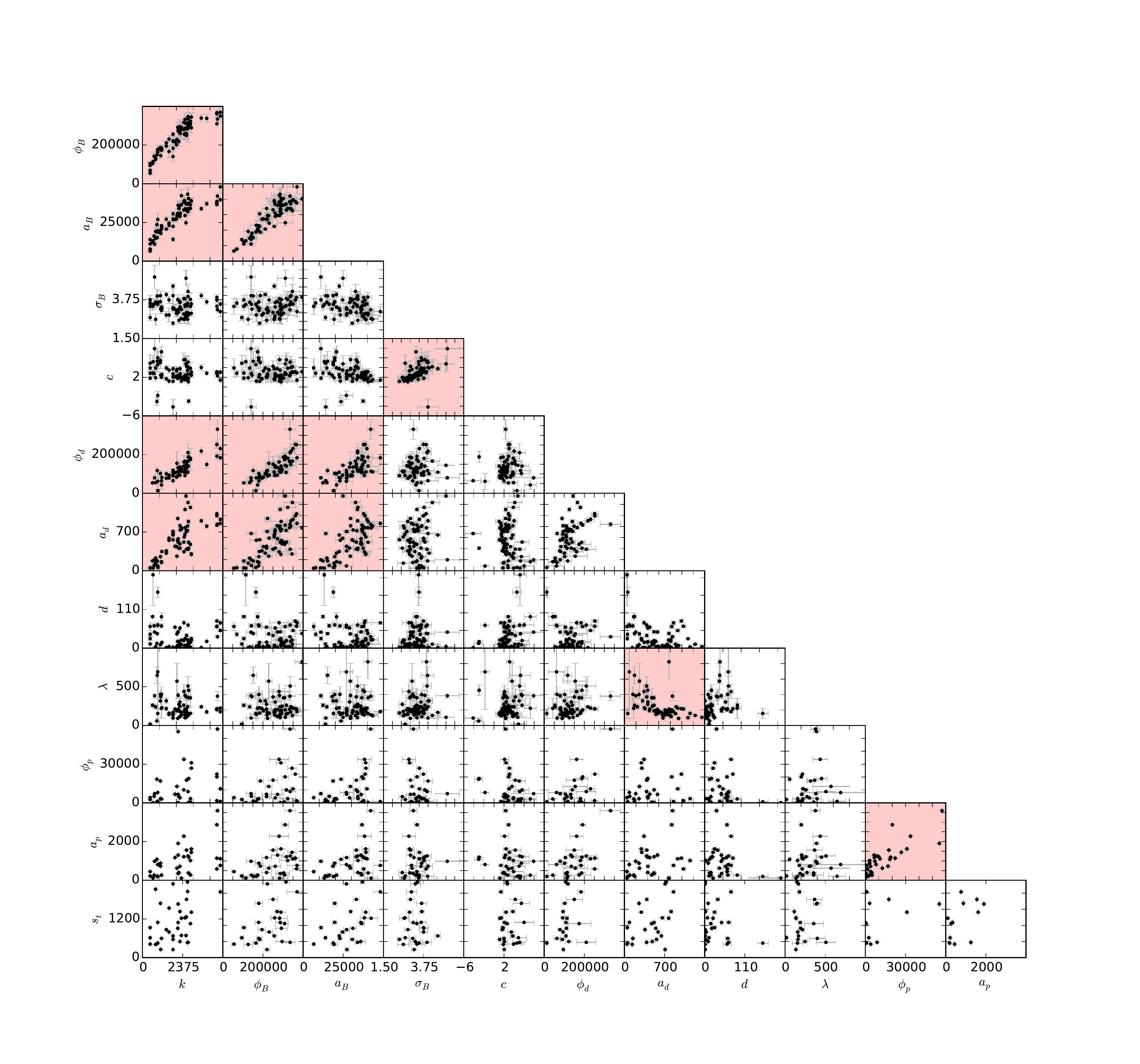}
  \caption{\small Covariance Matrix with a scatter plot of each of the 66 pairings of the 12 Normal Burst parameters listed in section \ref{sec:NormCorr}.  Pairings which show a correlation using the Spearman Rank metric with a significance $\geq5\,\sigma$ are highlighted in red.}
  \label{fig:corr}
\end{figure*}

The full correlation matrix can be found in Figure \ref{fig:corr}, in which these pairs with $>5\,\sigma$ correlations are highlighted.

\bsp	
\label{lastpage}
\end{document}